\documentclass[12pt,notitlepage,a4paper]{article}

\pdfoutput=1

\usepackage{a4wide}
\usepackage{epsfig}
\usepackage{color,graphicx}
\usepackage{cite}
\usepackage{amssymb,amsmath}
\usepackage{jheppub}

\newcommand{\be}{\begin{equation}}
\newcommand{\ee}{\end{equation}}
\newcommand{\bea}{\begin{eqnarray}}
\newcommand{\eea}{\end{eqnarray}}

\usepackage[dvipsnames]{xcolor}

\usepackage{graphics, color,tikz,mathtools, colortbl, xcolor,enumerate, tikz-3dplot, slashed,bm}

\usetikzlibrary{positioning}
\usetikzlibrary{chains}
\usetikzlibrary{arrows, arrows.meta ,fit,decorations.pathreplacing}
\tikzstyle{every picture}+=[remember picture]
\tikzstyle{na} = [baseline]

\usetikzlibrary{arrows, decorations.markings, calc, fadings, decorations.pathreplacing, patterns, decorations.pathmorphing, positioning}
\tikzset{>={Latex[width=1.5mm,length=1.5mm]}}
\usetikzlibrary{graphs,graphs.standard,quotes}

\tikzset{->-/.style={decoration={
  markings,
  mark=at position #1 with {\arrow{>}}},postaction={decorate}}}    
\tikzset{-<-/.style={decoration={
  markings,
  mark=at position #1 with {\arrow{<}}},postaction={decorate}}}  
  
\newcommand{\nn}{\nonumber}
\def\ga{\alpha}
\def\gb{\beta}
\def\gc{\gamma}

\begin{document}

\begin{center}  

\vskip 2cm 

\centerline{\Large {\bf On the 3d compactifications of 5d SCFTs}}
\centerline{\Large {\bf associated with SU(N+1) gauge theories}}

\vskip 1cm

\renewcommand{\thefootnote}{\fnsymbol{footnote}}

   \centerline{
    {\large \bf Matteo Sacchi${}^{a,b}$} \footnote{matteo.sacchi@maths.ox.ac.uk}{\large \bf , Orr Sela${}^{c,d}$} \footnote{osela@physics.ucla.edu} {\large \bf and Gabi Zafrir${}^{a,e,f}$} \footnote{gzafrir@scgp.stonybrook.edu}}
      
\vspace{1cm}
\centerline{{\it ${}^a$ Dipartimento di Fisica, Universit\`a di Milano-Bicocca \& INFN, Sezione di Milano-Bicocca,}}
\centerline{{\it I-20126 Milano, Italy}}
\centerline{{\it ${}^b$ Mathematical Institute, University of Oxford, Andrew-Wiles Building, Woodstock Road,}}
\centerline{{\it Oxford, OX2 6GG, United Kingdom}}
\centerline{{\it ${}^c$ Department of Physics, Technion, Haifa, 32000, Israel}}
\centerline{{\it ${}^d$ Mani L. Bhaumik Institute for Theoretical Physics, Department of Physics and Astronomy,}}
\centerline{{\it University of California, Los Angeles, CA 90095, USA}}
\centerline{{\it ${}^e$ C.~N.~Yang Institute for Theoretical Physics,  Stony Brook University, Stony Brook,}}
\centerline{{\it NY 11794-3840, USA}}
\centerline{{\it ${}^f$ Simons Center for Geometry and Physics, Stony Brook University, Stony Brook,}}
\centerline{{\it NY 11794-3840, USA}}
\vspace{1cm}

\end{center}

\vskip 0.3 cm

\setcounter{footnote}{0}
\renewcommand{\thefootnote}{\arabic{footnote}}   
   
\begin{abstract}

We study the $3d$ $\mathcal{N}=2$ theories resulting from the compactification of a family of $5d$ SCFTs on a torus with flux in the global symmetry. The family of $5d$ SCFTs used in the analysis is the one that UV completes the $5d$ $SU(N+1)$ gauge theories with Chern--Simons level $k$ and $N_f$ fundamental hypermultiplets, generalizing the previous investigation of the torus compactifications of the rank 1 Seiberg $E_{N_f+1}$ SCFT (which is the $N=1$ member of the family). This construction systematically yields three-dimensional theories presenting highly non-trivial non-perturbative phenomena such as infra-red dualities and enhanced symmetries, which we check using various methods.  

\end{abstract}
 
 \newpage
 
\tableofcontents

\section{Introduction}

The study of non-perturbative aspects of quantum field theories has been one of the most active field of research in theoretical physics for the last decades. Among these, of particular interest have been infra-red (IR) dualities and symmetry enhancements. The former occurs when two different quantum field theories flow to the same fixed point in the IR, while the latter refers to the situation in which the manifest symmetry of the microscopic theory gets enlarged to a bigger group once we flow at low energies. By now many such examples have been discussed in the literature, especially in three dimensions which is the case we will consider in this paper. This is due to the fact that the gauge coupling is dimensionful in $3d$, hence every gauge theory (with a sufficient number of flavors) is expected to flow to an interacting CFT in the IR which may exhibit several of such interesting non-perturbative effects.

In order to make some progress in the study of these phenomena, it is useful to consider theories which possess additional symmetries, such as supersymmetry. In particular, we will be mainly interested in $3d$ theories with $\mathcal{N}=2$ supersymmetry. This allows us to compute several quantities \emph{exactly}, that is including non-perturbative corrections, and which are invariant under the renormalization group (RG) flow. As such, they are extremely powerful tools since we can compute them in the ultra-violet (UV) where the theory is weakly coupled to extract information about the interacting theory in the IR, which is in general non-Lagrangian. Most notably, when the theory possesses at least four supercharges we can compute exactly its partition function on various compact manifolds using localization techniques (see \cite{Pestun:2016zxk} for a review and references therein). Some examples that will be relevant for us are the partition functions on $S^3$ and on $S^2\times S^1$, also known as the supersymmetric index. From the $S^3$ partition function one can also extract useful information about the low energy SCFT such as its central charges, that is the coefficients in two-point functions between conserved currents, and the superconformal R-charges, which in turn allow us to determine the dimensions of various protected operators.

Given the tremendous success in the study of dualities and symmetry enhancements of three-dimensional supersymmetric quantum fields theories, it is natural to wonder whether there is some general organizing principle behind them. One possible approach to this type of questions has been recently proposed in \cite{Sacchi:2021afk} following a similar idea that has proven to be very successful in the context of $4d$ $\mathcal{N}=1$ theories. Specifically, we can try to construct $3d$ $\mathcal{N}=2$ theories by compactifying $5d$ $\mathcal{N}=1$ SCFTs on Riemann surfaces with fluxes for their global symmetry through the surface. The $3d$ theory obtained in this way is typically a non-Lagrangian SCFT, but we can try to find a Lagrangian in $3d$ that at low energies flows to such SCFT\footnote{More generally, it may be that the $3d$ SCFT obtained from the $5d$ compactification sits on a point of a non-trivial conformal manifold $\mathcal{M}_c$ and that one is able to find a $3d$ UV Lagrangian that doesn't flow exactly to the same SCFT, but to another one which still lives on a different corner of $\mathcal{M}_c$.}. This phenomenon is sometimes referred to as \emph{across dimensional duality}. The power of this strategy relies in the fact that it allows us to predict symmetry enhancements and dualities for the $3d$ Lagrangian theories starting from known properties of the $5d$ SCFTs and from geometric considerations related to their compactifications. For example, the $5d$ global symmetry is broken to the subgroup that is preserved by the flux, but this subgroup may not be fully manifest in the $3d$ Lagrangian which means that it must get enhanced at low energies. Moreover, different fluxes to which we would associate distinct $3d$ Lagrangians may be related by an element of the Weyl group of the $5d$ global symmetry, meaning that they are actually equivalent and that they lead to the same SCFT in $3d$. Hence, the two different looking Lagrangians flow to this same SCFT in the IR, that is they are dual. This approach has been successfully applied to the rank 1 $E_{N_f+1}$ Seiberg SCFTs \cite{Seiberg:1996bd} with $N_f\le 6$\footnote{The case of the compactification of the $E_8$ SCFT is not understood yet.} in \cite{Sacchi:2021afk}, where many three-dimensional models with symmetry enhancements or related by dualities have been found. Among these, it was in particular possible to recover one instance of Aharony duality \cite{Aharony:1997gp}.

In this paper we extend this analysis to the study of compactifications of other known $5d$ SCFTs, which can be considered as a higher rank generalization of the findings of \cite{Sacchi:2021afk}. The rank 1 $E_{N_f+1}$ SCFTs for $N_f\le 7$ can be obtained by starting from the $6d$ rank 1 E-string theory \cite{Ganor:1996mu,Seiberg:1996vs}, compactifying it on a circle to obtain a $5d$ SCFT with $E_8$ global symmetry and then performing mass deformations that lead to lower values of $N_f$. One can then try to consider higher rank generalizations of this story. One possibility, which we will not pursue here and we will leave for future work, is to consider the rank $N$ E-string theory in $6d$. Another higher rank generalization can be found by considering the rank 1 E-string theory as the $N=1$ case of the family of the $(D_{N+3},D_{N+3})$ conformal matter theories \cite{Heckman:2015bfa,DelZotto:2014hpa}. Here we will consider the compactification of the series of $5d$ SCFTs obtained from the $(D_{N+3},D_{N+3})$ conformal matter theories on Riemann surfaces that are tubes and tori and for specific values of the flux.

The first step is to understand the compactifications on tubes. For these we will make a conjecture that is based on the knowledge of some gauge theory phases of the $5d$ SCFTs, as will be explained. Indeed, in five dimensions a gauge theory is always IR free, but it may be UV completed by an interacting SCFT. From this point of view, the gauge theory can be obtained as a specific mass deformation of the SCFT and different deformations may lead to distinct gauge theory phases of the same SCFT. For example, the $5d$ SCFTs we will be interested in always have a gauge theory description in terms of an $SU(N+1)_k$ gauge theory with $N_f$ hypermultiplets in the fundamental representation \cite{Hayashi:2015fsa}, where $k$ is the Chern--Simons (CS) level. From now on, we will use the notation $SU(N+1)_k+N_fF$ for brevity. Having understood tube compactifications we can then construct tori by gluing them together, which in field theory amounts to the operation of gauging a diagonal combination of the global symmetries carried by the punctures that we are connecting, plus possibly adding CS levels and monopole superpotentials.

We will then validate this conjecture by performing several tests. As we mentioned, our approach is very similar to the one that have been intensively used in the context of $4d$ $\mathcal{N}=1$ theories \cite{Benini:2009mz,Bah:2012dg,Gaiotto:2015usa,Razamat:2016dpl,Ohmori:2015pua,Ohmori:2015pia,Kim:2017toz,Bah:2017gph,Kim:2018bpg,Razamat:2018gro,Kim:2018lfo,Ohmori:2018ona,Chen:2019njf,Razamat:2019mdt,Pasquetti:2019hxf,Razamat:2019ukg,Razamat:2020bix,Sabag:2020elc,Hwang:2021xyw}. In that case, the four-dimensional theories are obtained by compactifications of $6d$ $\mathcal{N}=(1,0)$ SCFTs on Riemann surfaces with fluxes. Because of the similarity of the two set-ups, we will often draw many analogies between them and introduce concepts that will be useful in our $5d$ to $3d$ reductions starting from similar ones that have proven to be very useful in the study of $6d$ to $4d$ reductions. For example, the compactification of the $(D_{N+3},D_{N+3})$ conformal matter theories to $4d$ have been studied in \cite{Kim:2018bpg} and we will make use of their results to better understand the compactifications of the $5d$ SCFTs that we will consider. Moreover, several of the consistency checks that can be performed in the compactification of $6d$ theories also apply to our case and we shall now briefly summarize those that we will use.

First of all, we have mentioned that the $3d$ theories should possess the global symmetry that is preserved by the flux, up to possible additional accidental enhancements, and that this can be enhanced in the IR. The symmetry enhancement can be checked by computing the superconformal index of the theory and its central charges, which can be extracted from the $S^3$ partition function. Moreover, the operators of the $5d$ theory, such as the stress-energy tensor, the conserved currents and other Higgs branch chiral ring operators, are expected to lead to operators of the $3d$ theory and we can check their presence again by means of the superconformal index. Finally, from the $5d$ point of view we can predict the dimension of the conformal manifold and checking that this is the same as for our $3d$ models constitutes an additional consistency check.

There is another test that is possible to perform only in the compactification of $5d$ SCFTs and not of $6d$ SCFTs. Indeed, the $5d$ theories are related by various mass deformations which in the gauge theory description imply that some flavors are integrated out. In our case of $5d$ SCFTs that UV complete the $SU(N+1)_k+N_fF$ gauge theories, this amounts to lowering the value of $N_f$ and shifting the CS level $k$, while leaving the rank $N$ fixed. Checking that our $3d$ models can be connected by real mass deformations, reconstructing a pattern of flows that is identical to the $5d$ one, will provide additional evidence that they are the proper compactifications of the $5d$ SCFTs. We will also work out the precise relation between the $5d$ and $3d$ global symmetries and the corresponding mass deformations.

There are also other aspects that appear only in the $5d$ to $3d$ set-up and which are not present when studying four-dimensional theories coming from $6d$ SCFTs. One of these is the possibility of a Chern--Simons interaction for the gauge fields in three dimensions. In \cite{Sacchi:2021afk} it was observed that, even if the gauge theory description of the $5d$ SCFT doesn't have any CS interaction, it is in some cases necessary to turn on a non-trivial CS level for the gauge nodes of the resulting $3d$ quiver gauge theory in order for it to pass all the consistency checks. This is also required in order for the theory to be gauge invariant at the quantum level. Another peculiar feature of the three-dimensional case is the presence of monopole operators. In most of the models obtained from $5d$ compactifications it is necessary to turn on non-trivial monopole superpotentials in order for the theory to have the expected global symmetry\footnote{A similar phenomenon occurs in the compactifications of $4d$ $\mathcal{N}=1$ theories on a circle to $3d$ $\mathcal{N}=2$ \cite{Aharony:2013dha,Aharony:2013kma}. Here a monopole superpotential is dynamically generated leading to the breaking of some abelian symmetries that were anomalous in $4d$.}. Indeed, monopole superpotentials typically break some abelian symmetry, which in some cases would prevent the enhancement to the global symmetry expected from $5d$. 

The structure of the rest of the paper is as follows. In section \ref{5dto3dreview} we give a review of the general aspects of the compactification of $5d$ SCFTs to $3d$ $\mathcal{N}=2$ theories, drawing again many analogies with the six-dimensional case. In section \ref{5dproperties} we review some aspects of the $5d$ SCFTs we are interested in that will be useful for us. In section \ref{compactifications} we finally study the $3d$ models that are obtained by the compactification of the $5d$ SCFTs. We will present models with alternating $SU$ and $USp$ gauge groups and models with $SU$ gauge groups only, which are obtained by exploiting different gauge theory descriptions of the $5d$ SCFTs. In section \ref{additional} we consider gluing tubes of different types and in section \ref{conc} we conclude with some final remarks.

\section{Review of the compactification of $5d$ SCFTs to $3d$ $\mathcal{N}=2$ theories}
\label{5dto3dreview}

In this section we shall review the methods used to formulate and test conjectures for the compactification of $5d$ SCFTs on tori to get $3d$ $\mathcal{N}=2$ theories. This is based on the discussion on this subject in \cite{Sacchi:2021afk}, and also on similar ideas used to tackle the compactification of $6d$ SCFTs on tori to get $4d$ $\mathcal{N}=1$ theories \cite{Kim:2017toz,Kim:2018bpg,Kim:2018lfo}. We refer the reader to these references for more details.

\subsection{Conjecturing the $3d$ models}
\label{conjectures}

Our interest is in the compactification of $5d$ SCFTs on a torus with flux in the global symmetry. As this surface is flat, such compactifications are expected to preserve $3d$ $\mathcal{N}=4$ SUSY if the flux is not present. With the flux, only four supercharges are expected to be preserved, leading to $\mathcal{N}=2$ SUSY in $3d$. In order to systematically find the corresponding $3d$ UV Lagrangian theories that flow to the same IR fixed points as the $5d$ SCFTs on a torus, the first step in such constructions is to investigate the $3d$ model corresponding to a tube with flux. Once found, this tube theory can serve as a basic building block used to construct the theories associated with tori by gluing several tubes together. In the $3d$ field theory description, such gluing includes gauging a global symmetry, and may involve Chern--Simons terms and monopole superpotentials, as mentioned above. Notice that at the boundaries of the tube we will need to enforce boundary conditions, which in light of our discussion will be taken to preserve $\mathcal{N}=2$ SUSY in $3d$.

Our starting point in understanding torus compactifications with flux is therefore the study of tube compactifications. The general strategy used to tackle this, which we will explain in more detail below, is to first compactify on the circle to $4d$ and then reduce the resulting theory on the interval to $3d$. In the first compactification to $4d$, the flux in the global symmetry of the $5d$ SCFT is represented by a holonomy around the circle which is taken to be variable along the interval (forming a domain wall on the interval), such that at different places along the segment the resulting $4d$ gauge theory (obtained by compactifying the $5d$ SCFT on the circle with a non-trivial holonomy) might be different. In this way, we have in $4d$ two gauge theories separated by a domain wall on a compact line\footnote{To be precise, what we are realizing is a non-dynamical interface separating the two $4d$ theories. Nevertheless, with a little abuse of terminology we will still refer to this as a domain wall, since this term has become standard in the literature on compactification of six-dimensional theories.}. At this point, the second reduction to $3d$ can be done after specifying the boundary conditions at the ends of the line ({\it i.e.} the punctures of the tube), assuming we understand the behavior at the domain wall. The latter is usually filled by conjecture and by relying on the understanding gained in the study of similar cases, notably the compactifications of $6d$ SCFTs. Let us next turn to describe these steps in more detail.

Before we discuss the compactification process, we would first like to review some aspects of five dimensional SCFTs and gauge theories. We shall begin by discussing $5d$ gauge theories, which will play a prominent role here. The latter are non-renormalizable, and so naively do not correspond to microscopic theories. Nevertheless, many $5d$ gauge theories can be UV completed by either $5d$ or $6d$ SCFTs. In the former case the deformation leading to the $5d$ gauge theory is a mass deformation, while in the latter case it is a circle compactification, usually in the presence of flavor holonomies. These interesting deformation properties of $5d$ and $6d$ SCFTs play a prominent role in the study of their compactification.

To illustrate this, it is useful to consider an example, which we take to be the $5d$ gauge theory with gauge group $SU(2)$ and $N_f$ doublet hypermultiplets. For $N_f\leq 7$, this gauge theory can be UV completed by the $5d$ SCFTs known as the Seiberg $E_{N_f+1}$ theories \cite{Seiberg:1996bd}. These are $5d$ SCFTs with $E_{N_f+1}$ global symmetry. The deformation leading to the gauge theory here is a mass deformation breaking the $E_{N_f+1}$ global symmetry to $U(1)\times SO(2N_f)$, with the $SO(2N_f)$ part rotating the $N_f$ doublet hypers and the $U(1)$ being the topological symmetry associated with the instantons of the $SU(2)$ gauge group. When $N_f=8$, the gauge theory can be UV completed by a $6d$ $(1,0)$ SCFT known as the E-string theory \cite{Ganor:1996pc}. Here, the deformation leading to the $5d$ gauge theory is a circle compactification in the presence of a tuned holonomy. Cases with $N_f>8$ are believed not to possess a field theory UV completion. We can flow from one $SU(2)$ gauge theory with $N_f$ flavors to another one with less flavors by giving a mass to some of the flavors. This relation between the gauge theories implies a similar relation between their SCFT UV completions
\begin{equation}
6d \; \text{E-string}\rightarrow 5d \; E_8 \rightarrow 5d \; E_7 \rightarrow 5d \; E_6 \rightarrow ...
\end{equation} 
with the first arrow corresponding to circle compactification and the remaining ones corresponding to mass deformations. 

The structure we noted in the case of the $SU(2)$ gauge theory is quite generic and is present in many other cases. Here, we shall be mostly interested in higher rank generalizations of this class of theories. There are two notable types. One is the higher rank $E_{N_f+1}$ theories. These are the $5d$ SCFT UV completions of the $5d$ gauge theories with gauge group $USp(2N)$, an antisymmetric hypermultiplet and $N_f$ fundamental hypermultiplets. These behave similarly to the rank $1$ case, with the $N_f\leq 7$ cases UV completed by $5d$ SCFTs, while the $N_f=8$ case UV completed by a $6d$ SCFT, the rank $N$ E-string SCFT. Another interesting generalization is to the $5d$ gauge theories with gauge group $SU(N+1)$, $N_f$ fundamental hypermultiplets and a Chern--Simons term of level $k$. As here we have both the number of flavors and the Chern--Simons level as parameters, the space of $5d$ SCFTs in this case is more complicated. The case of $N_f=2N+6$ and $k=0$ is UV completed by a $6d$ SCFT, the $(D_{N+3},D_{N+3})$ minimal conformal matter theory \cite{Hayashi:2015fsa}. The cases with $N_f+2|k|<2N+6$ are UV completed by $5d$ SCFTs, and can be generated by integrating out flavors from the $6d$ lifting case. We shall say more about these theories in the next section.

The existence of these gauge theory deformations of what are otherwise strongly coupled SCFTs is quite useful in the study of the tube (and by gluing also torus) compactifications of these SCFTs. Specifically, recalling the first step in the compactification process we mentioned above, we should first compactify the SCFT on a circle with flavor holonomies (corresponding to the flux) such that we obtain a gauge theory in one lower dimension. For $6d$ SCFTs we use the fact that when compactified with a suitable holonomy, they lead to $5d$ gauge theories. The cases of $5d$ SCFTs are somewhat more involved as the mass deformations leading to gauge theories are now in $5d$. However, here we recall that in theories with eight supercharges, mass deformations are equivalent to a vev for a scalar in a background vector multiplet. When compactified on a circle, the component of the gauge field in such a multiplet along the circle becomes an additional scalar. This corresponds to the fact that holonomies are expected to become additional mass deformations, and in the specific case of compactifications to $4d$, both types of deformations (associated with the scalar and gauge field in the background vector multiplet) combine to form a single complex one in $4d$. As a result, we expect the $5d$ mass deformation and the one related to the holonomy along the circle to be equivalent, and so we again end up with a gauge theory in $4d$. 

Let us next discuss how these holonomies are chosen, and how they are related to the flux. For this, we shall need one more detail regarding the relationship between $5d$ gauge theories and SCFTs in five or six dimensions. The interesting point is that there can be more than one gauge theory deformation for a given $5d$/$6d$ SCFT. Specifically, there could be many different mass deformations of a given $5d$ SCFT leading to different $5d$ gauge theories or potentially to the same $5d$ gauge theories. In the case of $6d$ SCFTs, a similar thing happens, but with different flavor holonomies leading to the same or different $5d$ gauge theories.

We can then consider the following configuration. We consider the compactification of the higher dimensional theory, be it a $5d$ or $6d$ SCFT, on a tube. As discussed above, we first reduce on the circle where we include a variable holonomy, that is a holonomy that depends on the position on the line (the remaining direction of the tube). Explicitly, we take the holonomy to have the profile of a step function, {\it i.e.} it shall have one value at one end and another value at the other end, forming a domain wall. The holonomy is taken to be constant along all points on the line with the exception of one point where it jumps between the two values. There are next two interesting observations regarding this configuration. First, we note that the presence of a variable holonomy implies that there is a non-trivial flux supported on the surface, here the tube. Second, we can take the holonomies on the two sides of the line to be such that the theory flows to a gauge theory on both sides, as we previously outlined. We then expect to get the two gauge theories on the two sides of the line, separated by a domain wall which exists at the point where the holonomy jumps.

We see that when reducing in this way the $d$ dimensional theory on the circle of the tube in the presence of flux, we get two gauge theories separated by a domain wall living on a $(d-2)$-dimensional spacetime times a line. We note here that the flux is related to the jump in the holonomy and as such to the difference between the two gauge theories. Since not all deformations lead to gauge theories, the values of flux for which this applies might be limited.

Now that we have compactified on the circle in the presence of flux and got gauge theories in one lower dimension, we need to consider the boundaries of the line. As we mentioned, our focus is in the compactification on a tube, which is a sphere with two punctures. Here, the punctures are represented by the two boundaries where the line ends. On these boundaries we need to give boundary conditions, and these keep track of the type of punctures inserted. As we are interested in preserving supersymmetry, we shall only consider boundary conditions preserving four supercharges. More specifically, we will be interested in a special type of puncture, which generalizes the notion of a maximal puncture in theories of class $\mathcal{S}$ \cite{Gaiotto:2009we}. To define it, let us for concreteness consider the compactification of a $5d$ SCFT and use the description after the reduction to $4d$ with a variable holonomy such that we get $4d$ gauge theories. In that frame, close to the boundary, the theories are described by Lagrangians consisting of $4d$ $\mathcal{N}=2$ vector multiplets and hypermultiplets. Boundary conditions preserving $3d$ $\mathcal{N}=2$ supersymmetry can then be achieved by decomposing these multiplets near the boundary to representations of $3d$ $\mathcal{N}=2$ supersymmetry, and designate Dirichlet and Neumann boundary conditions to them. For the type of punctures we would be interested in here, the choice of boundary conditions is to give Dirichlet boundary conditions to the $3d$ $\mathcal{N}=2$ vector multiplet and Neumann boundary conditions to the adjoint chiral in the $4d$ $\mathcal{N}=2$ vector multiplet. In the case of the hypermultiplets, we break them to two ($3d$ $\mathcal{N}=2$) chiral fields with opposite charges, and give Dirichlet boundary conditions to one and Neumann to the other\footnote{Here we have a choice for which chiral multiplet receives which boundary condition. This choice exists for every hyper, and different choices lead to slightly different punctures, differing by the charges of the surviving chiral fields. This distinction is usually refereed to as the sign or color of the puncture.}.  

At this point, we can easily determine the remaining reduction on the line if the behavior at the domain wall is understood, since the theory is Lagrangian everywhere except potentially at the location of the wall. There are two possible sources for the matter content we expect to obtain after this reduction. The first is the bulk matter, where only fields with Neumann boundary conditions at both the corresponding end of the line and the domain wall will survive the reduction. Note that there are two such bulk pieces in the basic tube compactification that we are considering, corresponding to the two sides of the domain wall. The second source is the fields living on the domain wall, which may interact both among themselves and with the bulk fields. The main problem then in determining the resulting lower dimensional theory is understanding the domain wall theory and the behavior (that is, boundary conditions) of the bulk fields at this wall. 

This problem is tackled in various ways that were originally used in the study of the compactification of $6d$ theories to $4d$. One option is to rely on cases where the domain walls are relatively well understood. Another option is to try to conjecture the fields living on the domain walls, and then test the resulting theories. Here we shall use both methods. We will give more details on this for the specific cases we will be interested in later in section \ref{compactifications}, while here we will review the case considered in \cite{Sacchi:2021afk}.



In \cite{Sacchi:2021afk} the compactification of the $5d$ rank 1 Seiberg $E_{N_f+1}$ SCFTs was considered.
As discussed above, this set of theories UV completes the $5d$ $SU(2)$ gauge theories with $N_f$ flavors for $N_f\leq 7$. Moreover, upon compactifying these SCFTs on a circle to $4d$ with a suitable holonomy, one can get the analogous $4d$ theories, {\it i.e.} $\mathcal{N}=2$ $SU(2)$ gauge theories with $N_f$ hypermultiplets. In fact, there is more than one holonomy which results in this same gauge theory, and we will use it in the compactification procedure in the following way. Recall that the first step in the compactification is to reduce the SCFT to $4d$ with a variable holonomy (forming a domain wall) which represents the flux in the global symmetry. Choosing both of the holonomies on the two sides of the $4d$ compact direction to be of the kind corresponding to an $SU(2)$ gauge theory with $N_f$ hypers, results in $4d$ in two copies of this theory separated by a domain wall. Now, in the next step we need to specify boundary conditions at the ends of the line (the $4d$ compact direction). As outlined above, we choose them to be Dirichlet for all the vector fields and for one chiral multiplet inside each hypermultiplet, and Neumann for the adjoint chiral and for the other chiral multiplet inside each hypermultiplet. These boundary conditions preserve half of the $4d$ supersymmetry, resulting in $\mathcal{N}=2$ in $3d$. The final ingredient that we need to address in $4d$ is the domain wall. Assuming that it behaves similarly to the well-studied ones that appear in $6d$ compactifications to $4d$, specifically that of the $6d$ E-string SCFT \cite{Kim:2017toz}, it assigns Dirichlet boundary conditions for the adjoint chiral field and boundary conditions for the chirals in the hypermultiplets which are the same as the ones they have at the ends of the line. In addition, the fields living on the domain wall are several chiral multiplets that interact with each other and with the bulk fields through a cubic superpotential. 

\begin{figure}
	\center
	\includegraphics[width=0.65\textwidth]{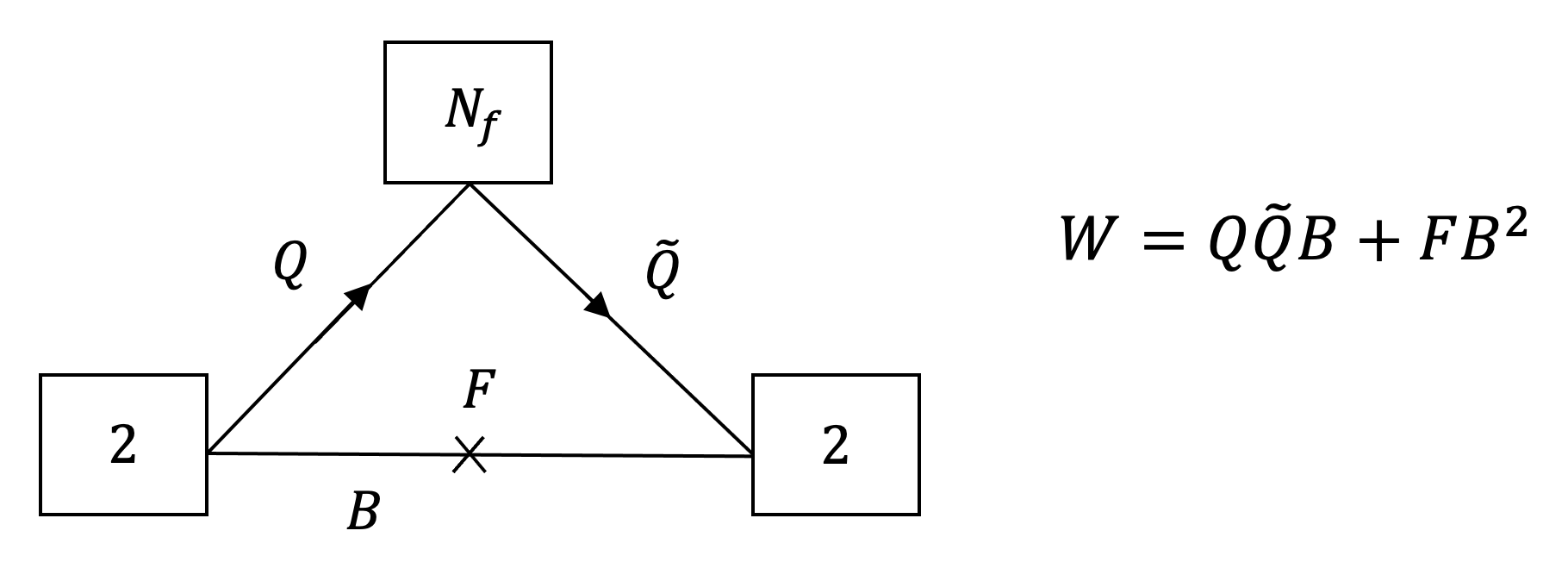} 
	\caption{The $3d$ model corresponding to the compactification of the $5d$ Seiberg $E_{N_f+1}$ SCFT on a tube with two maximal punctures and flux $\ensuremath{(\frac{1}{4}\sqrt{8-N_{f}};\frac{1}{4},\frac{1}{4},...,\frac{1}{4})}$, where the flux is specified using the $U(1)\times SO(2N_f)$ subgroup of $E_{N_f+1}$ with the first entry denoting the $U(1)$ flux. As customary, squares denote $SU$ type global symmetries, and lines connecting them denote bifundamental chiral fields under the connected groups. We also use crosses over fields to denote flipping, that is that there is an additional chiral field linearly coupled to the invariant made from the field being flipped under the non-abelian global symmetries. Finally, $W$ denotes the superpotentials, with the last term being the flipping one.}
	\label{BTube}
\end{figure}

Collecting all these pieces of information together, we can perform the remaining reduction to $3d$ where we are left with the following matter content. Among all the $4d$ bulk fields, only the chirals in the hypermultiplets which receive Neumann boundary conditions (at both the corresponding end of the line and the domain wall) survive in the $3d$ limit. Note that such chirals come from both sides of the domain wall. In addition to them, we also have the chirals living on the wall, along with the superpotential interaction mentioned before. All in all, we obtain the $3d$ model depicted in figure \ref{BTube}. 

Let us emphasize again the origin of the different ingredients appearing in figure \ref{BTube}. First, as we are assigning Dirichlet boundary conditions for the $4d$ gauge fields, the two $SU(2)$ gauge groups from the two sides of the domain wall become global symmetries, as appears in the figure. In addition, the $SU(N_f)$ symmetry is just the part of the global symmetry of the two $4d$ gauge theories preserved by the boundary conditions and the superpotential. Lastly, the $Q$ and $\tilde{Q}$ fields are the chirals inside the $4d$ hypermultiplets receiving Neumann boundary conditions (each from a different side of the domain wall), and the $B$ and $F$ fields are the chirals coming from the domain wall. 

The choices and assumptions we mentioned in the construction of the tube model of figure \ref{BTube} turn out to correspond to the flux $\ensuremath{(\frac{1}{4}\sqrt{8-N_{f}};\frac{1}{4},\frac{1}{4},...,\frac{1}{4})}$, where the various slots represent fluxes in the subgroup $U(1)\times SO(2)^{N_{f}}\subset U(1)\times SO(2N_{f})$ of $E_{N_f+1}$. This specific assignment of flux can be motivated by performing several tests, on which we will elaborate more in the next subsection. 

Once the basic tube model associated with compactifying a higher dimensional theory is found, it can be used to construct more general tube models and theories corresponding to torus compactifications. For example, gluing together several copies of the same tube model results in a new tube model corresponding to a higher value of flux. In general, when gluing tube models together we should sum the fluxes of the individual tubes in order to get the total flux on the resulting surface. Note that from the perspective of the lower-dimensional tube model, such a gluing of compactification surfaces at a common boundary of the tubes means undoing the boundary conditions we used in the construction of the individual models. In particular, this includes gauging the global symmetry associated with the glued tube boundaries and restoring the chirals which were given Dirichlet boundary conditions at the punctures that we are gluing. In the case of $3d$ theories resulting from the compactifications of $5d$ Seiberg $E_{N_f+1}$ SCFTs, such gauging was found to also involve Chern--Simons terms and monopole superpotentials for gauge groups which are adjacent in the quiver description of the theory.

\begin{figure}
	\center
	\includegraphics[width=0.75\textwidth]{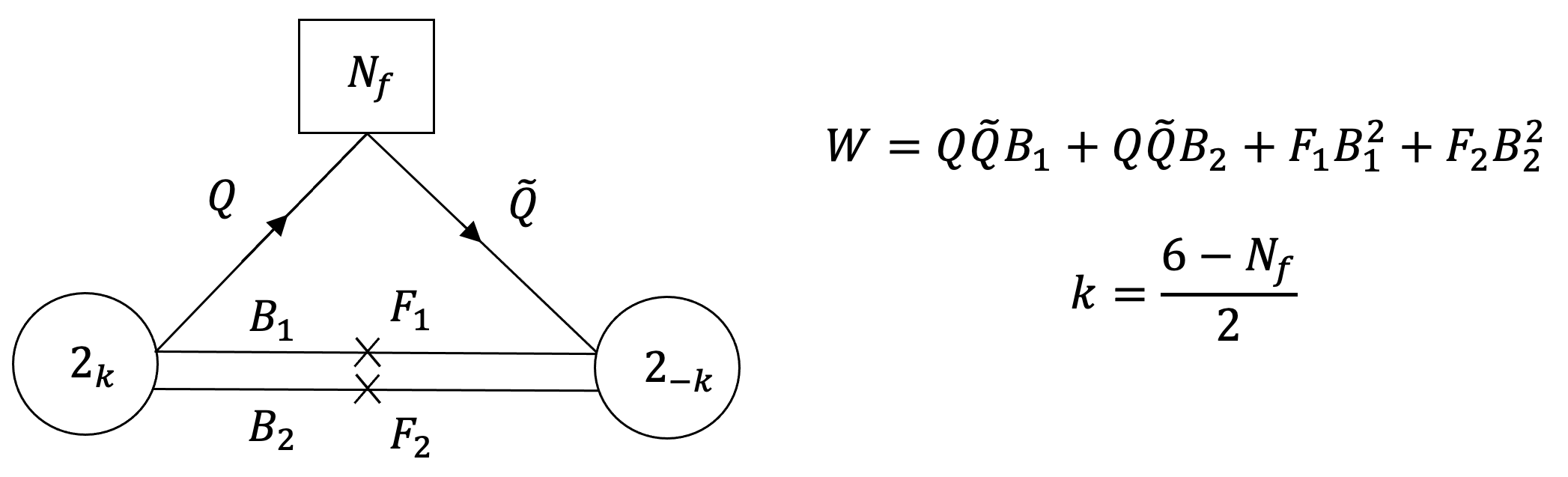} 
	\caption{The $3d$ theories associated with the compactification of the $5d$ $E_{N_f+1}$ SCFTs on a torus with flux $\frac{1}{2}(\sqrt{8-N_f};1,1,...,1)$, which is twice the flux of each of the constituent tube models. The theories also contain various superpotential interactions. Notably, there are cubic superpotentials running along the triangle as well as the flipping superpotentials. As is usual, circles denote $SU$ gauge groups (unless otherwise specified), and we use subscripts to denote the level of the corresponding Chern-Simons term. In cases where there is no such Chern-Simons term, we will simply omit the subscript.}
	\label{TorusSei}
\end{figure}

Let us note that in addition to gluing copies of the same tube model to form a tube theory with a higher flux (in the same subgroup of the global symmetry), one can construct more general tube theories by gluing tube models with a relative Weyl operation on the flux between them. Even though two tube models related by such an operation are equivalent to each other, gluing them together results in a new theory and flux. The way a certain Weyl element operates on the fields of the lower-dimensional theory is usually easy to understand and amounts to a simple action on them (such as swapping them), and gluing tube theories with such relative operations between them yields the lower-dimensional model associated with the new tube. We will present some examples of this in subsubsection \ref{moregeneral} and in section \ref{additional}.

Once the basic and general tube models are found, they can be used to construct theories corresponding to tori with various fluxes by gluing. An example for such a torus theory obtained by gluing (twice) two copies of the basic tube model of figure \ref{BTube} is presented in figure \ref{TorusSei}. In this way, we can systematically conjecture lower-dimensional models resulting from compactifying higher-dimensional theories on a surface. Our focus in this paper is in compactifications of $5d$ SCFTs to $3d$, and we will next turn to discuss how such conjectures can be tested.

\subsection{Tests}
\label{tests}

Once we arrive at our conjectured $3d$ theory, we need to test it and check whether its behavior is indeed consistent with it being the result of the compactification of the $5d$ SCFT. The main tests we shall employ are the identification of various operators expected from the $5d$ conserved current and energy-momentum tensor multiplets and the general consistency of the construction. Specifically, consider the compactification of the $5d$ SCFT on a Riemann surface to $3d$ with flux in its global symmetry. Various properties of the compactification imply various properties of the $3d$ theory. For instance, the $5d$ SCFT in general possesses some global symmetry, part of which might be broken by the flux, but the preserved part is expected to be a global symmetry of the resulting $3d$ theory.

In essence, in this example we have used the existence of a special class of operators in the $5d$ SCFT, those associated with conserved current multiplets, to say something about the properties of the $3d$ theory. This idea can be further extended to extract more detailed information about the operator spectrum of the $3d$ theory expected from the existence of special operators in the $5d$ SCFT. The specific information we would be interested in is the contribution to the $3d$ superconformal index of the $3d$ operators expected to descend from these $5d$ operators. The remarkable observation is that while the exact number of operators that descend from a given $5d$ operator may vary depending on geometric properties of the surface, their contribution to the $3d$ superconformal index, which counts the number of operators up to the possibility of merging into long multiplets, depends solely on topological properties of the surface. Unsurprisingly, this is related to the index of certain differential operators on the surface, which again counts the difference between two integers.

Here we shall simply state the result that we shall use. For the derivation and in-depth discussion, we refer the reader to \cite{BRZtoapp}, and for more applications of this formula for the study of the compactifications of $6d$ and $5d$ SCFTs on Riemann surfaces, we refer the reader to \cite{Kim:2017toz,Kim:2018bpg,Sacchi:2021afk}. The specific $5d$ operators that we would be interested in are those associated with conserved quantities, notably the conserved flavor current and energy-momentum tensor multiplets. This is as these are ubiquitous in $5d$ SCFTs.

Say we have a $5d$ SCFT with a global symmetry $G$, and we consider compactifying it on a Riemann surface of genus $g$ without punctures. We further introduce a flux $F$ supported on the Riemann surface, which for simplicity we shall take to be in a single $U(1)$ subgroup of $G$ (the generalization to cases of flux in multiple $U(1)$ subgroups being straightforward). This flux should break $G$ to $U(1)\times H$, under which we should have that the adjoint of $G$ decomposes as ${\bf Ad}_G\rightarrow \sum {\bf R}^{q_i}_i$, with ${\bf R}^{q_i}_i$ standing for the representation of $H$ of dimension ${\bf R}$ with $U(1)$ charge of $q_i$\footnote{\label{repconv}Throughout the paper, we use interchangeably two notations for representations of groups, depending on the situation. Sometimes, especially for groups of a definite rank, we use bold numbers to denote the representations of such dimensions. We additionally put bars for complex conjugate representations and primes to distinguish spinor representations of the same dimension, \emph{i.e.} $\bf 32$ and $\bf 32'$ denote the two spinor representations of opposite chirality of $SO(12)$. Another notation that we particularly use when dealing with groups of generic ranks, like in the current case, consists of denoting some important representations with bold letters that immediately recall them. For example, ${\bf Ad}$ stands for the adjoint representation, ${\bf \Lambda}^k$ for the rank $k$ totally antisymmetric representation, $\bf F$ ($\bf \bar{F}$) for the (anti-)fundamental representation of $SU$ groups, $\bf V$, $\bf S$ and $\bf C$ for the vector and two spinor representations of $SO$ groups.}. 

Additionally, the $5d$ SCFT has an $SU(2)_R$ symmetry. Its Cartan defines an R-symmetry in $3d$, and while it is in many cases not the superconformal one, the results for the contribution of $5d$ multiplets to the $3d$ index are expressed most clearly if we use it as the R-symmetry in its calculation. The statement then is that in that case the index of the $3d$ theory has the form:

\bea
& & \mathcal{I} = 1 + (\sum_{i|q_i>0} \alpha^{q_i} {\bf R}_i (g-1 + q_i F) ) x^2 + (3g-3 + (1+{\bf Ad}_H)(g-1) ) x^2 \nonumber \\ & & + (\sum_{i|q_i<0} \alpha^{q_i} {\bf R}_i (g-1 + q_i F) ) x^2 + ... , \label{FRel}
\eea     
where we use $\alpha$ as the fugacity for the $U(1)$.

Here, the term $3g-3$ comes from the contribution of the $5d$ energy-momentum tensor. The rest of the terms come from the conserved current multiplets, where we have split the contributions depending on whether their charge under the $U(1)$ is positive, negative or zero, which are the first, last and middle terms respectively. Here we have used the fact that the only terms in ${\bf Ad}_G$ that have zero charge under the $U(1)$ are the adjoint of the commutant $H$ of $U(1)$ inside $G$. We have chosen to separate the three as in most cases the $U(1)$ mixes with the R-symmetry, and the three then have very different physical properties, giving relevant, marginal and irrelevant operators. 

Let us consider the contribution associated with marginal operators in more detail. As our focus in this paper is on torus compactifications, $g=1$ and this contribution vanishes. At this point we recall that both marginal operators and ($3d$) conserved currents contribute to the index at the same order in its expansion, $x^2$ (where we use the superconformal R-symmetry), but with opposite signs \cite{Razamat:2016gzx,Beem:2012yn}. While marginal operators contribute with a positive sign, conserved currents come with a negative one, reflecting the fact \cite{Green:2010da} that marginal operators fail to be exactly marginal only if they combine with a current multiplet to form a long multiplet (and so the index, being invariant on conformal manifolds, is only sensitive to the difference between the numbers of the two). This appears in \eqref{FRel} in the following way. The $5d$ energy-momentum tensor contributes $3g-3$ marginal operators, which is zero in our case. The $5d$ conserved current multiplets, on the other hand, contribute $(1+{\bf Ad}_H)g$ marginal operators which appear in the index with a positive sign, and $(1+{\bf Ad}_H)$ $3d$ conserved currents that appear with a negative sign. These contributions cancel out for $g=1$, but correspond to a non-trivial conformal manifold. Indeed, following the prescription of \cite{Green:2010da} and our discussion here, the dimension of the resulting conformal manifold is $rank(H)+1$, where at a generic point of it the global symmetry is broken to its Cartan and is given by $U(1)^{rank(H)+1}$.

Finally, we would like to make several comments on the formula \eqref{FRel}. First, we note that the multiplicity of the contribution to the index for conserved current multiplets is always $g-1+q F$, where $q F$ is the flux felt by the operator. This naturally generalizes to $g-1 + \sum_i q_i F_i$ if there are multiple fluxes. We should also stress that this provides the contribution to the index from only the sector coming from the conserved current and energy-momentum tensor multiplets. However, the SCFT in general has many other local operators, as well as non-local operators that can wrap various cycles of the Riemann surface. As such, there would usually be other contributions to the index besides these. In some cases, these contributions can obscure this structure. For instance, they could lead to additional marginal operators or additional symmetries which would then also appear in the $x^2$ terms in the index. While this can occur for special low-values of the flux or genus, the behavior for generic values should be in accordance with \eqref{FRel}.

So far we have discussed the case of the conserved current multiplets, but some aspects of this formula can also be applied to other types of multiplets. Notably, the $5d$ conserved current multiplets belong to a family of BPS multiplets known as the Higgs branch chiral ring operators. These have a scalar in the ${\bf R}$ dimensional representation of $SU(2)_R$ as their primary and obey the maximal shortening condition possible, see \cite{Ferlito:2017xdq}. Their physical significance is that their vevs parameterize the Higgs branch of the $5d$ SCFT. The case of $R=3$ gives the conserved flavor current multiplet, with $R=2$ being a free hypermultiplet and $R=1$ being the vacuum. We can also consider Higgs branch chiral ring operators with $R>3$.

Incidentally, the formula we used can be extended to all Higgs branch chiral ring operators. The main differences are that first, the operator is now not necessarily in the adjoint of the group $G$, but rather in some representation ${\bf R}_G$. We can again decompose it to representations ${\bf R}_i$ of $H$ with charges $q_i$ under the $U(1)$. The multiplicity of each contribution is again $g-1 + q_i F$, with the generalization to more fluxes as previously outlined. The second main difference is that the operators appear in the index at order $x^{R-1}$ when we use the Cartan of $SU(2)_R$ to compute the index. As such, their contribution to the index is expected to be

\be
 \sum_i \alpha^{q_i} {\bf R}_i (g-1 + q_i F) x^{R-1}\,.
\ee    
Generically, they should all be irrelevant operators, barring extreme mixing. 

As discussed in this subsection so far, the main tests we will use to check the $3d$ models we obtain are given by identifying various operators expected from the $5d$ construction. In some cases, however, additional computations can be performed that test further the proposed symmetry enhancement in a $3d$ model. Suppose we have two $U(1)$ symmetries which are not related in a UV model, but that in the IR appear as two subgroups of the same larger symmetry group due to an enhancement of symmetry. Then, properties of the theory associated with these $U(1)$ symmetries, which are independent in the UV, are expected to be related to each other in the IR according to how these two symmetries are embedded in the larger group. Identifying such relations thus serves as a nontrivial check for the proposed symmetry enhancement. 

The property associated with such $U(1)$ symmetries that we will focus on is the central charge, defined as the coefficient appearing in the flat space two-point function (at separated points) of the corresponding current
\begin{equation}
	\label{JJ}
	\langle J^{\mu}\left(x\right)J^{\nu}\left(0\right)\rangle=\frac{C}{16\pi^{2}}\left(\delta^{\mu\nu}\partial^{2}-\partial^{\mu}\partial^{\nu}\right)\frac{1}{x^{2}}\,.
\end{equation}
The central charge of a given $U(1)_I$ symmetry can be computed from the second derivative of the real part of the free energy of the theory with respect to the mixing coefficient of this symmetry with the superconformal R-charge according to the following relation \cite{Closset:2012vg}:
\begin{equation}
	\label{d2F}
	\left.\left(\frac{\partial}{\partial t^{I}}\right)^{2}\textrm{Re}\,F\right|_{t=t_{SC}}=-\frac{\pi^{2}}{2}C_{I}\,.
\end{equation}
Here, $F=-\log|Z|$ is the free energy, $C_I$ and $t^{I}$ are the central charge and mixing coefficient of $U(1)_I$, and $t_{SC}$ is the set of mixing coefficients of all the $U(1)$'s in the theory corresponding to the superconformal value of the R-symmetry. 

Once the IR central charges of two $U(1)$ symmetries of the kind discussed above are computed, they can be employed to test the proposed symmetry enhancement since they are expected to be related by the corresponding embedding indices of these $U(1)$ symmetries in the larger symmetry group. We will indeed check such relations in some of the $3d$ models presented below, by explicitly computing the central charges using \eqref{d2F} when it is numerically achievable.

\section{Properties of $5d$ SCFTs}
\label{5dproperties}

\begin{figure}
\hspace{-0.7cm}
\includegraphics[width=1.1\textwidth]{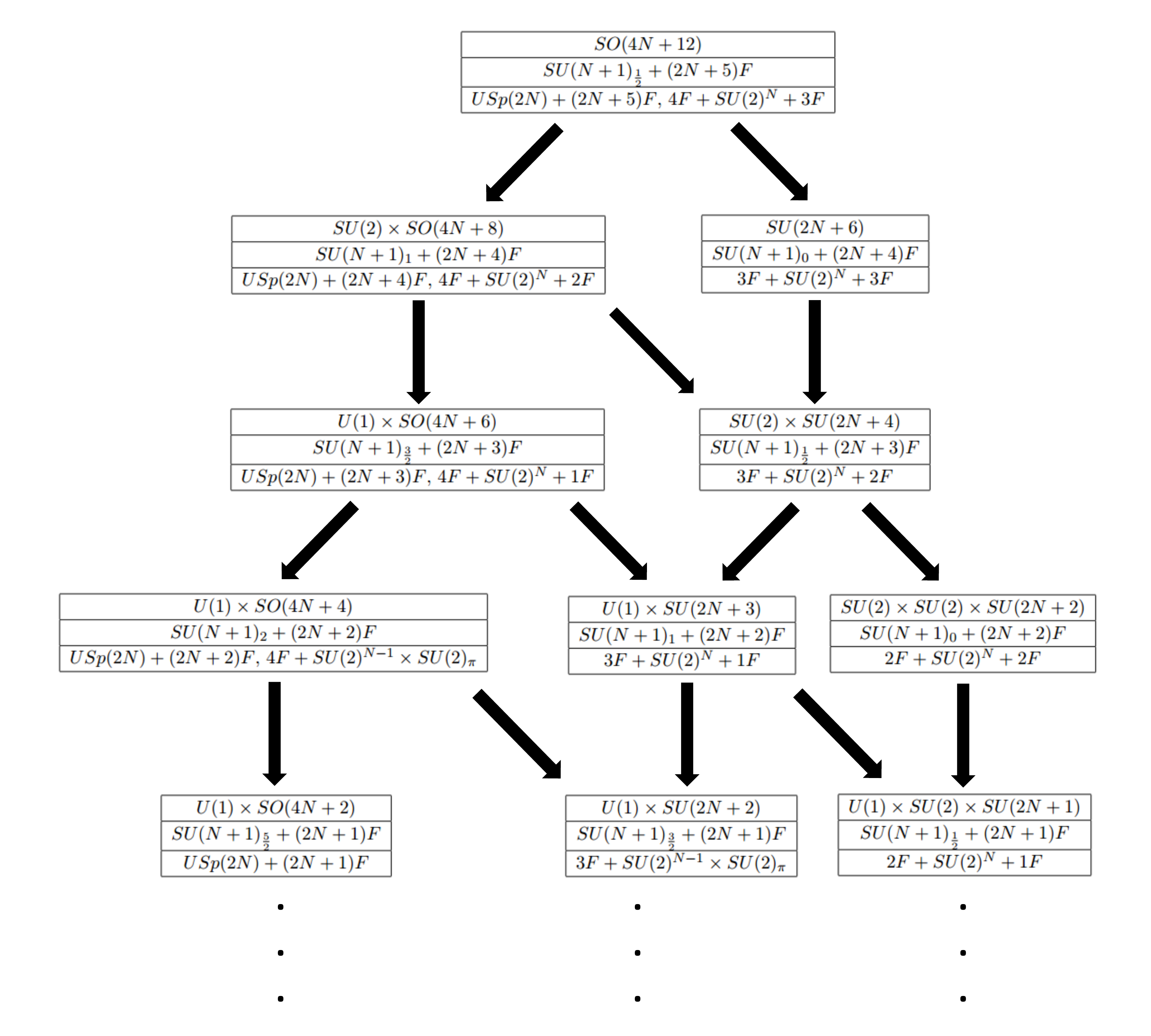} 
\caption{An illustration of the flow chart for $5d$ SCFTs. Here each table represents a $5d$ SCFT with the top entry denoting its global symmetry, the middle one giving its $SU$ type gauge theory description and the bottom one detailing additional gauge theory descriptions. Arrows between the different tables indicate mass deformations between the different $5d$ SCFTs. Here only the top cases are shown, and the flows continue to other $5d$ SCFTs that UV complete $SU(N+1)_k+N_fF$ gauge theories with $N_f<2N+1$.}
\label{5dFlowChart}
\end{figure}

In this section we shall consider some properties of $5d$ SCFTs that will be of use to us in this paper. We shall be mostly interested in the $5d$ SCFTs that UV complete $5d$ gauge theories of the type $SU(N+1)_k+N_fF$. These start with the case of the $SU(N+1)_0+(2N+6)F$ gauge theory, which is UV completed by a $6d$ SCFT, called the $(D_{N+3},D_{N+3})$ conformal matter \cite{Hayashi:2015fsa}. In addition to the $SU$ description, it also has gauge theory descriptions as a $USp(2N)+(2N+6)F$ and $4F+SU(2)^N+4F$ gauge theories\footnote{Here we use the shortened notation $SU(2)^N$ to denote a linear quiver of $N$ $SU(2)$ gauge groups connected by bifundamental hypermultiplets.}. Integrating flavors from these gauge theories leads to gauge theories that are UV completed by $5d$ SCFTs. This gives a family of $5d$ SCFTs that UV complete the $5d$ gauge theories $SU(N+1)_{\frac{M_p-M_n}{2}}+(2N+6-M_p-M_n)F$, where $M_p$ and $M_n$ are the number of flavors integrated with a positive or negative mass deformations, respectively. They also have $USp(2N)$ or $SU(2)^N$ gauge theory descriptions depending on the values of $M_p$ and $M_n$. We illustrate some of the top theories and the relations between them in figure \ref{5dFlowChart}.

The theories in this class have an interesting pattern of symmetry enhancement, see \cite{Bergman:2013ala,Bergman:2013aca,Bergman:2014kza,Tachikawa:2015mha,Hayashi:2015fsa,Yonekura:2015ksa,Gaiotto:2015una}. Specifically, the gauge theory always has a $U(1)^2 \times SU(N_f)$ global symmetry, with the two $U(1)$ groups associated with the baryonic and topological symmetries and the $SU(N_f)$ is the part rotating the flavors. For generic values of $N$ and $k$, these are also the symmetries of the $5d$ SCFT. However, the symmetry of the latter can be enhanced in certain cases, specifically, when $N_f + 2|k|$ is $2N+6$, $2N+4$ or $2N+2$. We shall be particularly interested in these cases.  

\subsubsection*{The case of $N_f + 2|k|=2N+6$}

The special feature of this parameter range is that the $5d$ SCFT has the enhanced symmetry $U(1)\times SU(N_f)\rightarrow SO(2N_f)$, where here the $U(1)$ is some combination of the topological and baryonic $U(1)$ groups which depends on the values of $N$ and $k$. For generic values in this range then, the symmetry of the $5d$ SCFT is $U(1)\times SO(2N_f)$. However, it is further enhanced for special cases. Specifically, for $N_f=2N+4$ we have the further enhancement of $U(1)\rightarrow SU(2)$, and the global symmetry of the $5d$ SCFT is $SU(2)\times SO(4N+8)$. This SCFT is realized prominently in the following discussion and we shall usually refer to it as the $SU(2)\times SO(4N+8)$ SCFT. Another special case is the $N_f=2N+5$ case where we have that $U(1)\times SO(4N+10)\rightarrow SO(4N+12)$, leading to a $5d$ SCFT with $SO(4N+12)$ global symmetry. As previously stated, the case of $N_f=2N+6$ does not have a UV completion as a $5d$ SCFT, but rather as a $6d$ SCFT.

For some values of the parameters, this class of $5d$ SCFTs has also an $USp(2N)$ or $SU(2)^N$ gauge theory descriptions. We shall be particularly interested in the first one here as it exists for the entire parameter range. 

Finally, we wish to make some comments regarding the operator content of this class of $5d$ SCFTs. As we previously mentioned, to test our proposed $3d$ models, we require some understanding of the operator content of the $5d$ SCFT, particularly, that of its Higgs branch chiral ring operators. As usual, all $5d$ SCFTs with a flavor symmetry have at least the Higgs branch chiral ring operators associated with their flavor symmetry currents. These are in the adjoint of the flavor symmetry, and their primary is a scalar in the ${\bf 3}$ of $SU(2)_R$.

Additionally, all the $5d$ SCFTs in this class possess an additional Higgs branch chiral ring generator, see \cite{Ferlito:2017xdq}. This generator is in the Dirac spinor representation of the $U(1)\times SO(2N_f)$ symmetry, that is it is the direct sum of a spinor of $SO(2N_f)$ with charge $+1$ under the $U(1)$ and the complex conjugate spinor with charge $-1$ under the $U(1)$. The primary is a scalar, as usual for Higgs branch operators, but now in the ${\bf N+2}$ of $SU(2)_R$. Finally, we note that for $N_f=2N+4$ and $2N+5$, the flavor representation of this operator becomes the $({\bf 2},{\bf S})$ of $SU(2)\times SO(4N+8)$ for $N_f=2N+4$ and the ${\bf S}$ of $SO(4N+12)$ for $N_f=2N+5$.

\subsubsection*{The case of $N_f + 2|k|=2N+4$}

The special feature of this parameter range is that the $5d$ SCFT has the enhanced symmetry $U(1)\times SU(N_f)\rightarrow SU(N_f+1)$, where here the $U(1)$ is some combination of the topological and baryonic $U(1)$ groups which depends on the values of $N$ and $k$. For generic values in this range then, the symmetry of the $5d$ SCFT is $U(1)\times SU(N_f+1)$. However, it is further enhanced for special cases. Specifically, for $N_f=2N+4$ we have the further enhancement of $U(1)\times SU(2N+5)\rightarrow SU(2N+6)$, and the global symmetry of the $5d$ SCFT is $SU(2N+6)$. This SCFT is realized prominently in the following discussion and we shall usually refer to it as the $SU(2N+6)$ SCFT. Another special case is the $N_f=2N+3$ case where we have that $U(1)\rightarrow SU(2)$, leading to a $5d$ SCFT with $SU(2)\times SU(2N+4)$ global symmetry.

Unlike the previous case, this class of $5d$ SCFTs do not have a $USp(2N)$ gauge theory description. Besides the $SU(N+1)$ description, which exists for all values of parameters, some values also have an $SU(2)^N$ gauge theory description, though that won't play a role here.

 Like in the previous case, we also wish to consider the Higgs branch chiral ring operators of this $5d$ SCFT. Beside the operator associated with the flavor symmetry currents, there can be additional Higgs branch chiral ring generators. Here we shall concentrate on the case of $N_f\geq N+1$, where there is one such Higgs branch chiral ring generator\footnote{Part of this generator is manifested in the gauge theory by the baryons which only exist when $N_f\geq N+1$, hence the limitation.}, see \cite{Ferlito:2017xdq}. This generator is in the $N+2$ antisymmetric representation of the $SU(N_f+1)$ global symmetry with charge $+1$ under the $U(1)$ plus the complex conjugate. Like in the previous case, the scalar primary here is in the ${\bf N+2}$ of $SU(2)_R$. Finally, we note that for $N_f=2N+3$ and $2N+4$, the flavor representation of this operator becomes $({\bf 2},{\bf \Lambda}^{N+2})$ of $SU(2)\times SU(2N+4)$ for $N_f=2N+3$ and $({\bf \Lambda}^{N+3})$ of $SU(2N+6)$ for $N_f=2N+4$.

\subsubsection*{The case of $N_f + 2|k|=2N+2$}

The special feature of this parameter range is that the $5d$ SCFT has the enhanced symmetry $U(1)\rightarrow SU(2)$, where here the $U(1)$ is some combination of the topological and baryonic $U(1)$ groups which depends on the values of $N$ and $k$. For generic values in this range then, the symmetry of the $5d$ SCFT is $U(1)\times SU(2) \times SU(N_f)$. However, it is further enhanced for the special case of $N_f=2N+2$. Then we have the further enhancement of $U(1)\rightarrow SU(2)$, and the global symmetry of the $5d$ SCFT is $SU(2)\times SU(2)\times SU(2N+2)$. This class of $5d$ SCFTs do not have a $USp(2N)$ gauge theory description, though an $SU(2)^N$ gauge theory description exists for some values of the parameters. 

Like in the previous cases, we also wish to consider the Higgs branch chiral ring operators of this $5d$ SCFT. First, we have the one associated with the flavor symmetry currents. Here we shall concentrate on the case of $N_f\geq N+1$, where there is one additional Higgs branch chiral ring generator, see \cite{Ferlito:2017xdq}. This generator is in the doublet representation of the $SU(2)$, the $N+1$ antisymmetric representation of the $SU(N_f)$ global symmetry and with charge $+1$ under the $U(1)$ plus the complex conjugate. Like in the previous case, the scalar primary here is in the ${\bf N+2}$ of $SU(2)_R$. Finally, we note that for $N_f=2N+2$ the flavor representation of this operator becomes $({\bf 2},{\bf 2},{\bf \Lambda}^{N+1})$ of $SU(2)\times SU(2)\times SU(2N+2)$.

\section{Compactifications of $5d$ SCFTs associated with $SU(N+1)_k+N_fF$ gauge theories}
\label{compactifications}

We will now discuss the $3d$ theories arising from the compactification on tubes and tori of the $5d$ SCFTs that UV complete some of the $SU(N+1)_k+N_fF$ gauge theories. As we explained in subsection \ref{conjectures}, the starting point is to understand the tube theories, which we can then glue to build tori. These will be conjectured by first reducing the $5d$ theory along the circle with some choice of holonomy at the two extrema of the interval, which induce the deformation to a gauge theory. We have seen that for a single $5d$ SCFT there might be various deformations that lead to different gauge theories. As a result, we can have different tube theories that correspond to domain walls interpolating between different gauge theories associated with the same SCFT. In addition to the $SU(N+1)$ gauge theory description we mentioned, some of the $5d$ SCFTs also admit a $USp(2N)$ gauge theory description. By considering different holonomies, we can then get domain walls between each of these choices leading to tubes with two $SU(N+1)$ punctures, two $USp(2N)$ punctures and a $SU(N+1)$ and $USp(2N)$ puncture\footnote{We have mentioned that in some cases a third $SU(2)^N$ description also exists. In those cases we can further consider tubes that have this description on one or both of the sides of the interval. Such tubes indeed exist in the case of the $4d$ tube compactification of the $6d$ $(D_{N+3},D_{N+3})$ conformal matter \cite{Kim:2018lfo}, though we shall not consider this here.}. In the case of two $USp(2N)$ punctures, we have not managed to find a candidate tube passing all the consistency checks in this case. It should be noted, that similarly, no such tube was found also in the closely related case of the tube compactification of the $6d$ $(D_{N+3},D_{N+3})$ conformal matter theory to $4d$ \cite{Kim:2018bpg,Kim:2018lfo}. As such we will only consider the two later cases, which will be denoted as the $SU-USp$ tubes and $SU-SU$ tubes. 

\subsection{Cases built from $SU-USp$ tubes}
\label{SUUSp}

We begin with the case where the holonomy is chosen such that on one side we have a $4d$ $SU(N+1)$ gauge theory, while on the other we have a $USp(2N)$ one. As such these tubes are associated with two inequivalent punctures, one carrying $SU(N+1)$ global symmetry and the other carrying a $USp(2N)$ one. Here the choice of $5d$ SCFTs is limited to the $SU(N+1)_k+N_fF$ ones with $N_f + 2|k|=2N+6$, where both gauge theory deformations exists.  

The case of $N_f=2N+6$ has the $6d$ SCFT known as the $(D_{N+3},D_{N+3})$ conformal matter as its UV completion. The $5d$ SCFTs associated with cases with $N_f<2N+6$ can then be generated from it by circle compactification with various flavor holonomies. Compactifications of this $6d$ SCFT were studied in \cite{Kim:2018bpg}, and we can employ their results to try to formulate conjectures for possible $3d$ compactifications of the $5d$ SCFTs.

 In what follows we shall concentrate on the case of the $5d$ SCFTs that UV complete the $5d$ gauge theories $SU(N+1)_1+(2N+4)F$, which are the highest $N_f$ for which we can find candidate tube theories passing the necessary consistency checks. Specifically, the conjecture that we shall soon introduce for the theory living on the domain wall does not seem to hold for $N_f=2N+5$. This is not surprising as the $N=1$ case is just the rank $1$ $E_8$ theory where a similar conjecture also fails \cite{Sacchi:2021afk}. We expect cases with lower $N_f$ to be given by real mass deformations of the tube with $N_f=2N+4$, as this was the case for $N=1$ \cite{Sacchi:2021afk}, though we shall not study this explicitly here. 

\begin{figure}
\hspace{-0.4cm}
\includegraphics[width=1.05\textwidth]{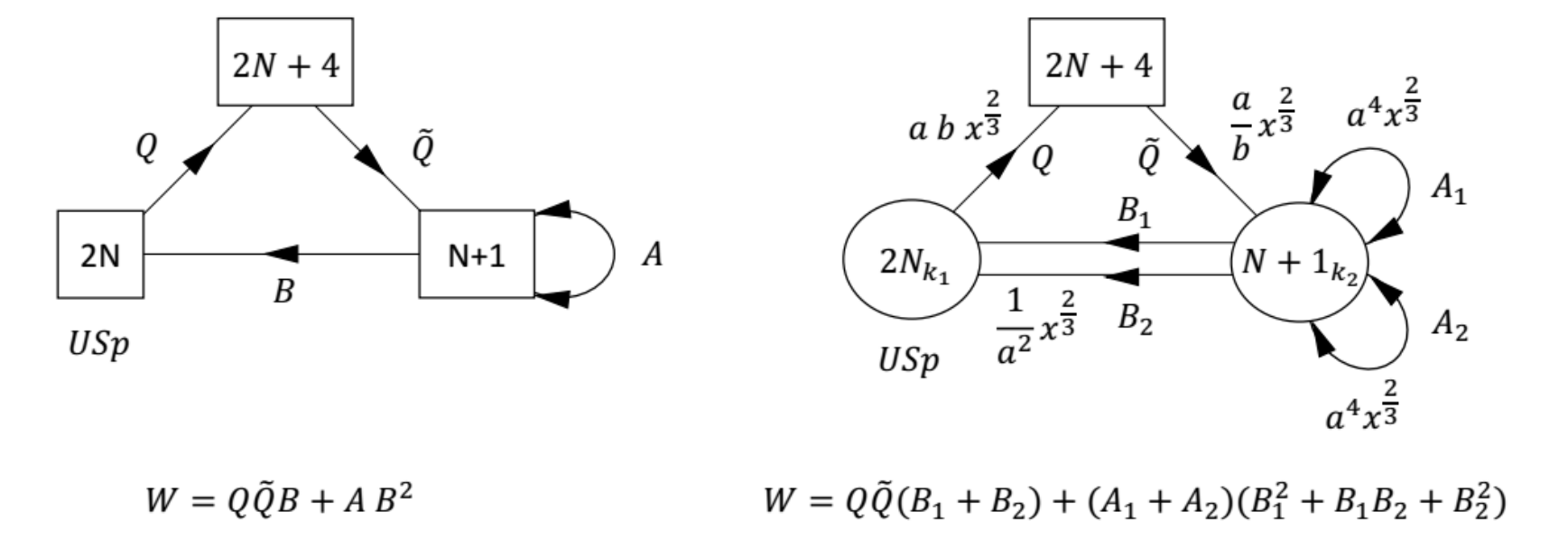} 
\caption{On the left the $3d$ quiver theory associated with the compactification of the $5d$ SCFT with $SU(2)\times SO(4N+8)$ global symmetry on a tube with some flux in its global symmetry. Similarly, on the right we have the case associated with compactification on a torus. The fields $A$, $A_1$ and $A_2$ stand for antisymmetric chiral fields, where we use the notation of arcs with two ingoing or outgoing arrows for antisymmetric or complex conjugate antisymmetric chiral fields. This should not lead to any confusion with possible symmetric representations as in all of our paper we only have antisymmetric representations and never symmetric ones. The subscript in the gauge nodes denotes the CS levels. Here we have momentarily allowed for a general CS term, though the case with the $5d$ origin has $k_1=k_2=0$. The charges of the fields under the two $U(1)$ global symmetries are displayed using the fugacities $a$ and $b$, with the fugacity $x$ displaying the charges under a chosen $U(1)$ R-symmetry.}
\label{HigherRankSU}
\end{figure}

An interesting property of the $5d$ SCFTs associated with the $SU(N+1)_1+(2N+4)F$ gauge theories is that they also have a deformation to a $USp(2N)+(2N+4)F$ gauge theory\footnote{In other words, both $5d$ gauge theories are UV completed by the same $5d$ SCFT.}. As such, we can now consider doing the reduction to $3d$ in the following way. We again consider first reducing to $4d$ with an holonomy, which we take to be constant along the entire $4d$ compact direction, save for a jump somewhere in the middle where a domain wall exists. Now, we take its value such that the theory on one side of the domain wall is a $4d$ $SU(N+1)+(2N+4)F$ gauge theory, while on the other side it is a $USp(2N)+(2N+4)F$ gauge theory. 

Such types of domain walls were studies in \cite{Kim:2018bpg}, and we can next use their results to analyze this case. Specifically, the case analyzed in \cite{Kim:2018bpg} was for $4d$ domain walls between the $5d$ gauge theories $SU(N+1)_0+(2N+6)F$ and $USp(2N)+(2N+6)F$ associated with the $6d$ lifting case. Here we generalize to the specific case of $3d$ domain walls between $4d$ gauge theories associated with $5d$ ones that are UV completed by $5d$ SCFTs. However, the important thing we need here are the fields expected to live on the domain wall, where we can use the results in \cite{Kim:2018bpg} for these fields to make progress, by assuming these are the same also in our case.

The resulting theory is shown in the left side of figure \ref{HigherRankSU}. This theory is associated with the compactification of a $5d$ SCFT, now the one with $SU(2)\times SO(4N+8)$ global symmetry that we just introduced, on a tube with some value of flux. Here the two punctures of the tube are different, one carrying a $USp(2N)$ global symmetry and one carrying an $SU(N+1)$ global symmetry. The theory is associated with a compactification with some value of flux in the $SU(2)\times SO(4N+8)$ global symmetry of the $5d$ SCFT, which needs to be determined by direct computation. However, assuming it works as in the $4d$-$6d$ case, we expect the associated flux to be $\frac{1}{4}(1,-1;1,1,...,1)$, where the first two numbers give the flux in the $SU(2)$, and the others give the flux in the $2N+4$ independent $SO(2)$ groups spanning $SO(4N+8)$. For the flux in the $SU(2)$, and also for other $SU(n)$ groups latter on, we use a $U(1)^n$ basis of fluxes restricted so that the fluxes sum to zero. For $SO(2n)$ type groups, however, we shall just use a $U(1)^n$ basis of fluxes given by the fluxes in its $n$ independent $SO(2)$ subgroups. 

We can glue two tubes together to get the theory on the right side of figure \ref{HigherRankSU}. This is the theory associated with the compactification of the $5d$ SCFT with $SU(2)\times SO(4N+8)$ global symmetry on a torus with flux. Here we have left the option for a Chern--Simons level for the two groups. We next need to check that this theory has the correct properties to indeed be the compactification of the chosen $5d$ SCFT.

\subsubsection{$N=2$, $N_f=8$}

For simplicity, we can consider first the $N=2$ case and assume that the necessary $USp(4)$ and $SU(3)$ Chern--Simons levels are $k_1=k_2=0$. On top of the non-abelian $SU(8)$ flavor symmetry, this model also possesses two abelian symmetries $U(1)_a$ and $U(1)_b$. We parameterize $U(1)_a$ such that the antisymmetrics have charge $4$, the bifundamentals have charge $-2$ and the fundamentals have charge 1. We parameterize the $U(1)_b$ symmetry, instead, such that the two fundamentals have opposite charges $\pm 1$, while all the other fields are uncharged. Moreover, we use a trial R-symmetry under which all the fields have R-charge $\frac{2}{3}$. With these normalizations, we can compute the index for that model finding
\be
\label{I2jij}
\mathcal{I} = 1 + \frac{a^4}{b^8} x^{\frac{2}{3}} + (\frac{a^8}{b^{16}} + 2\frac{a^5}{b} \bold{\bar{8}} + a^2 b^2 \bold{28} ) x^{\frac{4}{3}} + (\frac{a^{12}}{b^{24}} + 2\frac{a^9}{b^9} \bold{\bar{8}} + \frac{a^6}{b^6} \bold{28} + \frac{a^3}{b^3} \bold{\bar{56}} ) x^{2} + ...
\ee
This index is consistent with the theory being the compactification of the chosen $5d$ SCFT on a torus with flux $\frac{1}{2}(1,-1;1,1,...,1)$. This flux is the one breaking $SU(2)\rightarrow U(1)$ and $SO(16)\rightarrow U(1)\times SU(8)$, where the value of the flux is $\frac{1}{2}$ for both cases\footnote{Despite how it might look, this is actually consistent. This comes about as the actual $5d$ global symmetry is $\frac{SU(2)\times SO(4N+8)}{\mathbb{Z}_2}$, so there are no cases with odd charges under only one of the two $U(1)$ groups.}.
We indeed find states associated with the broken currents for the $SU(2)$ and $SO(16)$ symmetries, and further find states associated with higher order Higgs branch operators. Let us explain how these states are identified in \eqref{I2jij} in detail. 

We begin with the states coming from the $5d$ conserved currents in the adjoint representation of the $SU(2)\times SO(16)$ global symmetry, which split under the $U(1)^{2}\times SU(8)$ subgroup as follows:
\begin{equation}
	\mathbf{120}\rightarrow\mathbf{1}^{0}\oplus\mathbf{63}^{0}\oplus\mathbf{28}^{2}\oplus\mathbf{\overline{28}}^{-2}\,,\label{adso16}
\end{equation}
\begin{equation}
	\mathbf{3}\rightarrow\mathbf{1}^{0}\oplus\mathbf{1}^{2}\oplus\mathbf{1}^{-2}\,.\label{adjsu2}
\end{equation}
Under the Cartan of the $5d$ $SU(2)$ R-symmetry, the bifundamentals have R-charge 0, the fundamentals have R-charge 1 and the antisymmetric fields have R-charge 2. As a result, it is related to the R-symmetry $R_{\mathcal{I}}$ we used in our computation by 
\begin{equation}
R_{\mathcal{I}}=R_{5d}-\frac{1}{3}q_{a}\,,
\end{equation}
and we can identify the $5d$ conserved-current states, with multiplicities $g-1+qF$, as follows. The state $\mathbf{28}^{2}$ in \eqref{adso16} appears in the index as the term $\mathbf{28}\,a^{2}b^{2}x^{\frac{4}{3}}$ and corresponds to the $USp(4)$ baryon $Q^{2}$, and the state $\mathbf{1}^{2}$ in \eqref{adjsu2} appears in the index as the term $a^{4}b^{-8}x^{\frac{2}{3}}$ and corresponds to the basic monopole of $USp(4)$. Note that the flux is $\frac{1}{2}$ in both of the $U(1)$'s in $SU(2)\times SO(16)$, and they are related to the $U(1)$'s used in the computation of the index by $x_{SO(16)}=ab$ and $x_{SU(2)}=a^{2}b^{-4}$, where $x_{SO(16)}$ and $x_{SU(2)}$ are the fugacities of $U(1)_{SO(16)}$ and $U(1)_{SU(2)}$.  

We next turn to the states coming from the other Higgs branch chiral ring operator of the $5d$ SCFT, which is in the $(\mathbf{2},\mathbf{\overline{128}})$ of $SU(2)\times SO(16)$ and in the $\mathbf{4}$ of $SU(2)_{R}$ (and so contributes at order 3 with respect to $U(1)_{R_{5d}}$). Under the $U(1)^{2}\times SU(8)$ subgroup, we have the following decompositions:
\begin{equation}
\mathbf{\overline{128}}\rightarrow\mathbf{8}^{-3}\oplus\mathbf{\overline{8}}^{3}\oplus\mathbf{56}^{-1}\oplus\mathbf{\overline{56}}^{1}\,,
\end{equation}
\begin{equation}
\mathbf{2}\rightarrow\mathbf{1}^{1}\oplus\mathbf{1}^{-1}\,,
\end{equation}
leading to 
\begin{equation}
(\mathbf{2},\mathbf{\overline{128}})\rightarrow\mathbf{8}_{1}^{-3}\oplus\mathbf{8}_{-1}^{-3}\oplus\mathbf{\overline{8}}_{1}^{3}\oplus\mathbf{\overline{8}}_{-1}^{3}\oplus\mathbf{56}_{1}^{-1}\oplus\mathbf{56}_{-1}^{-1}\oplus\mathbf{\overline{56}}_{1}^{1}\oplus\mathbf{\overline{56}}_{-1}^{1}
\end{equation}
where we use the notation $\mathbf{r}_{q_{U(1)_{SU(2)}}}^{q_{U(1)_{SO(16)}}}$. We identify the state $\mathbf{\overline{56}}_{1}^{1}$ with the index term $\mathbf{\overline{56}}\,a^{3}b^{-3}x^2$ in \eqref{I2jij} which corresponds to the $SU(3)$ baryon $\widetilde{Q}^{3}$, and the state $\mathbf{\overline{8}}_{1}^{3}$ with the index terms $2\,\mathbf{\overline{8}}\,a^{5}b^{-1}$ that come from the operators $A_{2}\widetilde{Q}$ and $A_{1}\widetilde{Q}$. 

In addition to this identification of states based on the $5d$ picture, we find a conformal manifold of dimension $9$ on a generic point of which only a $U(1)^9$ subgroup is preserved. This comes about as we do not observe the conserved currents $U(1)^2\times SU(8)$ in the order $x^2$ of the index, implying that they are canceled by marginal operators transforming in the adjoint representation of this group. This is again in accordance with the $5d$ expectations, as explained in the previous sections.



\subsubsection{Higher $N$}

We can next consider the case of higher $N$, where the charges of the fields are the same as for $N=2$ since the superpotential is still the one given in figure \ref{HigherRankSU} for any $N$. Due to the complexity of the calculation we shall not consider the full index, but rather look at the contribution of specific operators. First, we consider the basic perturbative gauge invariant operators. These are the $USp$ and $SU$ baryons. The $SU$ baryons carry $U(1)^{5d}_R$ charge of $N+1$ and so are mapped to states coming from the spinor Higgs branch chiral ring operator, which we shall discuss later. The $USp$ baryons carry $U(1)^{5d}_R$ charge of $2$ and so are mapped to states coming from the broken currents. They carry charges $a^2 b^2$ and are in the antisymmetric representation, ${\bf \Lambda}^2$, of $SU(2N+4)$. Additionally, we have the operators associated with the triangle. These give two operators of $U(1)^{5d}_R$ charge of $2$, uncharged under $U(1)_a$, $U(1)_b$ and in the adjoint $+$ singlet of $SU(2N+4)$. The two singlets are already present in the superpotential, which leads to the breaking of $SU(2N+4)\times SU(2N+4)\rightarrow SU(2N+4)$ where in the left-hand side the two $SU(2N+4)$ global symmetry groups are the ones rotating the $SU(N+1)$ and $USp(2N)$ flavors individually. As such, one of the operators associated with the triangle in the adjoint of the remaining $SU(2N+4)$ gets eaten by the broken $SU(2N+4)$ currents. This leaves just one such operator in the adjoint of $SU(2N+4)$.   

Next, we consider the non-perturbative sector. Here we have the basic monopoles of the $USp$ and $SU$ groups. Specifically, the main ones of interest are the minimal $USp$, $SU$ and mixed monopoles. We first consider the minimal $USp$ monopole, which is the one inside an $SU(2)$ in $USp(2N)$ such that the commutant is $USp(2N-2)$. This corresponds to a magnetic flux vector of the form $(1,0,\cdots,0)$. This monopole carries $U(1)^{5d}_R$ charge of $2$ and so is mapped to a state coming from the broken $5d$ current. Additionally, it carries charges of $\frac{a^{2N}}{b^{2N+4}}$ and is a singlet under $SU(2N+4)$.

Similarly, the basic $SU$ monopole is the one with minimal charge, that is the one inside an $SU(2)$ in $SU(N+1)$ such that the commutant is $U(1)\times SU(N-1)$. This corresponds to a magnetic flux vector of the form $(1,0,\cdots,0,-1)$. This monopole carries $U(1)^{5d}_R$ charge of $2$, charges of $\frac{b^{2N+4}}{a^{2N-4}}$ under the abelian flavor symmetries and is a singlet under $SU(2N+4)$. However, it is charged under the unbroken gauge group inside $SU(N+1)$, and as such the basic invariant is the dressed monopole operator, here dressed by $n$ fundamentals and $(N-1-n)/2$ antisymmetrics for $n=1,\cdots,N-1$ such that $N-1-n$ is even, forming together the rank $N-1$ totally antisymmetric representation of $SU(N+1)$. This gives states in the rank $2N+4-k$ totally antisymmetric representation of $SU(2N+4)$, with charges $\frac{b^{2N+4-n}}{a^{n-2}}$ under the abelian global symmetry and with $U(1)^{5d}_R$ charge of $N+1$. 

Finally, we consider the basic mixed monopole, that is the monopole carrying minimal charge under both the $USp$ and $SU$ groups. This gives an operator with $U(1)^{5d}_R$ charge of $0$, with $\frac{1}{a^4}$ abelian charges and which is a singlet under $SU(2N+4)$. However, it is charged under the unbroken gauge group inside $SU(N+1)$. We can form a gauge invariant by dressing it with a component of the antisymmetric chirals. The end result is two gauge invariant operators with $U(1)^{5d}_R$ charge of $2$, and with no flavor charges.

Combining everything, we see that the results so far are consistent with this theory being the result of the compactification of the $SU(2)\times SO(4N+8)$ $5d$ SCFT with flux $\frac{1}{2}(1,-1;1,1,1,...,1)$ preserving $U(1)^2\times SU(2N+4)$. Specifically, we can identify the state coming from the basic $USp$ monopole with the broken currents of the $SU(2)$ part of the flavor symmetry, and similarly, the $USp$ baryons can be identified with the broken currents of the $SO(4N+8)$ part of the flavor symmetry. We also see that we expect to get $0$ at order $x^2$, as we have two singlet marginal operators and one in the adjoint of the $SU(2N+4)$ global symmetry, which exactly matches the global symmetry currents. This indeed leads to a $2N+5$ dimensional conformal manifold, on a generic point of which the preserved global symmetry is $U(1)^{2N+5}$.  

Finally, we can consider the states coming from the spinor Higgs branch chiral ring operator. As we mentioned, it is expected to have $U(1)^{5d}_R$ charge of $N+1$, be in the doublet of the $SU(2)$ and in the spinor of $SO(4N+8)$. We have seen that the $SU(N+1)$ baryons carry the correct $U(1)^{5d}_R$ charge and so it is natural to match it with them. These are given by baryons made from $i$ antisymmetric chirals and $N+1-2i$ fundamental chirals, for $i=0,1,2,...,\left \lfloor{\frac{N+1}{2}}\right \rfloor$. All of them carry $U(1)^{5d}_R$ charge of $N+1$, are in the ${\bf\Lambda}^{N+3+2i}$ of $SU(2N+4)$ and have charges $\frac{a^{N+1+2i}}{b^{N+1-2i}}$ under $U(1)_a$ and $U(1)_b$. As there are two antisymmetric chirals, and the product is done symmetrically, the number of such operators is $i+1$.

Now let us consider the spinor Higgs branch chiral ring operator. From our previous result, we expect the doublet of the $SU(2)$ to decompose to $(\frac{a^N}{b^{N+2}}+\frac{b^{N+2}}{a^N})$. Similarly, the spinors should decompose to $(a b)^{-(N+2-j)}{\bf \Lambda}^j$ for $j=0,1,2,...,2N+4$ and is even for one chirality but odd for the other. Finally, we note that (choosing the $\frac{a^N}{b^{N+2}}$ term in the decomposition of the doublet of $SU(2)$)\footnote{We recall the reader the conventions that we explained in footnote \ref{repconv} for representations of groups which we use in our paper.}
\be
\sum_j \frac{a^N}{b^{N+2}} (a b)^{-(N+2-j)}{\bf \Lambda}^j \stackrel{j\rightarrow N+3+2i}{\rightarrow} \sum_i \frac{a^N}{b^{N+2}} (a b)^{(2i+1)}{\bf \Lambda}^{N+3+2i} = \sum_i \frac{a^{N+1+2i}}{b^{N+1-2i}} {\bf \Lambda}^{N+3+2i}.
\ee  
This matches the charges we observed from the baryons, with $j$ being even (odd) for $N$ odd (even). Finally, their number is expected to be $\frac{1+j-N-2}{2}\rightarrow \frac{2+2i}{2} = i+1$.  

We noted that the basic monopole of the $SU(N+1)$ gauge group, properly dressed by the fundamentals and the antisymmetrics, also has the charges to match components of the spinor Higgs branch chiral ring operators. Specifically, choosing $j=2N+4-n$ in $\frac{b^{N+2}}{a^{N}}\left(ab\right)^{-(N+2-j)}{\bf \Lambda}^{j}$ (taking now the $\frac{b^{N+2}}{a^{N}}$ term in the decomposition of the doublet of $SU(2)$) we obtain $\frac{b^{2N+4-n}}{a^{n-2}}{\bf \Lambda}^{2N+4-k}$, which matches the charges of this operator.

Overall, while not an exact index calculation, we see that we can observe many of the operators we expect from the $5d$ realization from basic gauge invariant combinations of perturbative and non-perturbative fields. This supports our proposal regarding the $5d$ origin of this theory. 

\subsubsection{More general models}
\label{moregeneral}

In the previous subsection we considered the compactification of the $5d$ SCFT with $SU(2)\times SO(4N+8)$ global symmetry on a torus with two domain walls obtained by gluing two copies of the basic tube, see figure \ref{HigherRankSU}. The resulting total flux, $\frac{1}{2}(1,-1;1,1,...,1)$, breaks the $5d$ symmetry as follows: 
\begin{equation}
SU(2)\rightarrow U(1)_{SU(2)}\,,
\end{equation}
\begin{equation}
SO(4N+8)\rightarrow U(1)_{SO(4N+8)}\times SU(2N+4)\,.
\end{equation}
In this subsection we turn to consider the compactification of this $5d$ SCFT on a torus with four domain walls obtained by gluing two generalized tubes which we construct in the following way. Instead of gluing two basic tubes, each with flux $\frac{1}{4}(1,-1;1,1,...,1)$, to form a tube with flux $\frac{1}{2}(1,-1;1,1,...,1)$, we wish to glue the basic tube to a tube obtained from the basic one by acting on its flux with an element of the Weyl group of the symmetry. Even though these two constituent tubes are equivalent by themselves (since they are related by a Weyl group operation), their gluing results in a new tube carrying a novel flux. 

\begin{figure}
	\center
	\includegraphics[width=0.7\textwidth]{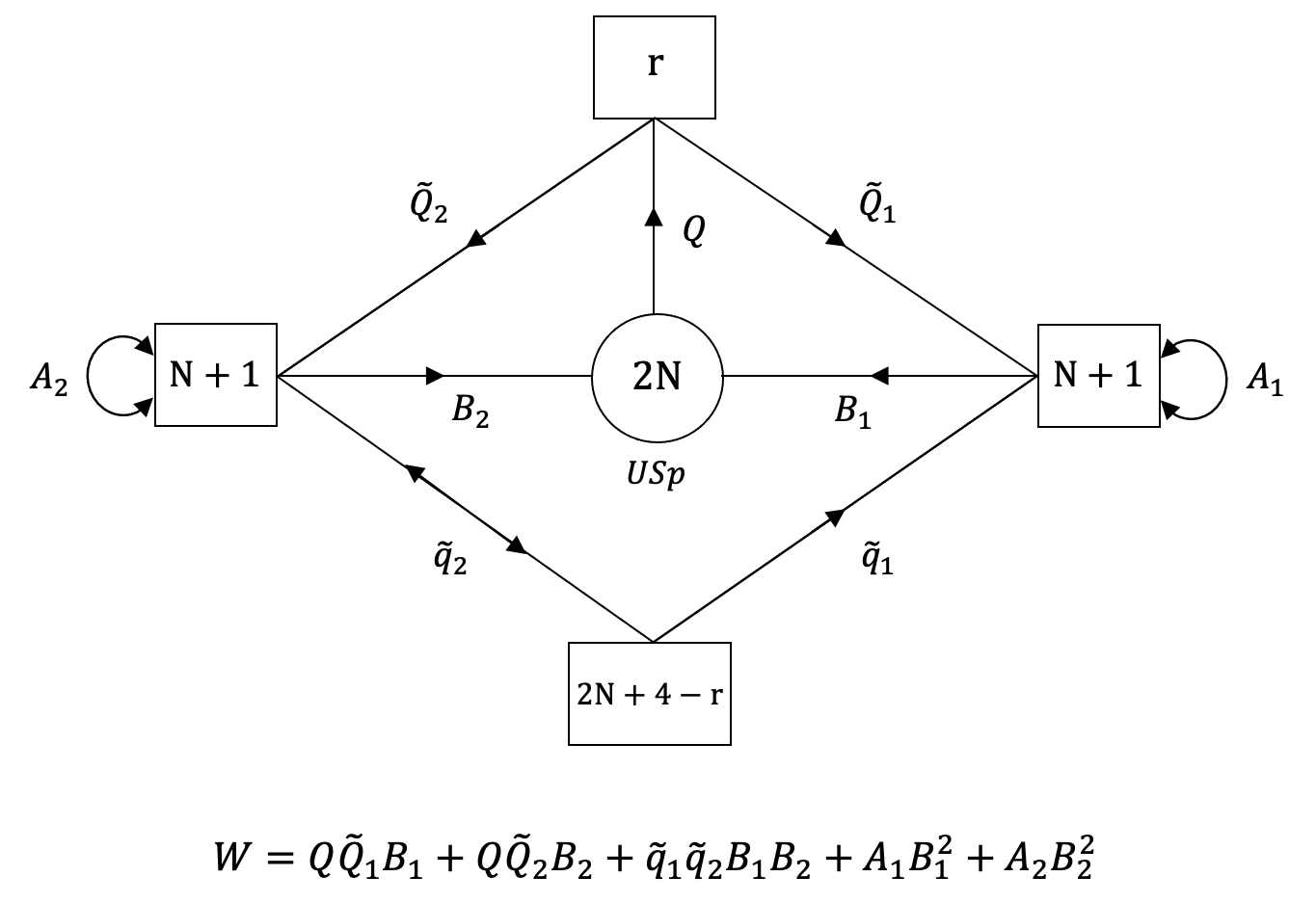} 
	\caption{The $3d$ quiver theory associated with the compactification of the $5d$ SCFT with $SU(2)\times SO(4N+8)$ global symmetry on a tube with flux preserving $U(1)^2\times SU(r)\times SO(4N+8-2r)$. }
	\label{gentube}
\end{figure}

Let us for concreteness consider the Weyl operation of $SO(4N+8)$ multiplying an even number of its flux elements by $-1$. Denoting the number of elements which remain unchanged by $r$ (such that $2N+4-r$ flux elements  are multiplied by $-1$) and acting on the basic tube, the flux of the resulting tube takes the form 
\begin{equation}
\frac{1}{4}(1,-1;\underbrace{1,1,...,1}_{r},\underbrace{-1,-1,...,-1}_{2N+4-r}).
\end{equation}
We can now glue it to a copy of the basic tube and obtain a new (generalized) tube with flux 
\begin{equation}
\frac{1}{2}(1,-1;\underbrace{1,1,...,1}_{r},\underbrace{0,0,...,0}_{2N+4-r}).
\end{equation}
In order to understand what the corresponding $3d$ theory looks like, we should examine the effect of the Weyl operation on the fields. In the $4d$ gauge theory description, the multiplication by $-1$ of a flux element corresponds to exchanging the two chiral multiplets inside the associated hypermultiplet. As a result, when gluing the Weyl-transformed tube to the original basic tube, the chiral multiplets of this kind that are present between the two domain walls have opposite boundary conditions on them. That is, they have Dirichlet boundary conditions on one domain wall and Neumann on the other, thus they do not survive the three dimensional limit. In the other segments of the new tube, on the other hand, the chiral fields have the same boundary conditions on both the corresponding domain wall and on the boundary of the tube. For these chirals the story remains as before, except for a new superpotential coupling between chirals coming from two different segments which is exerted by the chiral fields between the domain walls that do not survive in $3d$. As to the chirals in the segment between the domain walls that are not acted on by the Weyl operation, the gluing is exactly as in the case of two basic tubes. Overall, we end up in $3d$ with the model shown in figure \ref{gentube}. 

\begin{figure}
	\center
	\includegraphics[width=0.65\textwidth]{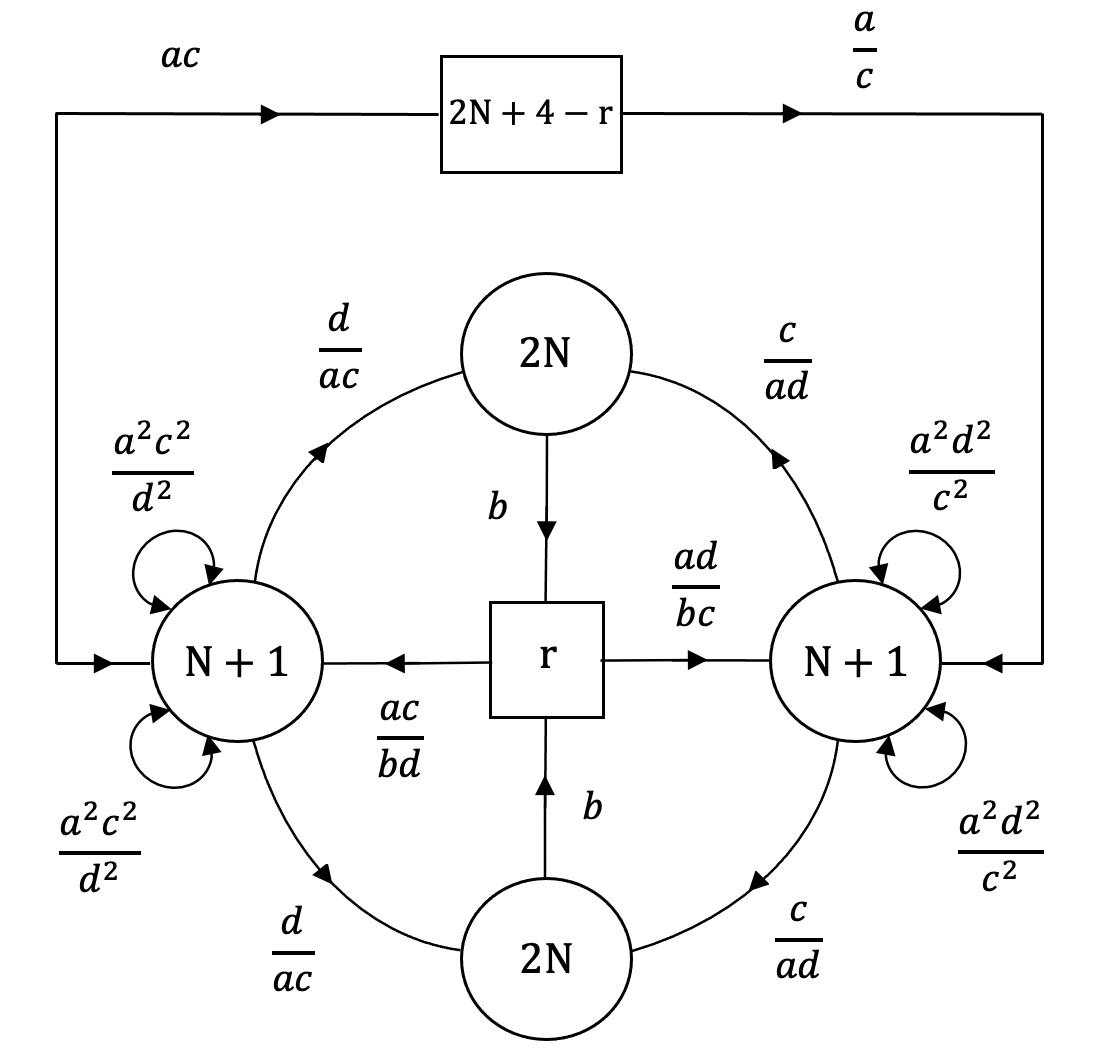} 
	\caption{The $3d$ quiver theory associated with the compactification of the $5d$ SCFT with $SU(2)\times SO(4N+8)$ global symmetry on a torus with flux preserving $U(1)^2\times SU(r)\times SO(4N+8-2r)$. There is a cubic superpotential for each triangle of the quiver and for each antisymmetric chiral coupling it to two bifundamentals. In addition, there is a quartic superpotential coupling the $SU(N+1)\times SU(2N+4-r)$ bifundamentals to the $USp(2N)\times SU(N+1)$ bifundamentals along the semi circles.}
	\label{generalSUSP}
\end{figure}

At this point we can take two copies of this tube and glue them together to form a torus with flux
\begin{equation}
(1,-1;\underbrace{1,1,...,1}_{r},\underbrace{0,0,...,0}_{2N+4-r}),
\end{equation}
see figure \ref{generalSUSP} (we take the Chern--Simons levels here to be zero). This flux breaks the $5d$ global symmetry as follows: 
\begin{equation}
SU(2)\rightarrow U(1)_{SU(2)}\,,
\end{equation}
\begin{equation}
SO(4N+8)\rightarrow U(1)_{SO(4N+8)}\times SU(r)\times SO(4N+8-2r)
\end{equation}
where we recall that $r$ is even. The symmetry group of the $3d$ UV Lagrangian contains a $U(1)\times SU(2N+4-r)$ part which is expected to enhance in the IR to $SO(4N+8-2r)$. Moreover, there are additional possible $U(1)$ symmetries that we take into account and that will guide us to the correct monopole superpotential we should add. Indeed, we will see that a monopole superpotential preserving only some combinations of them will be needed in order for the $3d$ theory to have the correct properties for being the compactification of the $5d$ SCFT. The charges under all these possible $U(1)$ symmetries are denoted in figure \ref{generalSUSP} using their corresponding fugacities. 

We next turn to check this torus compactification proposal and find the aforementioned monopole superpotential. As discussed in subsection \ref{tests}, we expect to find $3d$ operators coming from both the $5d$ conserved current and the Higgs branch chiral ring operator in the $(\mathbf{2},\mathbf{S})$ of $SU(2)\times SO(4N+8)$. Let us start with the conserved current, which yields $3d$ operators with $R_{5d}=2$. The current of the $5d$ SCFT transforms in the adjoint representation of its symmetry, $SU(2)\times SO(4N+8)$, which splits under the subgroup $U(1)^{2}\times SU(r)\times SO(4N+8-2r)$ preserved by the flux as follows:
\begin{equation*}
{\bf Ad}_{SO(4N+8)}\rightarrow(1,{\bf Ad}_{SO(4N+8-2r)})^{0}\oplus({\bf Ad}_{SU(r)},1)^{0}\oplus(1,1)^{0}\oplus({\bf \Lambda}_{SU(r)}^{2},1)^{2}\oplus
\end{equation*}
\begin{equation}
	\oplus({\bf \Lambda}_{SU(r)}^{r-2},1)^{-2}\oplus({\bf F}_{SU(r)},{\bf V}_{SO(4N+8-2r)})^{1}\oplus(\overline{\bf F}_{SU(r)},{\bf V}_{SO(4N+8-2r)})^{-1}\,,\label{adson}
\end{equation}
\begin{equation}
	\mathbf{3}\rightarrow\mathbf{1}^{0}\oplus\mathbf{1}^{2}\oplus\mathbf{1}^{-2}\,.\label{3su2}
\end{equation}
As we discussed, the $SO(4N+8-2r)$ part of the symmetry emerges only in the IR of the $3d$ theory, while in the UV this symmetry is $U(1)_{SO(4N+8-2r)}\times SU(2N+4-r)$. In terms of this UV symmetry, the representations in the second line of \eqref{adson} are given by 
\bea
	&& ({\bf F}_{SU(r)},{\bf V}_{SO(4N+8-2r)})^{1}\oplus(\overline{\bf F}_{SU(r)},{\bf V}_{SO(4N+8-2r)})^{-1}\rightarrow ({\bf F}_{SU(r)},{\bf F}_{SU(2N+4-r)})_{1}^{1} \nonumber \\ & \oplus & ({\bf F}_{SU(r)},\overline{\bf F}_{SU(2N+4-r)})_{-1}^{1} \oplus (\overline{\bf F}_{SU(r)},{\bf F}_{SU(2N+4-r)})_{1}^{-1}\oplus(\overline{\bf F}_{SU(r)},\overline{\bf F}_{SU(2N+4-r)})_{-1}^{-1} \nonumber \\ && \label{lkxns}
\eea
since 
\begin{equation}
{\bf V}_{SO(4N+8-2r)}\rightarrow({\bf F}_{SU(2N+4-r)})_{1}\oplus(\overline{\bf F}_{SU(2N+4-r)})_{-1}\,.
\end{equation}
We used in \eqref{lkxns} the representation notation $(R_{SU(r)},R_{SU(2N+4-r)})_{U(1)_{SO(4N+8-2r)}}^{U(1)_{SO(4N+8)}}$. 

Now let us identify the $3d$ operators with $R_{5d}=2$ that are in the representations that appear in \eqref{adson} and \eqref{lkxns}. First, we have the $SU(r)\times SU(2N+4-r)$ gauge invariant bifundamentals that we can build from the $SU(r)\times USp(2N)$, $USp(2N)\times SU(N+1)$ and $SU(N+1)\times SU(2N+4-r)$ bifundamentals. These are operators with $R_{5d}=2$ and in the representations $({\bf F}_{SU(r)},{\bf F}_{SU(2N+4-r)})bd$ and $({\bf F}_{SU(r)},\overline{\bf F}_{SU(2N+4-r)})\frac{b}{d}$, where we denoted the $U(1)$ charges here using their fugacities. Since these representations correspond to $({\bf F}_{SU(r)},{\bf F}_{SU(2N+4-r)})_{1}^{1}$ and $({\bf F}_{SU(r)},\overline{\bf F}_{SU(2N+4-r)})_{-1}^{1}$ of \eqref{lkxns}, this suggests the fugacity identification 
\begin{equation}
	x_{U(1)_{SO(4N+8)}}=b\;\,,\;\,x_{U(1)_{SO(4N+8-2r)}}=d\label{eq:idenu1}
\end{equation}
and the symmetry enhancement of $U(1)_{SO(4N+8-2r)}\times SU(2N+4-r)$ to $SO(4N+8-2r)$. As to the remaining representations $(\overline{\bf F}_{SU(r)},{\bf F}_{SU(2N+4-r)})_{1}^{-1}$ and $(\overline{\bf F}_{SU(r)},\overline{\bf F}_{SU(2N+4-r)})_{-1}^{-1}$ of \eqref{lkxns}, these can be identified with the operators built from the $SU(r)\times SU(N+1)$ and $SU(N+1)\times SU(2N+4-r)$ bifundamentals along with two copies of the $USp(2N)\times SU(N+1)$ bifundamental, with the same identification \eqref{eq:idenu1}. 

We are now left with the operators $({\bf \Lambda}_{SU(r)}^{2},1)^{2}$ and $({\bf \Lambda}_{SU(r)}^{r-2},1)^{-2}$ that appear in the first line of \eqref{adson}. The operator $({\bf \Lambda}_{SU(r)}^{2},1)^{2}$ is identified with the gauge invariant built from two copies of the $SU(r)\times USp(2N)$ bifundamental, while the operator $({\bf \Lambda}_{SU(r)}^{r-2},1)^{-2}$ is identified with the gauge invariant built from two copies of the $SU(r)\times SU(N+1)$ and $SU(N+1)\times USp(2N)$ bifundamentals. Both cases correspond to the same identification \eqref{eq:idenu1}. 

Next, we consider operators associated with the broken $SU(2)$ current in \eqref{3su2}. The representation $\mathbf{1}^{2}$ can be identified with the basic monopoles of the $USp(2N)$ gauge groups, $M_{USp(2N)}^{(1,0,...,0)}$, which have $R_{5d}=2$ and carry charges $a^{2(N+1)}b^{-r}$. This leads to the fugacity identification $x_{SU(2)}=a^{N+1}b^{-r/2}$.

This completes our discussion of the identification of states from the $5d$ conserved current, and we next turn to the states coming from the Higgs branch chiral ring operator in the $(\mathbf{2},\mathbf{S})$ of $SU(2)\times SO(4N+8)$ and with $R_{5d}=N+1$. Under 
\begin{equation}
SU(2)\rightarrow U(1)_{SU(2)}\,,
\end{equation}
\begin{equation}
SO(4N+8)\rightarrow U(1)_{SO(4N+8)}\times SU(r)\times SO(4N+8-2r)
\end{equation}
and 
\begin{equation}
SO(4N+8-2r)\rightarrow U(1)_{SO(4N+8-2r)}\times SU(2N+4-r)
\end{equation}
we have the following decompositions (correspondingly), 
\begin{equation}
\mathbf{2}_{SU(2)}\rightarrow\mathbf{1}^{1}\oplus\mathbf{1}^{-1}=a^{N+1}b^{-r/2}+a^{-\left(N+1\right)}b^{r/2}\,,
\end{equation}
\begin{equation*}
{\bf S}_{SO(4N+8)}\rightarrow(1,{\bf S}_{SO(4N+8-2r)})^{-\frac{r}{2}}\oplus({\bf F}_{SU(r)},{\bf C}_{SO(4N+8-2r)})^{-\frac{r}{2}+1}\oplus
\end{equation*}
\begin{equation*}
\oplus({\bf \Lambda}_{SU(r)}^{2},{\bf S}_{SO(4N+8-2r)})^{-\frac{r}{2}+2}\oplus\cdots=
\end{equation*}
\begin{equation*}
=(1,{\bf S}_{SO(4N+8-2r)})b^{-\frac{r}{2}}\oplus({\bf F}_{SU(r)},{\bf C}_{SO(4N+8-2r)})b^{-\frac{r}{2}+1}\oplus
\end{equation*}
\begin{equation}
\oplus(\varLambda_{SU(r)}^{2},S_{SO(4N+8-2r)})b^{-\frac{r}{2}+2}\oplus\cdots\,,
\end{equation}
\begin{equation}
{\bf S}_{SO(4N+8-2r)}\rightarrow\sum_{i=-(N+2-\frac{r}{2})\,,\,\textrm{\ensuremath{i} even}}^{N+2-\frac{r}{2}}({\bf \Lambda}_{SU(2N+4-r)}^{N+2-\frac{r}{2}+i})^{i}=\sum_{i=-(N+2-\frac{r}{2})\,,\,\textrm{\ensuremath{i} even}}^{N+2-\frac{r}{2}}({\bf \Lambda}_{SU(2N+4-r)}^{N+2-\frac{r}{2}+i})d^{i}\,,
\end{equation}
\begin{equation}
{\bf C}_{SO(4N+8-2r)}\rightarrow\sum_{i=-(N+2-\frac{r}{2})\,,\,\textrm{\ensuremath{i} odd}}^{N+2-\frac{r}{2}}({\bf \Lambda}_{SU(2N+4-r)}^{N+2-\frac{r}{2}+i})^{i}=\sum_{i=-(N+2-\frac{r}{2})\,,\,\textrm{\ensuremath{i} odd}}^{N+2-\frac{r}{2}}({\bf \Lambda}_{SU(2N+4-r)}^{N+2-\frac{r}{2}+i})d^{i}\,,
\end{equation}
and we can identify some of the states in the $3d$ theory. For example, we have the $SU(N+1)$ baryons made from $I$ $SU(r)\times SU(N+1)$ bifundamentals, $J$ $SU(2N+4-r)\times SU(N+1)$ bifundamentals and $K$ antisymmetrics, where $I+J+2K=N+1$. The baryons of the right $SU(N+1)$ group have $R_{5d}=N+1$, $U(1)$ charges $a^{N+1}b^{-I}c^{-(N+1)}d^{N+1-J}$ and are in the rank $r-I$ antisymmetric of $SU(r)$ and the rank $2N+4-r-J$ antisymmetric of $SU(2N+4-r)$. The baryons of the left $SU(N+1)$ group have $R_{5d}=N+1$, $U(1)$ charges $a^{N+1}b^{-I}c^{N+1}d^{J-(N+1)}$ and are in the rank $r-I$ antisymmetric of $SU(r)$ and the rank $J$ antisymmetric of $SU(2N+4-r)$. These baryonic representations can be matched with some of the representations appearing in the above decompositions, in which a general term is of the form (here we choose the $a^{N+1}b^{-r/2}$ term in the decomposition of $\mathbf{2}_{SU(2)}$)
\begin{equation}
\label{genterm1}
\left({\bf \Lambda}_{SU(r)}^{m},{\bf \Lambda}_{SU(2N+4-r)}^{N+2-\frac{r}{2}+n}\right)b^{-\frac{r}{2}+m}d^{n}a^{N+1}b^{-r/2}
\end{equation}
for some integers $m$ and $n$. In the case of the right $SU(N+1)$ baryons, we should choose $m=r-I$ and $n=N+2-\frac{r}{2}-J$ in order to match the nonabelian representations, obtaining 
\begin{equation}
\left({\bf \Lambda}_{SU(r)}^{r-I},{\bf \Lambda}_{SU(2N+4-r)}^{2N+4-r-J}\right)a^{N+1}b^{-I}d^{N+2-\frac{r}{2}-J}\,.
\end{equation}
As discussed above, the $U(1)$ charges of the right baryons are $a^{N+1}b^{-I}c^{-(N+1)}d^{N+1-J}$ and we will therefore get an agreement only if the following restriction on fugacities takes place:
\begin{equation}
	c^{2N+2}d^{2-r}=1.\label{restu1}
\end{equation}
Turning to the left $SU(N+1)$ baryons, we should choose $m=r-I$ and $n=-(N+2-\frac{r}{2})+J$ in order to match the nonabelian representations of the baryons and the Higgs-branch chiral-ring operator, obtaining
\begin{equation}
\left({\bf \Lambda}_{SU(r)}^{r-I},{\bf \Lambda}_{SU(2N+4-r)}^{J}\right)a^{N+1}b^{-I}d^{-(N+2-\frac{r}{2})+J}\,.
\end{equation}
As in the case of the right baryons, the $U(1)$ charges match only if the restriction \eqref{restu1} is applied. 

We now notice that there are four basic mixed monopole operators involving adjacent groups\footnote{By this we mean monopoles with minimal flux under any pair of $SU(N+1)$ and $USp(2N)$. In this paper we will sometimes call monopoles with minimal flux under adjacent gauge groups like these $(1,1)$ monopoles.}, two with $U(1)$ charges $c^{2N+2}d^{2-r}$ and two with charges $c^{-(2N+2)}d^{-(2-r)}$. As a result, turning on a monopole superpotential corresponding to these operators exactly yields the constraint on fugacities \eqref{restu1}. We see that we should add a monopole superpotential for adjacent groups in order to match the $3d$ spectrum with that expected from $5d$. 

So far we mentioned the basic monopoles of the $USp(2N)$ gauge groups and the basic mixed $USp-SU$ monopoles. Let us close this subsection with examining the remaining simple type of monopoles -- the basic $SU(N+1)$ ones. The $5d$ R-charge of such a bare monopole is $R_{5d}=2$ and its $U(1)$ charges are $a^{-2N}b^{r}c^{2N}d^{4-r}$ for the operator corresponding to the right $SU(N+1)$ node and $a^{-2N}b^{r}c^{-2N}d^{r-4}$ for the one corresponding to the left node. Since these are charged under the associated gauge groups, in order to construct gauge invariant operators we should dress them with fundamental and antisymmetric fields of the corresponding $SU(N+1)$ that overall form the $N-1$ antisymmetric representation (that is, ${\bf \Lambda}_{SU(N+1)}^{N-1}$). This results in dressed monopoles with $R_{5d}=N+1$, corresponding to operators that descend from the $5d$ Higgs branch chiral ring operator in the $(\mathbf{2},\mathbf{S})$. To keep track of the representations and charges of these operators under the various symmetries, let us denote as before the number of $SU(r)\times SU(N+1)$ bifundamentals used in the dressing by $I$ and the number of $SU(2N+4-r)\times SU(N+1)$ bifundamentals by $J$. Then, the number of $SU(N+1)$ antisymmetrics is $(N-1-I-J)/2$. For the dressed monopole corresponding to the left $SU(N+1)$ node, the representations and charges are therefore 
\begin{equation}
	\left({\bf \Lambda}_{SU(r)}^{r-I},{\bf \Lambda}_{SU(2N+4-r)}^{J}\right)a^{-(N+1)}b^{r-I}c^{-(N+1)}d^{r-3-N+J}\,.\label{lmrep}
\end{equation}
This can be matched with some of the representations appearing in the decomposition of $(\mathbf{2},\mathbf{S})$ above. Choosing the $a^{-\left(N+1\right)}b^{r/2}$ term in the decomposition of $\mathbf{2}_{SU(2)}$ (in contrast to the $a^{N+1}b^{-r/2}$ term used in \eqref{genterm1} above), a general term in the $(\mathbf{2},\mathbf{S})$ decomposition is of the form 
\begin{equation}
	\left({\bf \Lambda}_{SU(r)}^{m},{\bf \Lambda}_{SU(2N+4-r)}^{N+2-\frac{r}{2}+n}\right)a^{-(N+1)}b^{m}d^{n}\label{2Sdecom2}
\end{equation}
for some integers $m$ and $n$. Choosing $m=r-I$ and $n=-(N+2-\frac{r}{2})+J$ exactly yields \eqref{lmrep} when the constraint on fugacities \eqref{restu1}, imposed by the monopole superpotential, is taken into account. 

Turning to the dressed monopole corresponding to the right $SU(N+1)$ node, the representations and charges are 
\begin{equation}
	\left({\bf \Lambda}_{SU(r)}^{r-I},{\bf \Lambda}_{SU(2N+4-r)}^{2N+4-r-J}\right)a^{-(N+1)}b^{r-I}c^{N+1}d^{3-r+N-J}\,.\label{rmrep}
\end{equation}
Choosing $m=r-I$ and $n=N+2-\frac{r}{2}-J$ in \eqref{2Sdecom2} yields \eqref{rmrep} as expected, again under the constraint \eqref{restu1} imposed by the monopole superpotential.

\subsection{Cases built from $SU-SU$ tubes}

We now consider tubes with an holonomy such that we get a $4d$ $SU(N+1)$ gauge theory on both sides so that the tube has two $SU(N+1)$ punctures. As all the $5d$ SCFTs we consider here can be realized as the UV completions of the $5d$ $SU(N+1)$ gauge theories, such tubes can exist for all cases. Like the previous case, the $4d$ compactifications of the $6d$ SCFT parent of this family of $5d$ SCFTs were studied in \cite{Kim:2018bpg}, and we can employ their results to try to formulate conjectures for possible $3d$ compactifications of the $5d$ SCFTs.

\begin{figure}
\center
\includegraphics[width=0.95\textwidth]{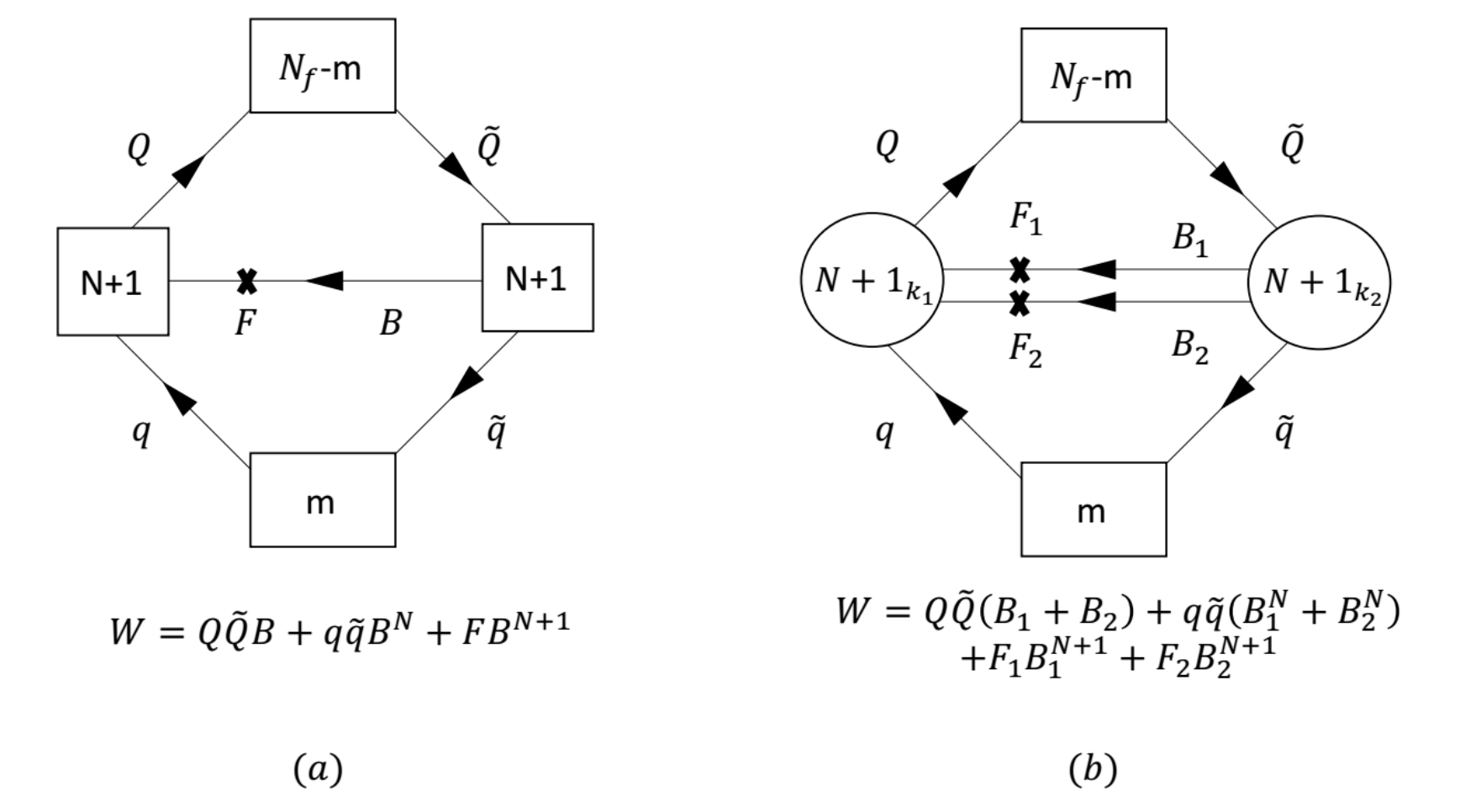} 
\caption{The $3d$ theories we conjectured are associated with the compactification of the $5d$ SCFTs we consider on (a) a tube with two $SU(N+1)$ maximal punctures and (b) a torus. Here the subscripts $k_1$, $k_2$ stand for the Chern-Simons level of the associated gauge groups.}
\label{SUSUtube}
\end{figure}

Using the methods employed previously, we are led to conjecture the model shown in figure \ref{SUSUtube} (a) for the tube. Here the two $SU(N+1)$ global symmetry groups come from the $4d$ gauge symmetry on the two sides of the interval. The $N_f$ chiral fields come from the components of the $N_f$ bulk hypermultiplets receiving Neumann boundary conditions. These chirals can be either in the fundamental or the antifundamental of the $SU(N+1)$ groups, depending on which component of the hyper receives the Neumann boundary conditions. Here we have allowed for $m$ of them to have one chirality and the other $N_f-m$ to have the other chirality. As we shall see, the flavor distribution distinguishes between $5d$ SCFTs associated with UV completions of the same gauge theories, but with different CS terms.    

Finally, we need to add the fields on the domain wall, which following the very similar $4d$ tubes in \cite{Kim:2018bpg}, we conjecture are just an $SU(N+1)\times SU(N+1)$ bifundamental and a singlet field coupling linearly to the baryon made from the bifundamental. The $SU(N+1)$ chiral fields on the two sides couple through a superpotential involving the bifundmental, which is cubic for one choice of chirality, but involves $N$ $SU(N+1)\times SU(N+1)$ bifundamentals for the other choice, due to the specific $SU(N+1)\times SU(N+1)$ representation of the bifundamental. As such the tube is not invariant under $m\rightarrow N_f-m$. These superpotentials ensure that the $5d$ global symmetry acting on each segment of the interval is properly identified.

Once we have a conjecture for the tube, we can take two copies of it, glue them together and get the theory associated with the torus compactification and twice the value of flux. This is shown in figure \ref{SUSUtube} (b). In the gluing process, we have the freedom of turning on Chern--Simons terms for the two groups, which we have denoted by $k_1$, $k_2$. We can next subject these tubes to several tests, mainly by computing their supersymmetric index and comparing against the $5d$ expectations. These can be used to uncover what choices of $m$, $k_1$ and $k_2$ correspond to compactification of $5d$ SCFTs and what is the associated flux. As the computation of the index for generic $N$ and $N_f$ is quite involved, it is convenient to first make an in-dept study of the simpler low $N$ cases, and then use the observations made there to understand the structure for generic $N$.  

For $N=1$, the tube just reduces to the one studied in \cite{Sacchi:2021afk}. Here $m$ is irrelevant as the fundamental representation of $SU(2)$ is self-conjugate. The tube appears to describe the compactifiction of $5d$ SCFTs for the cases of $N_f\leq 6$ with $k_1 = -k_2 = \frac{6-N_f}{2}$. The relation for the case of $N_f=7$ is less clear. The next case is the $N=2$ one, which is also the first case where $m$ is relevant, and is the case that we shall begin with. The highest value of $N_f$ possible for a $5d$ SCFT is $N_f=9$. However, like the $N_f=7$ case for $N=1$, this case does not appear to have a direct $5d$ interpretation.  

 The first case where we do find interesting theories that pass our tests is $N_f=8$, which is the case we shall start with. We will then move to theories with still $N=2$, but with lower $N_f$. These can be obtained in $5d$ by giving various real masses to the flavors. We will then understand the pattern of these deformations in the $3d$ theories. Later, we will perform the same analysis but for $N=3$, to validate the general pattern of deformations.

\begin{figure}
\center
\includegraphics[width=0.96\textwidth]{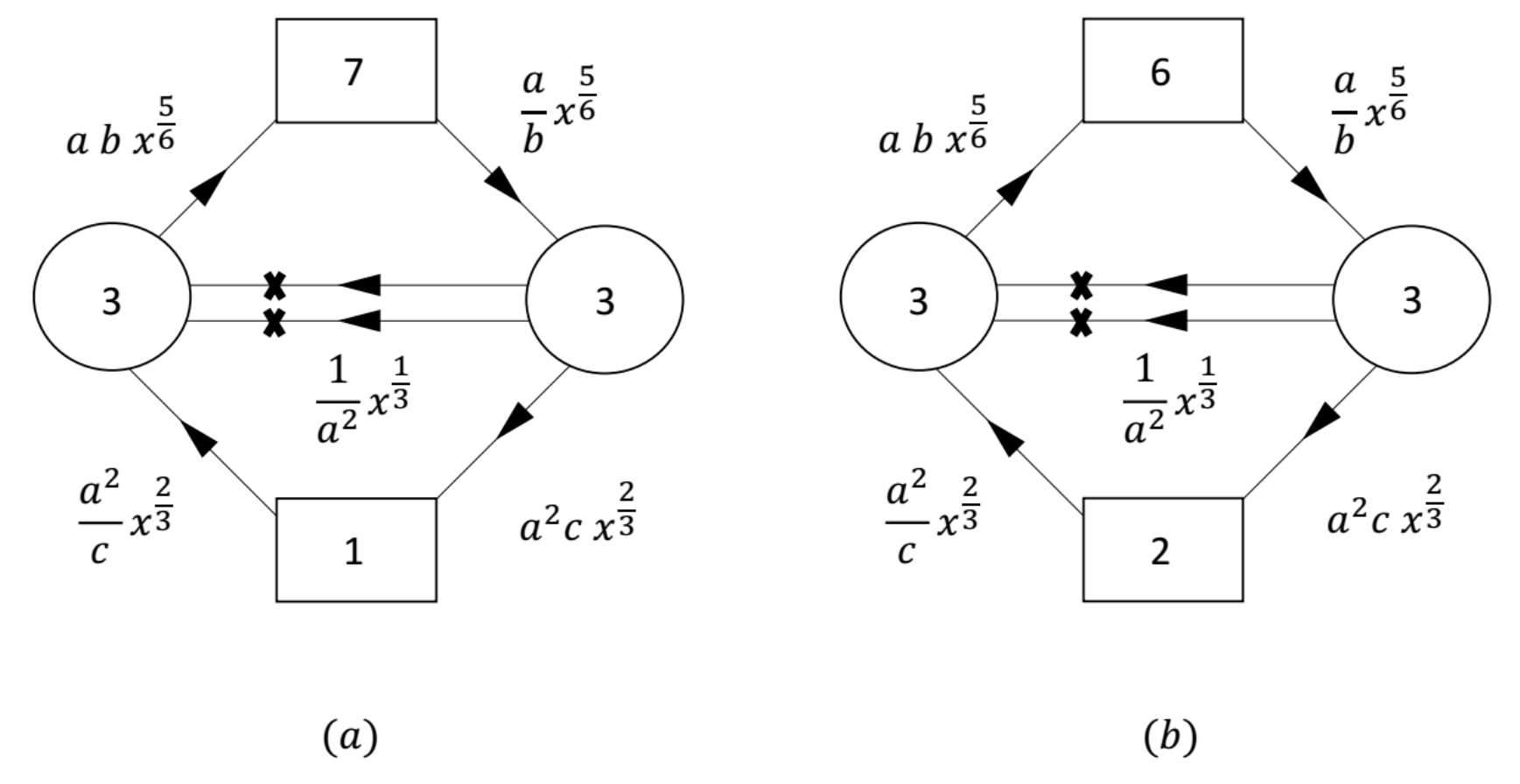} 
\caption{Two $3d$ models that are conjectured to be the compactification of the $5d$ SCFTs UV completing $SU(3)+8F$ gauge theories on a torus with flux. The charges of the various fields in these models under the flavor and R-symmetry $U(1)$ groups are illustrated using fugacities. There is also a superpotential that can be read from figure \ref{SUSUtube} or alternatively as the general one consistent with all the symmetries.}
\label{QuiversSU3X23d}
\end{figure}

\subsubsection{$N=2$, $N_f=8$}

Here we find two cases that have the right characteristic to be compactifications of the $5d$ SCFTs we are considering. These correspond to the choices $m=1$ and $m=2$, with $k_1=k_2=0$ for both cases. The two models are shown in figure \ref{QuiversSU3X23d}. We next perform a more detailed study of these models, highlighting their $5d$ origin and their properties that support such an identification. As the models differ by the $SU(N+1)$ representation of the $N_f$ chirals, we shall denote the different models by $(N_f-m,m)$, giving the split of the chiral fields between the two conjugate representations.  



\subsubsection*{The case of the $(6,2)$ split}
\label{susu62}

We begin by considering the $(6,2)$ split, which is the one shown in figure \ref{QuiversSU3X23d} (b). Here, the superconformal R-symmetry, which we denote by $U\left(1\right)_{\hat{R}}$, is obtained using F-maximization \cite{Jafferis:2010un} and is given by a mixing of $U(1)_R$ (appearing in the figure) and $U(1)_a$. Under this symmetry, the flip fields turn out to violate the unitarity bound (having R-charge smaller than 0.5), and the value of $U\left(1\right)_{\hat{R}}$ in the theory obtained after removing them is as follows:
\begin{equation}
	\hat{R}=R-0.184q_{a}\,.
\end{equation}

Let us compute the index of this model with no Chern--Simons terms. Keeping the flip fields and using $U(1)_R$ (shown in the figure), the index is given by
\bea
\label{ind62}
& & \mathcal{I} = 1 + (2 a^6 + \frac{4}{a^6}) x + a^3 \bold{2}_{SU(2)} (\frac{b}{c} \bold{6}_{SU(6)} + \frac{c}{b} \bold{\overline{6}}_{SU(6)}) x^{\frac{3}{2}} + (6 + 2\bold{3}_{SU(2)} + 3a^{12} + \frac{14}{a^{12}}) x^2 \nonumber \\  & + & ( \frac{1}{a^3} \bold{2}_{SU(2)}(\frac{b}{c} \bold{6}_{SU(6)} + \frac{c}{b} \bold{\overline{6}}_{SU(6)}) + a^3 (b^3 + \frac{1}{b^3})(\bold{20}_{SU(6)} + \frac{c^2}{b^2}\bold{6}_{SU(6)} + \frac{b^2}{c^2}\bold{\overline{6}}_{SU(6)}) \nonumber \\  & + & 2a^9 \bold{2}_{SU(2)}(\frac{b}{c} \bold{6}_{SU(6)} + \frac{c}{b} \bold{\overline{6}}_{SU(6)})) x^{\frac{5}{2}} + \cdots .
\eea
This index forms characters of $SU(2)^2\times SO(12)$, where the embedding is 
\bea
\label{embs012}
&&\bold{12}_{SO(12)}\rightarrow \frac{b}{c} \bold{6}_{SU(6)} + \frac{c}{b} \bold{\overline{6}}_{SU(6)}\nn\\
&&\bold{2}_{SU(2)_2}\rightarrow b^3 + \frac{1}{b^3}
\eea
and with $SU(2)_1$ being the $SU(2)$ already visible in the gauge theory. In terms of this symmetry, the index reads
\bea
& & \mathcal{I} = 1 + (2 a^6 + \frac{4}{a^6}) x + a^3 \bold{2}_{SU(2)_1} \bold{12}_{SO(12)} x^{\frac{3}{2}} + (6 + 2\bold{3}_{SU(2)_1} + 3a^{12} + \frac{14}{a^{12}}) x^2 \nonumber \\  & + & ( \frac{1}{a^3}\bold{2}_{SU(2)_1} \bold{12}_{SO(12)} + a^3 \bold{2}_{SU(2)_2} \bold{32}_{SO(12)} + 2a^9 \bold{2}_{SU(2)_1} \bold{12}_{SO(12)}) x^{\frac{5}{2}} + ... . \label{Index26split}
\eea

This index is mostly consistent with the theory being the result of the compactification of the $SU(2)\times SO(16)$ $5d$ SCFT on a torus with a unit flux in the $U(1)$ whose commutant in $SO(16)$ is $SU(2)\times SO(12)$. Specifically, the index forms characters of $U(1)\times SU(2)^2\times SO(12)$, which is the symmetry expected from the $5d$ picture. Furthermore, under the embedding of the $U(1)$ inside $SO(12)$, we have the following branching rules:
\bea
\bold{120} & \rightarrow & 1 + a^6 + \frac{1}{a^6} + \bold{3}_{SU(2)_1} + \bold{66}_{SO(12)} + (a^3 + \frac{1}{a^3}) \bold{2}_{SU(2)_1} \bold{12}_{SO(12)} , \\ \nonumber \bold{128} & \rightarrow & \bold{2}_{SU(2)_1} \bold{32'}_{SO(12)} + (a^3 + \frac{1}{a^3}) \bold{32}_{SO(12)} .
\eea
We previously mentioned that there are two types of Higgs branch chiral ring generators in this theory. One consists of the moment map operators associated with the conserved currents. The contribution of these to the index precisely matches the terms $2 a^6 x$ and $a^3 \bold{2}_{SU(2)_1} \bold{12}_{SO(12)} x^{\frac{3}{2}}$ in \eqref{Index26split}. The second one is the Higgs branch chiral ring operator in the $\bold{2}$ of the $SU(2)$, the $\bold{128}$ of $SO(16)$ and the $\bold{4}$ of $SU(2)_R$. Its contribution matches the $a^3 \bold{2}_{SU(2)_2} \bold{32}_{SO(12)} x^{\frac{5}{2}}$ term in the index.

However, there are two deviations from our general expectations. First the sign of the term $\frac{1}{a^3}\bold{2}_{SU(2)_1} \bold{12}_{SO(12)}x^{\frac{5}{2}}$ is opposite from what we expect. This term should come from the conserved current multiplet and we expect the coefficient to be $-1$. Another issue is the presence of a large number of marginal operators implying that the conformal manifold behaves differently than expected. We claim that this deviation from the expected results occurs only sporadically for low value of flux. 

We can consider for example the model corresponding to flux 2. This is obtained by gluing together four copies of the basic tube, so it will have four $SU(3)$ gauge nodes. In this case, we also turn on a monopole superpotential containing all the $(1,1)$ monopoles for adjacent gauge nodes\footnote{These are monopoles with unit magnetic flux under two adjacent gauge nodes. In the case of the $(6,2)$ split that we are considering, these monopole operators are gauge invariant. When this is not the case, as it will happen for different splits and for lower number of flavors, the monopoles must be properly dressed with powers of the bifundamentals so to make them gauge invariant.}. As it was observed already in \cite{Sacchi:2021afk} for the compactification of the rank 1 $E_{N_f+1}$ SCFTs, this is needed in order for the theory to possess the correct global symmetry expected from $5d$. Moreover, as we saw in the previous subsection in the case of theories constructed from $SU-USp$ tubes, this superpotential is needed in order to match the spectrum of the $3d$ theory with the one expected from $5d$.
The index of the model with flux 2 is then
\be\label{Index62splitflux2}
\mathcal{I}=1+4a^6x+2a^3{\bf 2}_{SU(2)}\left(\frac{b}{c}{\bf 6}_{SU(6)}+\frac{c}{b}\overline{\bf 6}_{SU(6)}\right)x^{\frac{3}{2}}+\left(10a^{12}+\frac{4}{a^{12}}\right)x^2+\cdots\,.\nn\\
\ee
We can see that the conformal manifold now conforms to our expectations. Also this index can be written in terms of $SU(2)^2\times SO(12)$ characters\footnote{Here the monopole superpotential is crucial, since otherwise we would have an additional $U(1)$ symmetry that would prevent us from re-arranging the terms in the index \eqref{Index62splitflux2} into $SO(12)$ characters.}
\be
\mathcal{I}=1+4a^6x+2a^3 \bold{2}_{SU(2)_1} \bold{12}_{SO(12)} x^{\frac{3}{2}}+\left(10a^{12}+\frac{4}{a^{12}}\right)x^2+\cdots\,.\nn\\
\ee
and again we can check the presence of operators expected from $5d$.

Let us finally comment that additional evidence for the proposed enhancement of the $SU(6)\times U(1)_{b/c}$ part of the symmetry to $SO(12)$ comes from examining the relation between the central charges of these symmetries. These can be computed numerically from the real part of the free energy by studying its dependence on the mixing coefficients of these symmetries with the R-symmetry. Specifically, calculating the second derivatives at the superconformal point (that is, where the mixing coefficients take the values corresponding to the IR R-symmetry) yields the central charges, see Eq. \eqref{d2F} and the discussion around it (see also \cite{Closset:2012vg} and the discussion in appendix B of \cite{Sacchi:2021afk} for more details). Applying this procedure for $U(1)_{b/c}$ and for the Cartan $\textrm{diag}\left(1,0,0,0,0,-1\right)$ of $SU(6)$, which we denote by $C$, we find the following values for the central charges:
\begin{equation}
	C_{b/c}=10.05\,\,\,,\,\,\,C_{C}=3.34\,.
\end{equation}
The value of the ratio of these charges is
\begin{equation}
	\label{ratioC}
	\frac{C_{b/c}}{C_{C}}=3.01,
\end{equation}
which exactly matches our expectations from the proposed symmetry enhancement and the embedding \eqref{embs012}. Indeed, the corresponding embedding indices are
\begin{equation}
	I_{SU\left(6\right)\rightarrow SO\left(12\right)}=1\,,\,\,\,I_{U\left(1\right)_{b/c}\rightarrow SO\left(12\right)}=12\,,\,\,\,I_{U\left(1\right)_{C}\rightarrow SU\left(6\right)}=4\,,
\end{equation}
implying the following relations between the various central charges:
\begin{equation}
	C_{SU\left(6\right)}=C_{SO\left(12\right)}\,,\,\,\,C_{b/c}=12C_{SO\left(12\right)}\,,\,\,\,C_{C}=4C_{SU\left(6\right)}\,.
\end{equation}
This, in turn, results in the ratio
\begin{equation}
	\label{CbCc12}
	\frac{C_{b/c}}{C_{C}}=3
\end{equation}
which agrees with our numerical result \eqref{ratioC} within the accuracy of the calculation.

\subsubsection*{The case of the $(7,1)$ split}

We next consider the case of the $(7,1)$ split, which is the one shown in figure \ref{QuiversSU3X23d} (a). The index of the model is given by
\bea
& & \mathcal{I} = 1 + (2 a^6 + \frac{5}{a^6}) x + a^3 (\frac{b}{c} \bold{7} + \frac{c}{b} \bold{\overline{7}} + \frac{c}{b^7} + \frac{b^7}{c}) x^{\frac{3}{2}} + (9 + 3a^{12} + \frac{20}{a^{12}}) x^2 \\ \nonumber & + & ( \frac{1}{a^3} (\frac{b}{c} \bold{7} + \frac{c}{b} \bold{\overline{7}} + \frac{c}{b^7} + \frac{b^7}{c}) + a^3 (b^3 \bold{35} + \frac{1}{b^3} \bold{\overline{35}}) + 2a^9 (\frac{b}{c} \bold{7} + \frac{c}{b} \bold{\overline{7}} + \frac{c}{b^7} + \frac{b^7}{c})) x^{\frac{5}{2}} + \cdots .
\eea
This index forms characters of $SU(8)$, where the embedding is 
\be
\bold{8}\rightarrow \frac{1}{b^{\frac{3}{4}}}\bold{7} + b^{\frac{21}{4}}\,.
\ee
In terms of this symmetry, the index reads
\bea \label{Index17split}
& & \mathcal{I} = 1 + (2 a^6 + \frac{5}{a^6}) x + a^3 (\frac{b^{\frac{7}{4}}}{c} \bold{8} + \frac{c}{b^{\frac{7}{4}}} \bold{\overline{8}}) x^{\frac{3}{2}} + 3(3 + a^{12} + \frac{6}{a^{12}}) x^2 \\ \nonumber & + & ( \frac{1}{a^3} (\frac{b^{\frac{7}{4}}}{c} \bold{8} + \frac{c}{b^{\frac{7}{4}}} \bold{\overline{8}}) + a^3 \bold{70} + 2a^9 (\frac{b^{\frac{7}{4}}}{c} \bold{8} + \frac{c}{b^{\frac{7}{4}}} \bold{\overline{8}})) x^{\frac{5}{2}} + ... .
\eea

This index is mostly consistent with the theory being the result of the compactifiation of the $SU(10)$ $5d$ SCFT on a torus with a unit flux in the $U(1)$ whose commutant in $SU(10)$ is $U(1)\times SU(8)$. Specifically, the index forms characters of $U(1)^2\times SU(8)$, which is the symmetry expected from the $5d$ picture. Furthermore, under the embedding of the $U(1)$ inside $SU(10)$, we have the following branching rules:

\bea
\bold{99} & \rightarrow & 2 + a^6 + \frac{1}{a^6} + \bold{63} + (a^3 + \frac{1}{a^3}) (\alpha^5 \bold{8} + \frac{1}{\alpha^5}\bold{\overline{8}}) , \\ \nonumber \bold{252} & \rightarrow & \alpha^5 \bold{\overline{56}} + \frac{1}{\alpha^5}\bold{56} + (a^3 + \frac{1}{a^3}) \bold{70} .
\eea
We previously mentioned that we we have two types of Higgs branch chiral ring generators in this theory. One consists of the moment map operators associated with the conserved currents. The contribution of these to the index precisely matches the terms $2 a^6 x$ and $a^3 (\frac{b^{\frac{7}{4}}}{c} \bold{8} + \frac{c}{b^{\frac{7}{4}}} \bold{\overline{8}})$ in \eqref{Index17split}, where here $\alpha^5=\frac{b^{\frac{7}{4}}}{c}$. The second one is the Higgs branch chiral ring operator in the $\bold{252}$ of $SU(10)$ and the $\bold{4}$ of $SU(2)_R$. Its contribution matches the $a^3 \bold{70} x^{\frac{5}{2}}$ term in the index.

However, there are two deviations from our general expectations. First the sign of the term $\frac{1}{a^3}(\frac{b^{\frac{7}{4}}}{c} \bold{8} + \frac{c}{b^{\frac{7}{4}}} \bold{\overline{8}}) x^{\frac{5}{2}}$ is opposite from what we expect. This term should come from the conserved current multiplet and we expect the coefficient to be $-1$. Another issue is the presence of a large number of marginal operators implying that the conformal manifold behaves differently than expected. Similarly to the previous case, we claim that this deviation from the expected results occurs only sporadically for low value of flux. Indeed, computing the index of the model with flux 2
\be
\mathcal{I}=1+4a^6x+2a^3\left(\frac{b}{c}{\bf 7}+\frac{c}{b}\overline{\bf 7}+\frac{b^7}{c}+\frac{c}{b^7}\right)x^{\frac{3}{2}}+\left(10a^{12}+\frac{5}{a^{12}}\right)x^2+\cdots
\ee
we can see that the conformal manifold now conforms to our expectations. Also this index can be written in terms of $SU(8)$ characters
\be
\mathcal{I}=1+4a^6x+2a^3 (\frac{b^{\frac{7}{4}}}{c} \bold{8} + \frac{c}{b^{\frac{7}{4}}} \bold{\overline{8}}) x^{\frac{3}{2}}+\left(10a^{12}+\frac{5}{a^{12}}\right)x^2+\cdots
\ee
and again we can check the presence of operators expected from $5d$.

%
%
%
%
%

As in the case of the $(6,2)$ split, we can further test the proposed symmetry enhancement by computing the IR central charges and checking the relation between them. Again the flip fields turn out to violate the unitarity bound, and the superconformal R-charge of the theory obtained after removing them is given by $\hat{R}=R-0.18q_{a}$. To test the expected enhancement, we compute at IR fixed point of the theory without the flip fields the central charges of $U(1)_{b}$ and of the Cartan $\textrm{diag}\left(1,0,0,0,0,0,-1\right)$ of $SU(7)$ (which we denote by $C$). We obtain $C_{b}\approx61$ and $C_{C}\approx4$ with the ratio between them being $15.25$. This is indeed the ratio expected from the enhancement within the accuracy of the calculation, since the embedding indices of these two $U(1)$ symmetries in $SU(8)$ are 
\begin{equation}
	I_{U\left(1\right)_{b}\rightarrow SU\left(8\right)}=63\,,\,\,\,I_{U\left(1\right)_{C}\rightarrow SU\left(8\right)}=4\,.
\end{equation}
These, in turn, imply the following relations between the central charges: 
\begin{equation}
	C_{b}=63C_{SU(8)}\,,\,\,\,C_{C}=4C_{SU\left(8\right)}\,,
\end{equation}
corresponding to the ratio $C_{b}/{C_{C}}=15.75$. This indeed agrees with our numerical result, again within the accuracy of the calculation.

\subsubsection{Integrating out flavors for $N=2$}

\begin{figure}
\center
\includegraphics[width=1\textwidth]{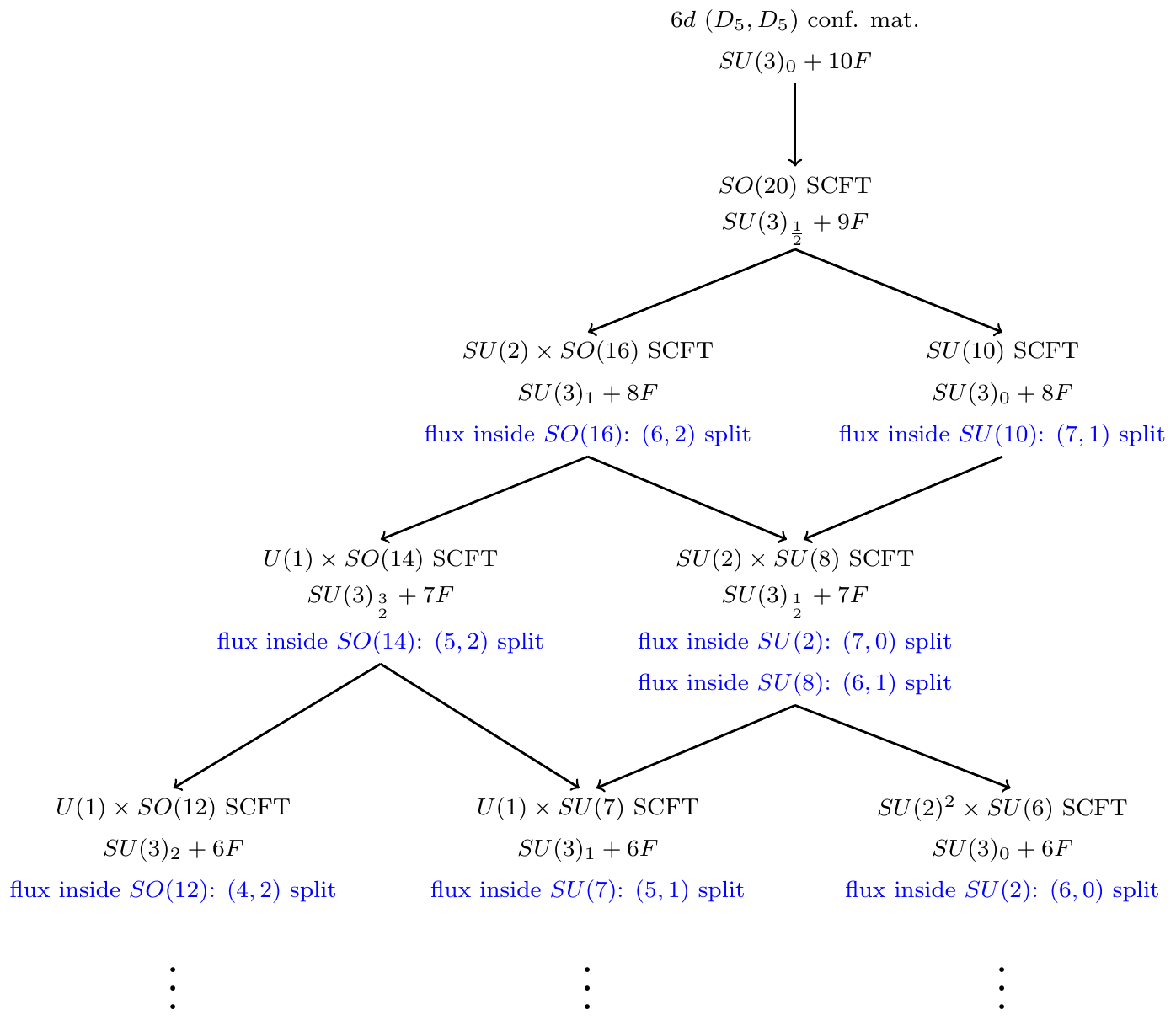} 
\caption{Flow pattern triggered by mass deformations for the $N=2$ case. At each step we specify the $5d$ SCFT ($6d$ SCFT for the first step), the $5d$ gauge theory based on $SU$ group that is UV completed by it and the split of the associated $3d$ $\mathcal{N}=2$ theory that is obtained from the torus compactifications. For some SCFTs we can have more splits and each one is associated with a different factor inside the $5d$ global symmetry for which we turn on the flux.}
\label{fig:N2flowchart}
\end{figure}

We can now consider models of the same structure but with less flavors. These can be obtained by integrating out flavors from the previous models and are thus expected to be associated with the compactification of $5d$ SCFTs that UV complete the $SU(3)_k$ gauge theories with lower $N_f$. We consider only the cases with $N_f\ge 6$, from which it will be clear that those with lower values of $N_f$ work out similarly. We summarize the findings of this subsection in figure \ref{fig:N2flowchart}. The parametrization of the abelian symmetries and of the R-symmetry that we will use is the same as the one we used for the models in figure \ref{QuiversSU3X23d}.

\subsubsection*{The $(7,0)$ split}

We consider first the case of the $(7,0)$ split. Here the number of flavors is too small and the quiver is bad\footnote{Here we use a definition of badness of a theory which is a slight modification of that of \cite{Gaiotto:2008ak} that applies to $3d$ $\mathcal{N}=2$ theories. Namely, for us a theory is bad when its partition functions, such as the supersymmetric index, is divergent. This typically happens when there is no choice of R-symmetry such that all the gauge invariant operators, including monopoles, have a non-negative R-charge.}. One way to make the theory good is to turn on Chern--Simons interactions. This can also be understood as integrating out one flavor from the theory with the $(7,1)$ split using a real mass deformation. We take the Chern--Simons levels to be $(-\frac{1}{2},+\frac{1}{2})$, where the level $\frac{1}{2}$ is for the $SU(3)$ node for which the bifundamental between the two gauge nodes is in the anti-fundamental representation and the level $-\frac{1}{2}$ is for the other\footnote{We point out that the $SU(3)$ gauge nodes must come with a non-trivial half-integer Chern--Simons level in order for the theory to be gauge invariant at the quantum level \cite{Aharony:1997bx,Intriligator:2013lca}.}. The index of the model for flux 1 is
\be
\mathcal{I}=1+\left(2a^6+\frac{12}{a^6}\right)x+\left(16+3a^{12}+\frac{64}{a^{12}}\right)x^2+a^3\left(b^3{\bf 35}+\frac{1}{b^3}\overline{\bf 35}\right)x^{\frac{5}{2}}+\cdots\,.
\ee
We can see that the index forms characters of $SU(8)$ where the embedding is such that, for example, ${\bf 70}_{SU(8)}\to b^3{\bf 35}+\frac{1}{b^3}\overline{\bf 35}$. The index in terms of $SU(8)$ characters is
\be
\mathcal{I}=1+\left(2a^6+\frac{12}{a^6}\right)x+\left(16+3a^{12}+\frac{64}{a^{12}}\right)x^2+a^3{\bf 70}x^{\frac{5}{2}}+\cdots\,.
\label{ind70split}
\ee
This result is compatible with the theory being the result of the torus compactification of the $5d$ $SU(2)\times SU(8)$ SCFT with a unit of flux for the $SU(2)$ factor. Notice that such $5d$ SCFT can be obtained as a mass deformation of the $SU(10)$ SCFT, which is compatible with the fact that the model with $(7,0)$ split can be obtained as a mass deformation of the model with $(7,1)$ split. We will discuss in more details the mapping of the mass deformations between $3d$ and $5d$ in the next subsection.

The spectrum of operators is also consistent with this claim. The set of the HB chiral ring generators of the $5d$ $SU(2)\times SU(8)$ SCFT contains the moment map for the $SU(2)$ global symmetry and an operator in the $\bf 2$ of $SU(2)$, the $\bf 70$ of $SU(8)$ and the $\bf 4$ $SU(2)_R$. Using the branching rules
\bea
&&{\bf 2}_{SU(2)}\to a^3+\frac{1}{a^3},\nn\\
&&{\bf 3}_{SU(2)}\to 1+a^6+\frac{1}{a^6}\,,
\eea
we can identify the contributions of some of these states in the index \eqref{ind70split}. Specifically, the term $2a^6x$ comes from a state contained in the $SU(2)$ current and the term $a^3{\bf 70}x^{\frac{5}{2}}$ comes from a state contained in the other HB operator, where we recall that the R-charge used in \eqref{ind70split} is related to the $5d$ R-charge by the shift $a\to a x^{\frac{1}{6}}$.

Simiarly to the previous cases, we encounter an issue with the conformal manifold of the theory, which seems to be larger than what expected from $5d$. Again we claim that this is an issue occurring only for low value of flux. Indeed, computing the index of the model with flux 2
\be
\mathcal{I}=1+4a^6x+\left(10a^{12}+\frac{12}{a^{12}}\right)x^2+\cdots\,,
\ee
we can see that the conformal manifold now conforms to our expectations. 

\subsubsection*{The $(6,1)$ split}

We next consider the case of the $(6,1)$ split, where again we take the Chern--Simons levels to be $(-\frac{1}{2},+\frac{1}{2})$. This can also be understood as integrating out one flavor in the theory with the $(7,1)$ split or in the one with the $(6,2)$ split. The index of the model for flux 1 is
\bea
\mathcal{I}&=&1+\left(2a^6+\frac{8}{a^6}\right)x+a^3\left(\frac{b}{c}{\bf 6}+\frac{c}{b}\overline{\bf 6}\right)x^{\frac{3}{2}}+\left(14+3a^{12}+\frac{36}{a^{12}}\right)x^2+\nn\\
&+&\left(a^3\left(b^3+\frac{1}{b^3}\right){\bf 20}+2\left(a^9+\frac{1}{a^3}\right)\left(\frac{b}{c}{\bf 6}+\frac{c}{b}\overline{\bf 6}\right)\right)x^{\frac{5}{2}}+\cdots\,.
\eea
We can see that the index forms characters of $SU(2)$, where the embedding is ${\bf 2}_{SU(2)}\to b^3+\frac{1}{b^3}$
\bea
\mathcal{I}&=&1+\left(2a^6+\frac{8}{a^6}\right)x+a^3\left(\frac{b}{c}{\bf 6}_{SU(6)}+\frac{c}{b}\overline{\bf 6}_{SU(6)}\right)x^{\frac{3}{2}}+\left(14+3a^{12}+\frac{36}{a^{12}}\right)x^2+\nn\\
&+&\left(a^3{\bf 2}_{SU(2)}{\bf 20}_{SU(6)}+2\left(a^9+\frac{1}{a^3}\right)\left(\frac{b}{c}{\bf 6}_{SU(6)}+\frac{c}{b}\overline{\bf 6}_{SU(6)}\right)\right)x^{\frac{5}{2}}+\cdots\,.
\label{ind61split}
\eea
Remember that this theory can be either obtained as a mass deformation of the $(7,1)$ theory corresponding to the $5d$ $SU(10)$ SCFT or as a mass deformation of the $(6,2)$ theory corresponding to the $5d$ $SU(2)\times SO(16)$. Both of these SCFTs have a mass deformation to the $5d$ $SU(2)\times SU(8)$ SCFT, so the natural interpretation is that the $(6,1)$ theory corresponds to the torus compactification of such SCFT. In contrast with the previous case of the $(7,0)$ split, though, the flux is in the $SU(8)$ part, specifically it is of the form $(0;1,-1,0,\cdots,0)$, and preserves $U(1)^2\times SU(6)\subset SU(8)$.

The spectrum of operators is also consistent with this claim. Recall from the previous section that the set of the HB chiral ring generators of the $5d$ $SU(2)\times SU(8)$ SCFT, on top of the moment map for the $SU(8)$ global symmetry, contains also an operator in the $\bf 2$ of $SU(2)$, the $\bf 70$ of $SU(8)$ and the $\bf 4$ of $SU(2)_R$. This time the flux breaks $SU(8)\to U(1)^2\times SU(6)$, so we need to use the branching rules
\bea
&&{\bf 63}_{SU(8)}\to 2+a^6+\frac{1}{a^6}+{\bf 35}_{SU(6)}+\ga^4\left(a^3+\frac{1}{a^3}\right){\bf 6}_{SU(6)}+\ga^{-4}\left(a^3+\frac{1}{a^3}\right)\overline{\bf 6}_{SU(6)},\nn\\
&&{\bf 70}_{SU(8)}\to \left(a^3+\frac{1}{a^3}\right){\bf 20}_{SU(6)}+\ga^4\overline{\bf 15}_{SU(6)}+\ga^{-4}{\bf 15}_{SU(6)}\,.
\eea
Using these, we can identify the contributions of some of these states in the index \eqref{ind61split}. Specifically, the terms $2a^6x$ and $a^3\left(\frac{b}{c}{\bf 6}_{SU(6)}+\frac{c}{b}\overline{\bf 6}_{SU(6)}\right)x^{\frac{3}{2}}$ come from states contained in the $SU(8)$ current and the term $a^3{\bf 2}_{SU(2)}{\bf 20}_{SU(6)}x^{\frac{5}{2}}$ comes from a state contained in the other HB operator, where we recall that the R-charge used in \eqref{ind61split} is related to the $5d$ R-charge by the shift $a\to a x^{\frac{1}{6}}$ and we should identify $\ga=\frac{b}{c}$.

Simiarly to the previous cases, we encounter an issue with the conformal manifold of the theory, which seems to be larger than what expected from $5d$. Again we claim that this is an issue occuring only for low value of flux. Indeed, computing the index of the model with flux 2
\bea
\mathcal{I}&=&1+4a^6x+2a^3\left(\frac{b}{c}{\bf 6}+\frac{c}{b}\overline{\bf 6}\right)x^{\frac{3}{2}}+\left(10a^{12}+\frac{8}{a^{12}}\right)x^2+\cdots\,,
\eea
we can see that the conformal manifold now conforms to our expectations. Moreover, we can again check the presence of operators expected from $5d$.

\subsubsection*{The $(5,2)$ split}

We next consider the case of the $(5,2)$ split, where again we take the Chern--Simons levels to be $(-\frac{1}{2},+\frac{1}{2})$. This can also be understood as integrating out one flavor in the theory with the $(6,2)$ split.  The index of the model for flux 1 is
\bea
\mathcal{I}&=&1+\left(2a^6+\frac{6}{a^6}\right)x+a^3{\bf 2}_{SU(2)}\left(\frac{b}{c}{\bf 5}_{SU(5)}+\frac{c}{b}\overline{\bf 5}_{SU(5)}\right)x^{\frac{3}{2}}+\nn\\
&+&\left(9+3a^{12}+\frac{23}{a^{12}}+3{\bf 3}_{SU(2)}\right)x^2+\nn\\
&+&\left(a^3\left(b^3\overline{\bf 10}_{SU(5)}+\frac{1}{b^3}{\bf 10}_{SU(5)}+b\,c^2{\bf 5}_{SU(5)}+\frac{1}{b\,c^2}\overline{\bf 5}_{SU(5)}+\frac{b^5}{c^2}+\frac{c^2}{b^5}\right)+\right.\nn\\
&+&\left.2\left(a^9+\frac{1}{a^3}\right){\bf 2}_{SU(2)}\left(\frac{b}{c}{\bf 5}_{SU(5)}+\frac{c}{b}\overline{\bf 5}_{SU(5)}\right)\right)x^{\frac{5}{2}}+\cdots\,.
\eea
The index is consistent with the enhancement of $U(1) \times SU(5)\rightarrow SO(10)$, where the embedding is
\bea
\label{emb52}
& & {\bf 10}_{SO(10)} \rightarrow \frac{b}{c}{\bf 5}_{SU(5)}+\frac{c}{b}\overline{\bf 5}_{SU(5)} , \\ \nonumber
& & {\bf 16}_{SO(10)} \rightarrow \frac{b^{\frac{5}{2}}}{c^{\frac{5}{2}}}+\frac{b^{\frac{1}{2}}}{c^{\frac{1}{2}}}\overline{\bf 10}_{SU(5)}+\frac{c^{\frac{3}{2}}}{b^{\frac{3}{2}}}{\bf 5}_{SU(5)} . 
\eea
In terms of characters of $U(1)^2\times SU(2)\times SO(10)$, the index can be written as
\bea
\mathcal{I}&=&1+\left(2a^6+\frac{6}{a^6}\right)x+a^3{\bf 2}_{SU(2)}{\bf 10}_{SO(10)}x^{\frac{3}{2}}+\left(9+3a^{12}+\frac{23}{a^{12}}+3{\bf 3}_{SU(2)}\right)x^2+\nn\\
&+&\left(a^3\left( c^{\frac{1}{2}} b^{\frac{5}{2}}{\bf 16}_{SO(10)} + \frac{1}{c^{\frac{1}{2}} b^{\frac{5}{2}}}\overline{\bf 16}_{SO(10)}\right)+2\left(a^9+\frac{1}{a^3}\right){\bf 2}_{SU(2)}{\bf 10}_{SO(10)}\right)x^{\frac{5}{2}}+ \nn\\ &+& \cdots\,.
\label{ind52split}
\eea
This result is compatible with the theory being the result of the torus compactification of the $5d$ $U(1)\times SO(14)$ SCFT with a unit of flux for a $U(1)$ whose commutant in $SO(14)$ is $U(1)\times SU(2)\times SO(10)$. Notice that such $5d$ SCFT can be obtained as a mass deformation of the $SU(2)\times SO(16)$ SCFT, which is compatible with the fact that the model with $(5,2)$ split can be obtained as a mass deformation of the model with $(6,2)$ split. 

The spectrum of operators is also consistent with this claim. The set of the HB chiral ring generators of the $5d$ $U(1)\times SO(14)$ SCFT contains the moment map for the $SO(14)$ global symmetry and a pair of operators in the $\bf 4$ of $SU(2)_R$, the first one transforming in the $\bf 64$ of $SO(14)$ and having charge $+1$ under the $U(1)$, while the second one transforming in the $\overline{\bf 64}$ and having charge $-1$. Using the branching rules
\bea
&&{\bf 64}_{SO(14)}\to\left(a^3+\frac{1}{a^3}\right){\bf 16}_{SO(10)}+{\bf 2}_{SU(2)}\overline{\bf 16}_{SO(10)},\nn\\
&&{\bf 91}_{SO(14)}\to 1+a^6+\frac{1}{a^6}+{\bf 3}_{SU(2)}+{\bf 45}_{SO(10)}+\left(a^3+\frac{1}{a^3}\right){\bf 10}_{SO(10)}\,,
\eea
we can identify the contributions of some of these states in the index \eqref{ind52split}. Specifically, the terms $2a^6x$ and $a^3{\bf 2}_{SU(2)}{\bf 10}_{SO(10)}x^{\frac{3}{2}}$ come from states contained in the $SO(14)$ current and the terms $a^3\left( c^{\frac{1}{2}} b^{\frac{5}{2}}{\bf 16}_{SO(10)} + \frac{1}{c^{\frac{1}{2}} b^{\frac{5}{2}}}\overline{\bf 16}_{SO(10)}\right)x^{\frac{5}{2}}$ come from states contained respectively in the HB operators in the $\bf 64$ and in the $\overline{\bf 64}$ of $SO(14)$, where we recall that the R-charge used in \eqref{ind52split} is related to the $5d$ R-charge by the shift $a\to a x^{\frac{1}{6}}$ and we should identify the fugacity $q$ for the $5d$ $U(1)$ symmetry with $q=c^{\frac{1}{2}}b^{\frac{5}{2}}$.

Similarly to the previous cases, we encounter an issue with the conformal manifold of the theory, which seems to be larger than what expected from $5d$. Again we claim that this is an issue occurring only for low value of flux. Indeed, computing the index of the model with flux 2
\be
\mathcal{I}=1+4a^6x+2a^3{\bf 2}_{SU(2)}\left(\frac{b}{c}{\bf 5}_{SU(5)}+\frac{c}{b}\overline{\bf 5}_{SU(5)}\right)x^{\frac{3}{2}}+\left(10a^{12}+\frac{6}{a^{12}}\right)x^2+\cdots\,,
\ee
we can see that the conformal manifold now conforms to our expectations. Also this index can be written in terms of $SO(10)$ characters
\be
\mathcal{I}=1+4a^6x+2a^3{\bf 2}_{SU(2)}{\bf 10}_{SO(10)}x^{\frac{3}{2}}+\left(10a^{12}+\frac{6}{a^{12}}\right)x^2+\cdots\,,
\ee
and again we can check the presence of operators expected from $5d$.

The IR central charges of two $U(1)$ symmetries which are different subgroups of the enhanced IR symmetry are also consistent with the enhancement we observe in the index. For example, let us consider the two symmetries $U(1)_{b/c}$ and $U(1)_{C}$, where $C$ is the Cartan $\textrm{diag}\left(1,0,0,0,-1\right)$ of $SU(5)$. Their embedding indices in $SO(10)$, following the embedding \eqref{emb52}, are 
\begin{equation}
	I_{U\left(1\right)_{b/c}\rightarrow SO\left(10\right)}=10\,,\,\,\,I_{U\left(1\right)_{C}\rightarrow SO(10)}=4\,.
\end{equation}
As a result, their central charges are expected to be related in the following way:
\begin{equation}
\label{relc52}
	C_{b/c}=10C_{SO(10)}\,,\,\,\,C_{C}=4C_{SO(10)}\,,
\end{equation}
To test that, we compute $C_{b/c}$ and $C_{C}$ numerically from the real part of the free energy (see Eq. \eqref{d2F} and the discussion around it) and find $C_{b/c}=7.6$, $C_{C}=2.9$. Their ratio is $2.6$, which matches within the accuracy of the computation the result $2.5$ expected from the relations \eqref{relc52}. Let us comment that as in previous cases, the flip fields violate the unitarity bound and we perform the computations in the theory obtained after removing them (in which the superconformal R-charge is given by $\hat{R}=R-0.17q_{a}$).

\subsubsection*{The $(6,0)$ split}

We next consider the case of the $(6,0)$ split. Here the number of flavors is again too small and the quiver is bad. One way to make the theory good is to turn on Chern--Simons interactions, which we now take to be at levels $(-1,+1)$. This theory can be obtained by integrating out one flavor in the theory with the $(6,1)$ split or in the one with the $(7,0)$ split. The index for flux 1 is
\bea
\mathcal{I}&=&1+\left(2a^6+\frac{18}{a^6}\right)x+\left(24+3a^{12}+\frac{102}{a^{12}}\right)x^2+a^3\left(b^3+\frac{1}{b^3}\right){\bf 20}_{SU(6)}x^{\frac{5}{2}}+\nn\\
&+&\left(4a^{18}+32a^6+\frac{46}{a^6}+\frac{318}{a^{18}}\right)x^3+\cdots\,.
\eea
We can see that the index appears consistent with the enhancement of $U(1)_b \rightarrow SU(2)$, where the embedding is ${\bf 2}_{SU(2)}\to b^3+\frac{1}{b^3}$. The index written in terms of $SU(2)$ characters is
\bea
\mathcal{I}&=&1+\left(2a^6+\frac{18}{a^6}\right)x+\left(24+3a^{12}+\frac{102}{a^{12}}\right)x^2+a^3{\bf 2}_{SU(2)}{\bf 20}_{SU(6)}x^{\frac{5}{2}}+\nn\\
&+&\left(4a^{18}+32a^6+\frac{46}{a^6}+\frac{318}{a^{18}}\right)x^3+\cdots\,.
\label{ind60split}
\eea
This result is compatible with the theory being the result of the torus compactification of the $5d$ $SU(2)\times SU(2)\times SU(6)$ SCFT with a unit of flux for one of the $SU(2)$ factors. Notice that such $5d$ SCFT can be obtained as a mass deformation of the $SU(2)\times SU(8)$ SCFT, which is compatible with the fact that the model with $(6,0)$ split can be obtained as a mass deformation of the model with either $(6,1)$ or $(7,0)$ split. 

The spectrum of operators is also consistent with this claim. The set of the HB chiral ring generators of the $5d$ $SU(2)\times SU(2)\times SU(6)$ SCFT contains the moment map for one of the $SU(2)$ global symmetries and an operator in the $\bf 2$ of each $SU(2)$, the $\bf 20$ of $SU(6)$ and the $\bf 4$ of $SU(2)_R$. Using the branching rules
\bea
&&{\bf 2}_{SU(2)}\to a^3+\frac{1}{a^3},\nn\\
&&{\bf 3}_{SU(2)}\to 1+a^6+\frac{1}{a^6}\,,
\eea
we can identify the contributions of some of these states in the index \eqref{ind60split}. Specifically, the term $2a^6x$ comes from a state contained in the $SU(2)$ current and the term $a^3{\bf 2}_{SU(2)}{\bf 20}_{SU(6)}x^{\frac{5}{2}}$ comes from a state contained in the other HB operator, where we recall that the R-charge used in \eqref{ind60split} is related to the $5d$ R-charge by the shift $a\to a x^{\frac{1}{6}}$.

Simiarly to the previous cases, we encounter an issue with the conformal manifold of the theory, which seems to be larger than what expected from $5d$. Again we claim that this is an issue occuring only for low value of flux. Indeed, computing the index of the model with flux 2
\be
\mathcal{I}=1+4a^6x+\left(10a^{12}+\frac{18}{a^{12}}\right)x^2+\cdots
\ee
we can see that the conformal manifold now conforms to our expectations. 

\subsubsection*{The $(5,1)$ split}

We consider now the case of the $(5,1)$ split, where again we take the Chern--Simons levels to be $(-1,+1)$. This can also be understood as integrating out one flavor in the theory with the $(6,1)$ split or in the one with the $(5,2)$ split. The index of the model for flux 1 is
\bea
\mathcal{I}&=&1+\left(2a^6+\frac{11}{a^6}\right)x+a^3\left(\frac{b}{c}{\bf 5}+\frac{c}{b}\overline{\bf 5}\right)x^{\frac{3}{2}}+\left(19+3a^{12}+\frac{53}{a^{12}}\right)x^2+\nn\\
&+&\left(a^3\left(b^3\overline{\bf 10}+\frac{1}{b^3}{\bf 10}\right)+\left(2a^9+\frac{3}{a^3}\right)\left(\frac{b}{c}{\bf 5}+\frac{c}{b}\overline{\bf 5}\right)\right)x^{\frac{5}{2}}+\cdots\,.
\label{ind51split}
\eea
This result is compatible with the theory being the result of the compactification of the $5d$ $U(1)\times SU(7)$ SCFT with flux $(0;1,-1,0,\cdots,0)$, which preserves $U(1)^2\times SU(5)\subset SU(7)$. Notice that such $5d$ SCFT can be obtained as a mass deformation of the $SU(2)\times SU(8)$ SCFT, which is compatible with the fact that the model with $(5,1)$ split can be obtained as a mass deformation of the model with $(6,1)$ split. 

The spectrum of operators is also consistent with this claim. The set of the HB chiral ring generators of the $5d$ $U(1)\times SU(7)$ SCFT contains the moment map for the $SU(7)$ global symmetry and a pair of operators in the $\bf 4$ of $SU(2)_R$, the first one transforming in the $\bf 35$ of $SU(7)$ and having charge $+1$ under the $U(1)$, while the second one transforming in the $\overline{\bf 35}$ and having charge $-1$. Using the branching rules
\bea
&&{\bf 48}_{SU(7)}\to 2+a^6+\frac{1}{a^6}+{\bf 24}_{SU(5)}+\ga^7\left(a^3+\frac{1}{a^3}\right){\bf 5}_{SU(5)}+\ga^{-7}\left(a^3+\frac{1}{a^3}\right)\overline{\bf 5}_{SU(5)}\nn\\
&&{\bf 35}_{SU(7)}\to \ga^6\overline{\bf 10}_{SU(5)}+\frac{1}{\ga}\left(a^3+\frac{1}{a^3}\right){\bf 10}_{SU(5)}+\frac{1}{\ga^8}{\bf 5}_{SU(5)}\,,
\eea
we can identify the contributions of some of these states in the index \eqref{ind51split}. Specifically, the terms $2a^6x$ and $a^3\left(\frac{b}{c}{\bf 5}+\frac{c}{b}\overline{\bf 5}\right)x^{\frac{3}{2}}$ come from states contained in the $SU(7)$ current and the terms $a^3\left(b^3\overline{\bf 10}+\frac{1}{b^3}{\bf 10}\right)x^{\frac{5}{2}}$ come from states contained respectively in the HB operators in the $\bf 35$ and in the $\overline{\bf 35}$ of $SU(7)$, where we recall that the R-charge used in \eqref{ind51split} is related to the $5d$ R-charge by the shift $a\to a x^{\frac{1}{6}}$. Moreover, we should identify $\ga^7=\frac{b}{c}$ and also the fugacity $q$ for the $5d$ $U(1)$ symmetry with $\frac{\ga}{q}=b^3$.

Simiarly to the previous cases, we encounter an issue with the conformal manifold of the theory, which seems to be larger than what expected from $5d$. Again we claim that this is an issue occuring only for low value of flux. Indeed, computing the index of the model with flux 2
\be
\mathcal{I}=1+4a^6x+2a^3\left(\frac{b}{c}{\bf 5}+\frac{c}{b}\overline{\bf 5}\right)x^{\frac{3}{2}}+\left(10a^{12}+\frac{11}{a^{12}}\right)x^2+\cdots\,.
\ee
we can see that the conformal manifold now conforms to our expectations. Again we can check the presence of operators expected from $5d$.

\subsubsection*{The $(4,2)$ split}

We consider now the case of the $(4,2)$ split, where again we take the Chern--Simons levels to be $(-1,+1)$. This can also be understood as integrating out two flavors in the theory with the $(6,2)$ split or one flavor in the theory with the $(5,2)$ split. The index of the model for flux 1 is
\bea
\mathcal{I}&=&1+\left(2a^6+\frac{8}{a^6}\right)x+a^3{\bf 2}_{SU(2)}\left(\frac{b}{c}{\bf 4}_{SU(4)}+\frac{c}{b}\overline{\bf 4}_{SU(4)}\right)x^{\frac{3}{2}}+\nn\\
&+&\left(12+3a^{12}+\frac{33}{a^{12}}+4{\bf 3}_{SU(2)}\right)x^2+\left(a^3\left(b^2c+\frac{1}{b^2c}\right)\left(\frac{c}{b}{\bf 4}_{SU(4)}+\frac{b}{c}\overline{\bf 4}_{SU(4)}\right)+\right.\nn\\
&+&\left.\left(2a^9+\frac{3}{a^3}\right)\left(\frac{b}{c}{\bf 4}_{SU(4)}+\frac{c}{b}\overline{\bf 4}_{SU(4)}\right)\right)x^{\frac{5}{2}}+\cdots\,.
\eea
We can see that the index forms characters of $SO(8)$ where the embedding is ${\bf 8}_{v,SO(8)}\to \frac{b}{c}{\bf 4}_{SU(4)}+\frac{c}{b}\overline{\bf 4}_{SU(4)}$. The index in terms of $SO(8)$ characters is
\bea
\mathcal{I}&=&1+\left(2a^6+\frac{8}{a^6}\right)x+a^3{\bf 2}_{SU(2)}{\bf 8}_{v,SO(8)}x^{\frac{3}{2}}+\left(12+3a^{12}+\frac{33}{a^{12}}+4{\bf 3}_{SU(2)}\right)x^2+\nn\\
&+&\left(a^3\left(b^2c+\frac{1}{b^2c}\right){\bf 8}_{s,SO(8)}+\left(2a^9+\frac{3}{a^3}\right){\bf 8}_{v,SO(8)}\right)x^{\frac{5}{2}}+\cdots\,.
\label{ind42split}
\eea
This result is compatible with the theory being the result of the compactification of the $5d$ $U(1)\times SO(12)$ SCFT with a unit of flux for a $U(1)$ whose commutant in $SO(12)$ is $SU(2)\times SO(8)$. Notice that such $5d$ SCFT can be obtained as a mass deformation of the $SU(2)\times SO(16)$ SCFT, which is compatible with the fact that the model with $(4,2)$ split can be obtained as a mass deformation of the model with $(6,2)$ split. 

The spectrum of operators is also consistent with this claim. The set of the HB chiral ring generators of the $5d$ $U(1)\times SO(12)$ SCFT contains the moment map for the $SO(12)$ global symmetry and a pair of operators in the $\bf 4$ of $SU(2)_R$, in the $\bf 32$ of $SO(12)$ and having charges respectively $\pm1$ under the $U(1)$ Using the branching rules
\bea
&&{\bf 66}_{SO(12)}\to 1+a^6+\frac{1}{a^6}+{\bf 3}_{SU(2)}+{\bf 28}_{SO(8)}+\left(a^3+\frac{1}{a^3}\right){\bf 8}_{v,SO(8)}\nn\\
&&{\bf 32}_{SO(12)}\to \left(a^3+\frac{1}{a^3}\right){\bf 8}_{s,SO(8)}+{\bf 2}_{SU(2)}{\bf 8}_{c,SO(8)}\,,
\eea
we can identify the contributions of some of these states in the index \eqref{ind42split}. Specifically, the terms $2a^6x$ and $a^3{\bf 2}_{SU(2)}{\bf 8}_{v,SO(8)}x^{\frac{3}{2}}$ come from states contained in the $SO(12)$ current and the terms $a^3\left(b^2c+\frac{1}{b^2c}\right){\bf 8}_{s,SO(8)}$ come from states contained in the HB operators in the $\bf 32$ of $SO(12)$, where we recall that the R-charge used in \eqref{ind42split} is related to the $5d$ R-charge by the shift $a\to a x^{\frac{1}{6}}$ and we should identify the fugacity $q$ for the $5d$ $U(1)$ symmetry with $q=b^2c$.

Once again we see that the conformal manifold is larger than what is expected from $5d$. Similarly to the previous cases, we find that this is just a deviation occurring for low value of flux. The index of the model of flux 2 indeed reads
\bea
\mathcal{I}&=&1+4a^6x+2a^3{\bf 2}_{SU(2)}\left(\frac{b}{c}{\bf 4}_{SU(4)}+\frac{c}{b}\overline{\bf 4}_{SU(4)}\right)x^{\frac{3}{2}}+\left(10a^{12}+\frac{2}{a^{12}}\right)x^2+\cdots\,.
\eea
This can be rewritten in terms of $SO(8)$ characters as
\bea
\mathcal{I}&=&1+4a^6x+2a^3{\bf 2}_{SU(2)}{\bf 8}_{SO(8)}x^{\frac{3}{2}}+\left(10a^{12}+\frac{2}{a^{12}}\right)x^2+\cdots\,,
\eea
and we can again check for the presence of the operators expected from $5d$.

\subsubsection{Structure of the flow pattern for $N=2$}

Here we shall try to understand the structure of the flow pattern. Specifically, we shall consider the embedding of the $5d$ global symmetry into the $3d$ global symmetry. Since the mass deformations in both $5d$ and $3d$ are associated with global symmetries, we expect them to map to one another in accordance with the mapping of the global symmetries. An interesting question is what happens to the flux under these deformations. Naively, we would expect the flux to just join along for the ride, as long as it is not in the global symmetry receiving the deformation. There is an interesting question what happens if we try to turn on a real mass in a symmetry containing flux, but we shall not consider this for now.

We begin with the case of the $SU(10)$ SCFT. From our previous discussion regarding the $(7,1)$ split model, we see that the $SU(10)$ symmetry of the $5d$ SCFT is embedded in the $3d$ theory as

\be \label{SU10GSMap}
{\bf 10}_{SU(10)} \rightarrow \frac{c^{\frac{4}{5}}}{b^{\frac{7}{5}}} (a^3 + \frac{1}{a^3}) + \frac{b^{\frac{7}{20}}}{c^{\frac{1}{5}}} \left(\frac{1}{b^{\frac{3}{4}}}{\bf 7}_{SU(7)} + b^{\frac{21}{4}} \right) ,
\ee
where here the flux is in $U(1)_a$.

Now let us consider the real mass deformations that we can implement in $3d$. Here we have two choices, integrating either the upper or lower flavor. Integrating out the lower one corresponds to a real mass in $U(1)_c$, while integrating out one of the upper flavors corresponds to breaking ${\bf 7}_{SU(7)} \rightarrow y{\bf 6}_{SU(6)}+\frac{1}{y^6}$ and turning on a mass deformation in $U(1)_y - U(1)_b$. Let's consider how these are mapped in the $5d$ symmetry. Specifically, according to \eqref{SU10GSMap} the mass deformation in $U(1)_c$ maps to the $5d$ mass deformation breaking $SU(10) \rightarrow U(1)_c \times SU(2)_a \times SU(8)$. This latter deformation is expected to initiate a flow from the $SU(10)$ to the $SU(2)\times SU(8)$ $5d$ SCFT. Note that as the flux is in $U(1)_a$, we expect it to go over to the flux in the $SU(2)$ factor of the global symmetry of the $SU(2)\times SU(8)$ $5d$ SCFT.

We can next consider the real mass in the $U(1)_y - U(1)_b$ symmetry. For this, it is convenient to break ${\bf 7}_{SU(7)} \rightarrow \frac{y}{b}{\bf 6}_{SU(6)}+\frac{b^6}{y^6}$, so that the real mass can be taken to be solely in $U(1)_b$. We can then rewrite \eqref{SU10GSMap} as:

\bea
& & {\bf 10}_{SU(10)} \rightarrow \frac{c^{\frac{4}{5}}}{b^{\frac{7}{5}}} (a^3 + \frac{1}{a^3}) + \frac{b^{\frac{7}{20}}}{c^{\frac{1}{5}}} \left(\frac{1}{b^{\frac{3}{4}}}{\bf 7}_{SU(7)} + b^{\frac{21}{4}} \right) \\ \nonumber & \rightarrow &  \frac{y^{\frac{3}{4}} c^{\frac{1}{20}}}{b^{\frac{7}{5}}} \left(\frac{c^{\frac{3}{4}}}{y^{\frac{3}{4}}}(a^3 + \frac{1}{a^3}) + \frac{y^{\frac{1}{4}}}{c^{\frac{1}{4}}} {\bf 6}_{SU(6)} \right) + \frac{b^{\frac{28}{5}}}{y^3 c^{\frac{1}{5}}} (y^3 + \frac{1}{y^3}).
\eea
We then see that this mass deformation is also mapped to one breaking $SU(10) \rightarrow U(1)_b \times SU(2)_y \times SU(8)$, so we again expect it to initiate a flow from the $SU(10)$ to the $SU(2)\times SU(8)$ $5d$ SCFT. However, note that now the flux is embedded in a $U(1)$ subgroup of $SU(8)$ whose commutant is $U(1)^2 \times SU(6)$. These observations precisely match what we observed when we studied the individual theories.

We can next consider the other cases. First we have the $(7,0)$ split. Here we only have one mass deformation we can consider, associated with integrating out an upper flavor. This seems to match the fact that as the flux is in the $SU(2)$, we expect only mass deformations in the $SU(8)$ to be interesting. This is compatible with the fact that we can't seem to flow from this model to one describing the compactification of the $U(1)\times SU(7)$ SCFT as such deformations involve breaking the $SU(2)$. This deformation acts exactly as in the previous case, but now we have a breaking of $SU(8)\rightarrow U(1)\times SU(2)\times SU(6)$. On the $5d$ side, we expect this deformation to initiate a flow from the $5d$ $SU(2)\times SU(8)$ SCFT to the $5d$ $SU(2)\times SU(2)\times SU(6)$ SCFT. As the flux is in the $SU(2)$ factor in the initial $SU(2)\times SU(8)$ SCFT, we expect it to go to a flux in one of the $SU(2)$ factors of the resulting $SU(2)\times SU(2)\times SU(6)$ SCFT.

The case of the $(6,1)$ split is more involved. We can again integrate out the lower flavor, which works in the same way as in the previous case. Specifically, it is mapped to the breaking of $SU(8)\rightarrow U(1)\times SU(2)\times SU(6)$, such that the flux is in the $SU(2)$. Thus, we end up with the same model we got from the $(7,0)$ split, as expected from the commutativity of mass deformations. The analysis of the integrating out of an upper flavor is more involved. For this we first consider the embedding of the symmetries in the $5d$ global symmetry

\bea \label{SU2SU8GSMap1}
& & {\bf 8}_{SU(8)} \rightarrow \frac{c^{\frac{3}{4}}}{b^{\frac{3}{4}}} (a^3 + \frac{1}{a^3}) + \frac{b^{\frac{1}{4}}}{c^{\frac{1}{4}}} {\bf 6}_{SU(6)} , \\ \nonumber
& & {\bf 2}_{SU(2)} \rightarrow b^3 + \frac{1}{b^3} .
\eea
The deformation is again given by breaking ${\bf 6}_{SU(6)} \rightarrow \frac{y}{b}{\bf 5}_{SU(5)} + \frac{b^5}{y^5}$ and taking the deformation to be in $U(1)_b$. We then have that

\bea \label{SU2SU8GSMap2}
 & & {\bf 8}_{SU(8)} \rightarrow \frac{c^{\frac{3}{4}}}{b^{\frac{3}{4}}} (a^3 + \frac{1}{a^3}) + \frac{b^{\frac{1}{4}}}{c^{\frac{1}{4}}} {\bf 6}_{SU(6)} \rightarrow \frac{b^{\frac{21}{4}}}{c^{\frac{1}{4}}y^5} + \frac{c^{\frac{1}{28}}y^{\frac{5}{7}}}{b^{\frac{3}{4}}} \left( \frac{c^{\frac{5}{7}}}{y^{\frac{6}{7}}}(a^3 + \frac{1}{a^3}) + \frac{y^{\frac{2}{7}}}{c^{\frac{2}{7}}} {\bf 5}_{SU(5)} \right). \nonumber \\ && 
\eea
From \eqref{SU2SU8GSMap1} and \eqref{SU2SU8GSMap2} we see that this mass deformation corresponds to the one breaking $SU(2)\rightarrow U(1)$ and $SU(8)\rightarrow U(1)\times SU(7)$. This is expected to initiate a flow from the $SU(2)\times SU(8)$ $5d$ SCFT to the $U(1)\times SU(7)$ one. We also see that the flux should go to a flux inside the $SU(7)$. All these observations match what we have seen in the individual models.




We next move to consider the case of the $(6,2)$ split, associated with the compactification of the $SU(2)\times SO(16)$ $5d$ SCFT. First, we consider the embedding of the $3d$ global symmetry in the $5d$ one, which is

\bea
& & {\bf 16}_{SO(16)} \rightarrow (a^3 + \frac{1}{a^3}) {\bf 2}_{SU(2)_{3d}} + \frac{b}{c} {\bf 6}_{SU(6)} + \frac{c}{b}\overline{\bf 6}_{SU(6)} , \nonumber \\ 
& & {\bf 2}_{SU(2)_{5d}} \rightarrow b^3 + \frac{1}{b^3} .
\eea 
We can again consider the real mass deformations given by integrating out one of the lower or upper flavors. Let's first consider the case of the lower flavors. This corresponds to breaking ${\bf 2}_{SU(2)_{3d}} \rightarrow \frac{z}{c} + \frac{c}{z}$, and turning on the mass deformation for $U(1)_c$. We then have that

\bea
& & {\bf 16}_{SO(16)} \rightarrow (a^3 + \frac{1}{a^3}) {\bf 2}_{SU(2)_{3d}} + \frac{b}{c} {\bf 6}_{SU(6)} + \frac{c}{b}\overline{\bf 6}_{SU(6)} \\ \nonumber & & \rightarrow \frac{b^{\frac{3}{4}}z^{\frac{1}{4}}}{c} \left( \frac{b^{\frac{1}{4}}}{z^{\frac{1}{4}}} {\bf 6}_{SU(6)} + \frac{z^{\frac{3}{4}}}{b^{\frac{3}{4}}} (a^3 + \frac{1}{a^3}) \right) + \frac{c}{b^{\frac{3}{4}}z^{\frac{1}{4}}} \left( \frac{z^{\frac{1}{4}}}{b^{\frac{1}{4}}} \overline{\bf 6}_{SU(6)} + \frac{b^{\frac{3}{4}}}{z^{\frac{3}{4}}} (a^3 + \frac{1}{a^3}) \right) .  
\eea 
As such, we see that this mass deformation should correspond to the $5d$ mass deformation breaking $SO(16)\rightarrow U(1)\times SU(8)$. This mass deformation is then expected to cause an RG flow from the $SU(2)\times SO(16)$ $5d$ SCFT to the $SU(2)\times SU(8)$ one. From the decomposition we further see that the flux should be mapped to one inside the $SU(8)$ factor such that its commutant is $U(1)^2\times SU(6)$.

Consider instead the upper flavor. We can continue as done so far and analyze this case by taking ${\bf 6}_{SU(6)} \rightarrow \frac{y}{b}{\bf 5}_{SU(5)} + \frac{b^5}{y^5}$, and turn on the mass deformation for $U(1)_b$. We then have that

\bea
& & {\bf 16}_{SO(16)} \rightarrow (a^3 + \frac{1}{a^3}) {\bf 2}_{SU(2)_{3d}} + \frac{b}{c} {\bf 6}_{SU(6)} + \frac{c}{b}\overline{\bf 6}_{SU(6)} \\ \nonumber & & \rightarrow (a^3 + \frac{1}{a^3}) {\bf 2}_{SU(2)_{3d}} + \frac{y}{c} {\bf 5}_{SU(5)} + \frac{c}{y}\overline{\bf 5}_{SU(5)} + \frac{b^6}{c y^5} + \frac{c y^5}{b^6} .  
\eea 
The mass deformation then is the one breaking $SO(16)\rightarrow U(1)\times SO(14)$ and $SU(2)\rightarrow U(1)$, and so is expected to map to the $5d$ mass deformation initiating the RG flow from the $SU(2)\times SO(16)$ SCFT to the $U(1)\times SO(14)$ one. Furthermore, the flux here is mapped to one in $SO(14)$ whose commutant is $U(1)\times SU(2)\times SO(10)$. This is in agreement with what we observed in the individual models. 

We can continue and study the other $(N_f,2)$ splits, but at this point the behavior should be apparent from the previous discussion. Specifically, the masses associated with the lower flavors should correspond to the $5d$ mass terms leading to the $U(1)\times SU(N_f+2)$ SCFTs, by the previous discussion and the commutativity of mass deformations. The masses associated with the upper flavors should correspond to the $5d$ mass terms leading to the $U(1)\times SO(2N_f+2)$ SCFTs, with flux preserving their $U(1)\times SU(2)\times SO(2N_f-2)$ subgroup. 

Finally, we can consider the implication of our discussion so far on the remaining cases. Consider first the case of the $(6,0)$ split, related to the $SU(2)\times SU(2)\times SU(6)$ SCFT. As we are looking for mass deformations that are not in the symmetries with flux, the only interesting option is the one giving the $U(1)\times SU(2)\times SU(5)$ SCFT from the $SU(2)\times SU(2)\times SU(6)$ one. In terms of the gauge theory descriptions, this corresponds to the one leading from $SU(3)_0+6F$ to $SU(3)_{\frac{1}{2}}+5F$. Indeed, in the $3d$ model, we seem to only have one interesting deformation, the one associated with integrating out one of the upper flavors. In fact, we have already seen, when discussing the $(6,1)$ split, that this deformation should act as $SU(2)\times SU(6)\rightarrow U(1)\times SU(5)$, which is precisely the one leading to the $U(1)\times SU(2)\times SU(5)$ SCFT from the $SU(2)\times SU(2)\times SU(6)$ one.

We can next consider the $(5,1)$ split, which should be associated with the $U(1)\times SU(7)$ SCFT. Here we can integrate both the upper and lower flavors. The lower one is expected to give the model associated with the $U(1)\times SU(2)\times SU(5)$ SCFT with flux in the $SU(2)$. However, integrating out the upper one is expected to break $SU(7)\rightarrow U(1)\times SU(6)$ and as such we expect it to be the one associated with the $5d$ mass deformation between the $U(1)\times SU(7)$ and $U(1)\times SU(6)$ SCFTs. Similarly, for the $(4,2)$ split, we expect integrating out the lower flavor to lead to the model associated with the $U(1)\times SU(6)$ SCFT, but the one integrating out the upper flavor is expected instead to be related to the $5d$ mass deformation between the $U(1)\times SO(12)$ and $U(1)\times SO(10)$ SCFTs. 

Overall, we see that for $N_f<6$, we expect the $(N_f,0)$ split to give a model associated with the compactification of the $U(1)\times SU(2)\times SU(N_f)$ $5d$ SCFT, which is the UV completion of the $5d$ gauge theory $SU(3)_{\frac{6-N_f}{2}}+N_f F$. Here the flux is in the $SU(2)$ factor. Alternatively, the $(N_f-1,1)$ split is expected to give a model associated with the compactification of the $U(1)\times SU(N_f+1)$ $5d$ SCFT, which is the UV completion of the $5d$ gauge theory $SU(3)_{\frac{8-N_f}{2}}+N_f F$. Here the flux is in the $SU(N_f+1)$ factor such that the commutant is $U(1)^2\times SU(N_f-1)$. Finally, the $(N_f-2,2)$ split is expected to give a model associated with the compactification of the $U(1)\times SO(2N_f)$ $5d$ SCFT, which is the UV completion of the $5d$ gauge theory $SU(3)_{\frac{10-N_f}{2}}+N_f F$. Here the flux is in the $SO(2N_f)$ factor such that the commutant is $U(1)\times SU(2)\times SO(2N_f-4)$. This exhasts the options for mass deformations that we seem to have in $3d$. It is interesting that the $5d$ SCFTs UV completing the $5d$ gauge theories $SU(3)_k+N_f F$ for $|k|<\frac{6-N_f}{2}$ appear to be inaccessible by this class of models. 

\subsubsection{$N=3$, $N_f=10$}

\begin{figure}
\center
\includegraphics[width=0.5\textwidth]{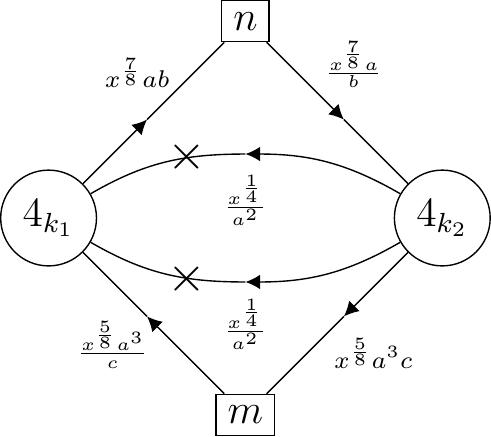}
\caption{Quivers for the models with $N=3$ and split of the flavors $(n,m)$. We also summarize our conventions for the abelian symmetries and the R-symmetry. Notice that these are consistent with a cubic superpotential for the upper triangle involving the $n$ flavors and one copy of the bifundamentals and a quintic superpotential for the lower triangle involving the $m$ flavors and three copies of the bifundamentals.}
\label{QuiversSU4}
\end{figure}

\begin{figure}
\center
\includegraphics[width=1\textwidth]{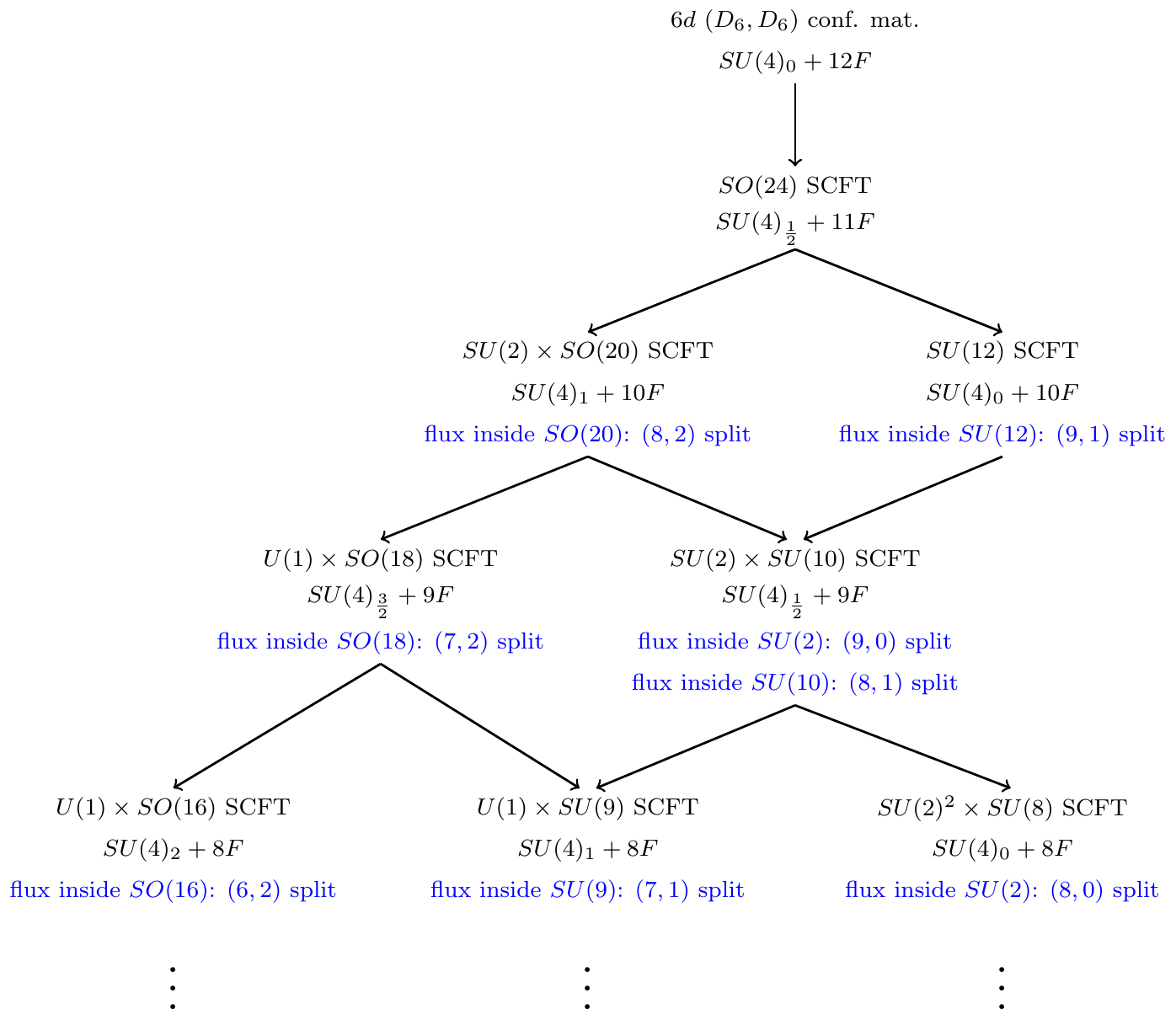} 
\caption{Flow pattern triggered by mass deformations for the rank $N=3$ case. At each step we specify the $5d$ SCFT ($6d$ SCFT for the first step), the $5d$ gauge theory based on $SU$ group that is UV completed by it and the split of the associated $3d$ $\mathcal{N}=2$ theory that is obtained from the torus compactifications. For some SCFTs we can have more splits and each one is associated with a different factor inside the $5d$ global symmetry for which we turn on the flux.}
\label{fig:N3flowchart}
\end{figure}

We now consider the higher rank case of $N=3$. We use the assignment of charges for the abelian global symmetries and the R-symmetry as summarized in Figure \ref{QuiversSU4}, where we also specify our convention for the Chern--Simons levels $(k_1,k_2)$. We will start considering cases with splits of the form $(n,m)$ with $n+m=10$, which should correspond to the possible mass deformations of the $5d$ $SO(24)$ SCFT, and then we will consider cases with less flavors corresponding to further mass deformations. We summarize the findings of this subsubsection in figure \ref{fig:N3flowchart}. 

\subsubsection*{The $(9,1)$ split}

The index for $N=3$ and flux 1 reads
\be
\mathcal{I}=1+\left(2a^8+\frac{9}{a^8}\right)x+a^4\left(\frac{b}{c}{\bf 9}+\frac{c}{b}\overline{\bf 9}+\frac{b^9}{c}+\frac{c}{b^9}\right)x^{\frac{3}{2}}+\left(22+3a^{16}+\frac{60}{a^{16}}\right)x^2+\cdots\,.
\ee
We can see that the index forms characters of $SU(10)$ where the embedding is ${\bf 10}\to\frac{1}{b^{\frac{4}{5}}}{\bf 9}\oplus b^{\frac{36}{5}}$. The index in terms of $SU(10)$ characters is
\be
\mathcal{I}=1+\left(2a^8+\frac{9}{a^8}\right)x+a^4\left(\frac{b^{\frac{9}{5}}}{c}{\bf 10}+\frac{c}{b^{\frac{9}{5}}}\overline{\bf 10}\right)x^{\frac{3}{2}}+\left(22+3a^{16}+\frac{60}{a^{16}}\right)x^2+\cdots\,.
\ee
This result is compatible with the theory being the result of the compactification of the $5d$ $SU(12)$ SCFT with a unit of flux for a $U(1)$ whose commutant in $SU(12)$ is $U(1)\times SU(10)$. Another evidence for this is that one can check the presence of operators that are expected to descend from the broken $5d$ currents (states coming from Higgs branch operators contribute to higher order compared to the one to which we evaluated the index). The analysis carries out exactly as in the $N=2$ case, so we avoid reporting it here. Moreover, again as in the $N=2$ case, the extra exactly marginal operators are expected to disappear for higher values of flux.

\subsubsection*{The $(8,2)$ split}

The index for $N=3$ and flux 1 reads
\bea
\mathcal{I}&=&1+\left(2a^8+\frac{6}{a^8}\right)x+a^4{\bf 2}_{SU(2)}\left(\frac{b}{c}{\bf 8}_{SU(8)}+\frac{c}{b}\overline{\bf 8}_{SU(8)}\right)x^{\frac{3}{2}}+\nn\\
&+&\left(12+3a^{16}+\frac{30}{a^{16}}+4{\bf 3}_{SU(2)}\right)x^2+\cdots\,.
\eea
We can see that the index forms characters of $SO(16)$ where the embedding is ${\bf 16}_{SO(16)}\to\frac{b}{c}{\bf 8}_{SU(8)}+\frac{c}{b}\overline{\bf 8}_{SU(8)}$. The index in terms of $SO(16)$ characters is
\be
\mathcal{I}=1+\left(2a^8+\frac{6}{a^8}\right)x+a^4{\bf 2}_{SU(2)}{\bf 16}_{SO(16)}x^{\frac{3}{2}}+\left(12+3a^{16}+\frac{30}{a^{16}}+4{\bf 3}_{SU(2)}\right)x^2+\cdots\,.
\ee
This result is compatible with the theory being the result of the compactification of the $5d$ $SU(2)\times SO(20)$ SCFT with a unit of flux for a $U(1)$ whose commutant in $SO(20)$ is $SU(2)\times SO(16)$. One can also check the presence of operators expected from $5d$. The extra exactly marginal operators are expected to disappear for higher values of flux, as in the $N=2$ case.


\subsubsection{Integrating out flavors for $N=3$}

\subsubsection*{The $(7,2)$ split}

We can now consider models of the same structure but with less flavors. We consider first the case of the $(7,2)$ split. Here the number of flavors is too small and the quiver is bad. One way to make the theory good is to turn on Chern--Simons interactions. This can also be understood as integrating out one flavor in the theory with $(8,2)$ flavors. We take the Chern--Simons levels to be $(-\frac{1}{2},+\frac{1}{2})$, where the level $\frac{1}{2}$ is for the $SU(4)$ node for which the bifundamental between the two gauge nodes is in the anti-fundamental representation and the level $-\frac{1}{2}$ is for the other. The index of the model is
\bea
\mathcal{I}&=&1+\left(2a^8+\frac{9}{a^8}\right)x+a^4{\bf 2}_{SU(2)}\left(\frac{b}{c}{\bf 7}_{SU(7)}+\frac{c}{b}\overline{\bf 7}_{SU(7)}\right)x^{\frac{3}{2}}+\nn\\
&+&\left(18+3a^{16}+\frac{51}{a^{16}}+6{\bf 3}_{SU(2)}\right)x^2+\cdots\,.
\eea
We can see that the index forms characters of $SO(14)$ where the embedding is ${\bf 14}_{SO(14)}\to \frac{b}{c}{\bf 7}_{SU(7)}+\frac{c}{b}\overline{\bf 7}_{SU(7)}$. The index in terms of $SO(14)$ characters is
\bea
\mathcal{I}&=&1+\left(2a^8+\frac{9}{a^8}\right)x+a^4{\bf 2}_{SU(2)}{\bf 14}_{SO(14)}x^{\frac{3}{2}}+\left(18+3a^{16}+\frac{51}{a^{16}}+6{\bf 3}_{SU(2)}\right)x^2+\cdots\,.
\eea
This result is compatible with the theory being the result of the compactification of the $5d$ $U(1)\times SO(18)$ SCFT with a unit of flux for a $U(1)$ whose commutant in $SO(18)$ is $SU(2)\times SO(14)$. Notice that such $5d$ SCFT can be obtained as a mass deformation of the $SU(2)\times SO(20)$ SCFT, which is compatible with the fact that the model with $(7,2)$ split can be obtained as a mass deformation of the model with $(8,2)$ split. One can also check the presence of operators expected from $5d$. Again, the extra exactly marginal operators are expected to disappear for higher values of flux, as in the $N=2$ case. 

\subsubsection*{The $(8,1)$ split}

We next consider the case of the $(8,1)$ split. Again the number of flavors is too small and the quiver is bad, but we can make the theory good by turning on Chern--Simons interactions. This can also be understood as integrating out one of the lower flavors in the theory with $(8,2)$ split or one of the upper flavors in the theory with $(9,1)$ split. We take the Chern--Simons levels to be $(-\frac{1}{2},+\frac{1}{2})$. The index of the model is
\be
\mathcal{I}=1+\left(2a^8+\frac{15}{a^8}\right)x+a^4\left(\frac{b}{c}{\bf 8}_{SU(8)}+\frac{c}{b}\overline{\bf 8}_{SU(8)}\right)x^{\frac{3}{2}}+\left(36+3a^{16}+\frac{120}{a^{16}}\right)x^2+\cdots\,.
\ee
This result is compatible with the theory being the result of the compactification of the $5d$ $SU(2)\times SU(10)$ SCFT with a unit of flux for a $U(1)$ whose commutant in $SU(10)$ is $U(1)\times SU(8)$. Notice that such $5d$ SCFT can be obtained as a mass deformation either of the $SU(2)\times SO(20)$ SCFT or of the $SU(12)$ SCFT, which is compatible with the fact that the model with $(8,1)$ split can be obtained as a mass deformation of both the models with $(8,2)$ and $(9,1)$ split. One can also check the presence of operators expected from $5d$. Again, the extra exactly marginal operators are expected to disappear for higher values of flux, as in the $N=2$ case.


\subsubsection*{The $(9,0)$ split}

From the $(9,1)$ split one can also flow with a mass deformation to the $(9,0)$ split, which we are now going to consider. Again we take the Chern--Simons levels to be $(-\frac{1}{2},+\frac{1}{2})$. The index of the model is
\be
\mathcal{I}=1+\left(2a^8+\frac{40}{a^8}\right)x+\left(70+3a^{16}+\frac{520}{a^{16}}\right)x^2+\cdots\,.
\ee
This result is compatible with the theory being the result of the compactification of the $5d$ $SU(2)\times SU(10)$ SCFT, but this time with flux inside the $SU(2)$ and not inside the $SU(10)$ as for the $(8,1)$ split that we just saw. Notice that such $5d$ SCFT can be obtained as a mass deformation of the $SU(12)$ SCFT, which is compatible with the fact that the model with $(9,0)$ split can be obtained as a mass deformation of the model with $(9,1)$ split. One can also check the presence of operators expected from $5d$. Again, the extra exactly marginal operators are expected to disappear for higher values of flux, as in the $N=2$ case.

This suggests that the $SU(9)$ symmetry visible in the Lagrangian and a $U(1)$ global symmetry should enhance to $SU(10)$ somewhere on the conformal manifold of the $3d$ theory. As such we expect the index to form characters of $SU(10)$, though unfortunately, these appear at orders higher than $x^2$ making their computation more involved.

\subsubsection*{The $(6,2)$ split}

We next consider the case of the $(6,2)$ split. Again the number of flavors is too small and the quiver is bad, but we can make the theory good by turning on Chern--Simons interactions. This can also be understood as integrating out one flavor in the theory with $(7,2)$ flavors. We take the Chern--Simons levels to be $(-1,+1)$, where the level $-1$ is for the $SU(4)$ node for which the bifundamental between the two gauge nodes is in the anti-fundamental representation and the level $+1$ is for the other. The index of the model is
\bea
\mathcal{I}=1+\left(2a^8+\frac{12}{a^8}\right)x &+& a^4{\bf 2}_{SU(2)} \left(\frac{b}{c}{\bf 6}_{SU(6)}+\frac{c}{b}\overline{\bf 6}_{SU(6)}\right)x^{\frac{3}{2}} \\ \nonumber &+& \left(24+3a^{16}+\frac{75}{a^{16}}+8{\bf 3}_{SU(2)}\right)x^2+\cdots\,.
\eea
We can see that the index forms characters of $SO(12)$ where the embedding is ${\bf 12}_{SO(12)}\to \frac{b}{c}{\bf 6}_{SU(6)}+\frac{c}{b}\overline{\bf 6}_{SU(6)}$. The index in terms of $SO(12)$ characters is
\be
\mathcal{I}=1+\left(2a^8+\frac{12}{a^8}\right)x+a^4{\bf 2}_{SU(2)}{\bf 12}_{SO(12)}x^{\frac{3}{2}}+\left(24+3a^{16}+\frac{75}{a^{16}}+8{\bf 3}_{SU(2)}\right)x^2+\cdots\,.
\ee
This result is compatible with the theory being the result of the compactification of the $5d$ $U(1)\times SO(16)$ SCFT with a unit of flux for a $U(1)$ whose commutant in $SO(16)$ is $SU(2)\times SO(12)$. Notice that such $5d$ SCFT can be obtained as a mass deformation of the $U(1)\times SO(18)$ SCFT, which is compatible with the fact that the model with $(6,2)$ split can be obtained as a mass deformation of the model with $(7,2)$ split. One can also check the presence of operators expected from $5d$. Again, the extra exactly marginal operators are expected to disappear for higher values of flux, as in the $N=2$ case.

\subsubsection*{The $(7,1)$ split}

We next consider the case of the $(7,1)$ split. Again the number of flavors is too small and the quiver is bad, but we can make the theory good by turning on Chern--Simons interactions. This can also be understood as integrating out one of the lower flavors in the theory with $(7,2)$ split or one of the upper flavors in the theory with $(8,1)$ split. We take the Chern--Simons levels to be $(-1,+1)$. The index of the model is
\be
\mathcal{I}=1+\left(2a^8+\frac{21}{a^8}\right)x+a^4\left(\frac{b}{c}{\bf 7}_{SU(7)}+\frac{c}{b}\overline{\bf 7}_{SU(7)}\right)x^{\frac{3}{2}}+\left(50+3a^{16}+\frac{186}{a^{16}}\right)x^2+\cdots\,.
\ee
This result is compatible with the theory being the result of the compactification of the $5d$ $U(1)\times SU(9)$ SCFT with a unit of flux for a $U(1)$ whose commutant in $SU(9)$ is $U(1)\times SU(7)$. Notice that such $5d$ SCFT can be obtained as a mass deformation either of the $U(1)\times SO(18)$ SCFT or of the $SU(2)\times SU(10)$ SCFT, which is compatible with the fact that the model with $(7,1)$ split can be obtained as a mass deformation of both the models with $(7,2)$ and $(8,1)$ split. One can also check the presence of operators expected from $5d$. Again, the extra exactly marginal operators are expected to disappear for higher values of flux, as in the $N=2$ case.

\subsubsection*{The $(8,0)$ split}

From the $(8,1)$ split or the $(9,0)$ split one can also flow with a mass deformation to the $(8,0)$ split. Again we take the Chern--Simons levels to be $(-1,+1)$. The index of the model is
\be
\mathcal{I}=1+\left(2a^8+\frac{68}{a^8}\right)x+\left(120+3a^{16}+\frac{968}{a^{16}}\right)x^2+\cdots\,.
\ee
This result is compatible with the theory being the result of the compactification of the $5d$ $SU(2)^2\times SU(8)$ SCFT with flux inside one of the $SU(2)$ groups. Notice that such $5d$ SCFT can be obtained as a mass deformation of the $SU(2)\times SU(10)$ SCFT, which is compatible with the fact that the model with $(9,0)$ split can be obtained as a mass deformation of the model with $(8,1)$ or $(9,0)$ split. One can also check the presence of operators expected from $5d$. Again, the extra exactly marginal operators are expected to disappear for higher values of flux, as in the $N=2$ case.

\subsubsection{Analysis of the generic case}

\begin{figure}
\center
\includegraphics[width=0.9\textwidth]{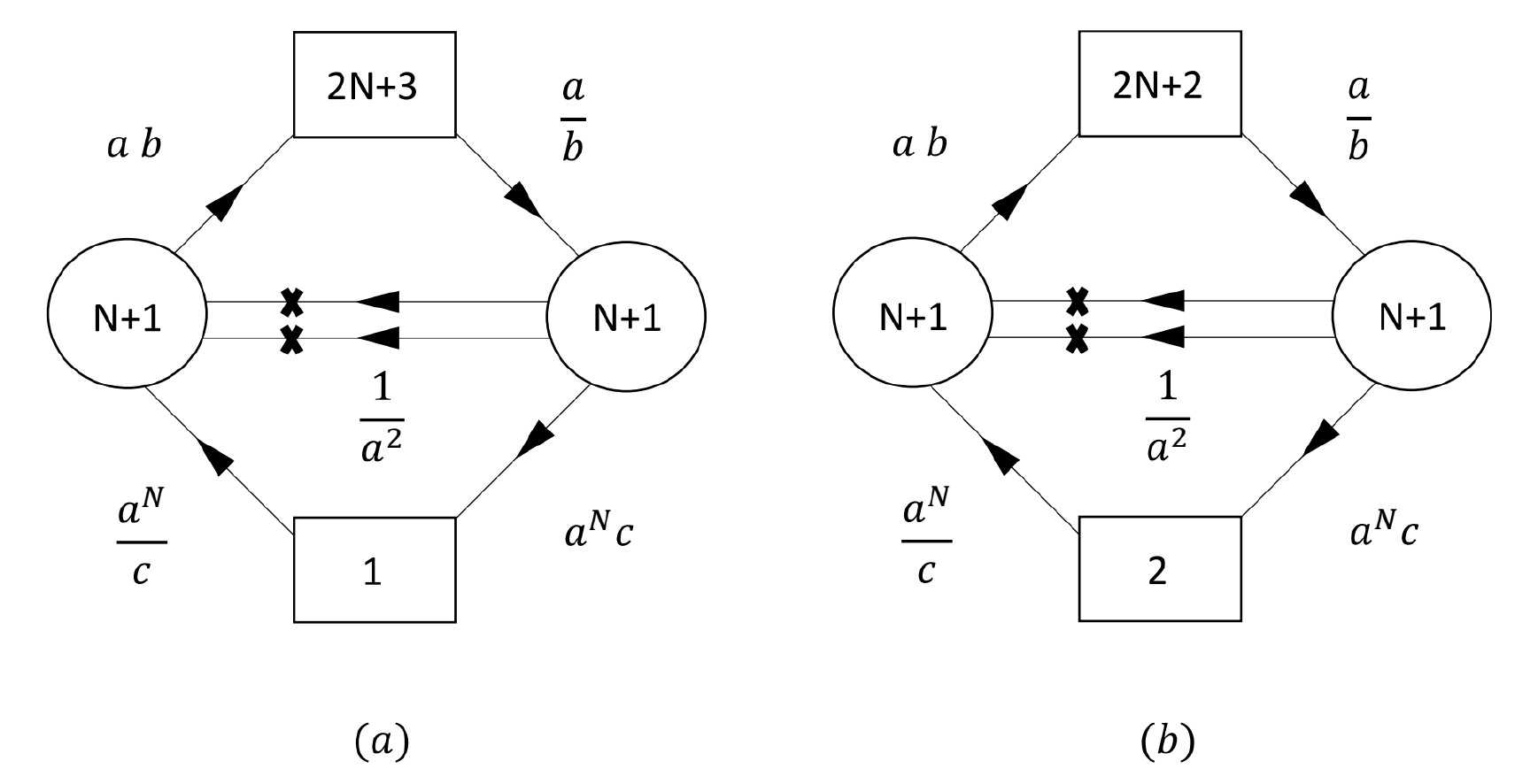} 
\caption{The $3d$ theories associated with the compactification of the $5d$ SCFTs UV completing the $5d$ gauge theories $SU(N+1)_k+(2N+4)F$ on a torus with flux. (a) is for the case of $k=1$, corresponding to the $SU(2N+6)$ SCFT, with flux $(1,-1,0,0,...,0)$. (b) is for the case of $k=2$, corresponding to the $SU(2)\times SO(4N+8)$ SCFT, with flux $(0,0;1,1,0,0,...,0)$. Additionally, the charges of the fields under the non-R symmetries are shown. In all cases there are no Chern-Simons terms.}
\label{QuiverSUNX23d}
\end{figure}

Finally, we can consider the general case. As we did previously when analyzing generic cases, we shall not perform index computations, but instead look at selected multiplets. As in the low $N$ cases explicitly studied previously, the first case in which we find theories with promising properties to be compactifications of the $5d$ SCFTs we consider is when $N_f=2N+4$, where we have the two models associated with the choices $m=1$ and $m=2$ with $k_1=k_2=0$. The resulting quivers are shown in figure \ref{QuiverSUNX23d} together with the charges of the fields under all non-R symmetries. As we are not going to compute the index, we can use the $5d$ R-charge under which the bifundamentals have R-charge $0$, the flip fields have R-charge $2$ and all other fields have R-charge $1$. 

We next consider the possible gauge invariant operators in these models, beginning with the perturbative sector of the model in \ref{QuiverSUNX23d} (a). Here, we first have the flip fields, which contribute to the index the term $2 a^{2(N+1)} x$. Second we have the gauge invariants made from the lower and upper flavors, which contribute\footnote{We recall the reader the conventions that we explained in footnote \ref{repconv} for representations of groups which we use in our paper.}: $a^{N+1} \left( \frac{b}{c} {\bf F}_{SU(2N+3)} + \frac{c}{b} \overline{\bf F}_{SU(2N+3)} \right)x^2$. Finally, we have the $SU(N+1)$ baryons which contribute: $a^{N+1} \left( b^{N+1} {\bf \Lambda}^{N+1}_{SU(2N+3)} + \frac{1}{b^{N+1}} \overline{\bf \Lambda}^{N+1}_{SU(2N+3)} \right)x^{N+1}$.

We next move to the non-perturbative part. Here we shall not discuss all possibilities, but rather concentrate on the basic ones. These are the minimal ones of $SU(N+1)$, carrying magnetic fluxes of $(1,0,0,...,0,-1)$, under one or both of the groups. We first consider the case where the monopole is only in one $SU(N+1)$ group. In this case, the  monopole is gauge invariant and contribute to the index the terms: $a^{N+1} ( \frac{b^{2N+3}}{c} + \frac{c}{b^{2N+3}} )x^2$. 

The monopole with minimal charges under both groups has $5d$ R-charge $0$, $U(1)_a$ charge $2N-6$, uncharged under $U(1)_b$ and $U(1)_c$, but carry gauge charges and so is not gauge invariant. Specifically, here the magnetic charge sits in a $U(1)$ whose commutant is $U(1) \times SU(N-1)$, and the chiral matter causes the monopole to acquire charges under the $U(1)$ part of the commutant. It is possible to build an invariant by taking the $SU(N-1)\times SU(N-1)$ part of the $SU(N+1)\times SU(N+1)$ bifundamental, build an invariant under the $SU(N-1)\times SU(N-1)$ unbroken gauge symmetry and dress the monopole by the square of this invariant. Calling the monopole $M_{(1,1)}$ and the $SU(N-1)\times SU(N-1)$ part of the $SU(N+1)\times SU(N+1)$ bifundamental $B_{(N-1,N-1)}$, then the invariant is given by $ M_{(1,1)} (B^{N-1}_{(N-1,N-1)})^2$. This invariant carries $5d$ R-charge $0$, $U(1)_a$ charge $-(2N+2)$ and is uncharged under the remaining symmetries. Multiplying with the flip fields then gives the marginal operators that are neutral under all symmetries. Finally, we note that we have two $SU(N+1)\times SU(N+1)$ bifundamentals leading to two $SU(N-1)\times SU(N-1)$ bifundamentals. As such there are actually many operators of this type depending on which one of them we use for the $2(N-1)$ $SU(N-1)\times SU(N-1)$ bifundamentals in the invariant. This explains the large number of such operators that appears to increase with $N$ as we observed fror $N=2,3$. This further suggests that their number is proportional to $(N!)^2$.

The spectrum we observed is consistent with this theory being the result of the compactification of the $5d$ $SU(2N+6)$ SCFT on a torus with flux $(1,-1,0,0,0,...,0)$, which is a unit flux in a $U(1)$ whose commutant in $SU(2N+6)$ is $U(1)^2\times SU(2N+4)$. Specifically, we see characters of $SU(2N+4)$, where ${\bf F}_{SU(2N+4)} \rightarrow \frac{1}{b^{\frac{N+1}{N+2}}}{\bf F}_{SU(2N+3)} + b^{\frac{(N+1)(2N+3)}{N+2}}$. This comes about as the contribution of the gauge invariant made from the upper and lower flavors and the basic $SU(N+1)$ monopole can be combined to form
\bea
& & a^{N+1} \left( \frac{b}{c} {\bf F}_{SU(2N+3)} + \frac{c}{b} \overline{\bf F}_{SU(2N+3)} \right)x^2 + a^{N+1} ( \frac{b^{2N+3}}{c} + \frac{c}{b^{2N+3}} )x^2 \nonumber \\  & & \rightarrow a^{N+1} \left( \frac{b^{\frac{2N+3}{N+2}}}{c} {\bf F}_{SU(2N+4)} + \frac{c}{b^{\frac{2N+3}{N+2}}}\overline{\bf F}_{SU(2N+4)} \right) x^2 .
\eea   
Furthermore, the two $SU(N+1)$ baryons can also be merged to form 
\be
a^{N+1} \left( b^{N+1} {\bf \Lambda}^{N+1}_{SU(2N+3)} + \frac{1}{b^{N+1}} \overline{\bf \Lambda}^{N+1}_{SU(2N+3)} \right)x^{N+1} \rightarrow a^{N+1} {\bf \Lambda}^{N+2}_{SU(2N+4)} x^{N+1}.
\ee

We also note that the operators we observed are consistent with our $5d$ expectations. Specifically, under the decomposition of $SU(2N+6)$ into $U(1)^2\times SU(2N+4)$ we have that
\bea
& & {\bf F}_{SU(2N+6)} \rightarrow \frac{1}{z} {\bf F}_{SU(2N+4)} + z^{N+2} (y + \frac{1}{y}) , \\ \nonumber
& & {\bf Ad}_{SU(2N+6)} \rightarrow 2 + y^2 + \frac{1}{y^2} + {\bf Ad}_{SU(2N+4)} \\ \nonumber & + & \frac{1}{z^{N+3}} (y + \frac{1}{y}) {\bf F}_{SU(2N+4)} + z^{N+3} (y + \frac{1}{y}) \overline{\bf F}_{SU(2N+4)}, \\ \nonumber
& & {\bf \Lambda}^{N+3}_{SU(2N+6)} \rightarrow z^{N+3} {\bf \Lambda}^{N+1}_{SU(2N+4)} + \frac{1}{z^{N+3}} \overline{\bf \Lambda}^{N+1}_{SU(2N+4)} + (y + \frac{1}{y}) {\bf \Lambda}^{N+2}_{SU(2N+4)} ,
\eea
where here the flux is in $U(1)_y$. We indeed see that most of the operators we noted precisely match the contribution expected from the broken current multiplets and the Higgs branch chiral ring operator if we identify: $y=a^{N+1}$, $z^{N+3} = \frac{c}{b^{\frac{2N+3}{N+2}}}$.  

We can use this to determine the embedding of the $3d$ global symmetry in the $5d$ global symmetry. Specifically, we see that:
\be \label{SUdecomp}
{\bf F}_{SU(2N+6)} \rightarrow \frac{c^{\frac{N+2}{N+3}}}{b^{\frac{2N+3}{N+3}}} (a^{N+1} + \frac{1}{a^{N+1}}) + \frac{b^{\frac{2N+3}{(N+2)N+3}}}{c^{\frac{1}{N+3}}} \left( \frac{1}{b^{\frac{N+1}{N+2}}}{\bf F}_{SU(2N+3)} + b^{\frac{(N+1)(2N+3)}{N+2}} \right) ,
\ee
which we can put to use in analyzing the real mass deformations of this theory. Like the previous cases, there are two cases of most interest, corresponding to integrating out a lower or upper flavor. Let's first consider the case of integrating out the lower flavor. This corresponds to a mass deformation in $U(1)_c$. From \eqref{SUdecomp} we see that it should map to the $5d$ mass deformation breaking $SU(2N+6)\rightarrow U(1)_c \times SU(2)_a \times SU(2N+4)$. This mass deformation is expected to lead to a flow from the $5d$ $SU(2N+6)$ SCFT to the $5d$ $SU(2)\times SU(2N+4)$ SCFT. As such we expect the resulting $3d$ theory to be the result of the compactification of the $SU(2)\times SU(2N+4)$ SCFT, where here the flux is in the $SU(2)$ factor.

We can next consider integrating out one of the upper flavors. For this we again take ${\bf F}_{SU(2N+3)} \rightarrow \frac{y}{b} {\bf F}_{SU(2N+2)} + \frac{b^{2N+2}}{y^{2N+2}}$ so that the deformation sits in $U(1)_b$. We then have that:

\bea
{\bf F}_{SU(2N+6)} & \rightarrow & \frac{c^{\frac{1}{(N+2)(N+3)}} y^{\frac{(N+1)}{(N+2)}}}{b^{\frac{2N+3}{N+3}}} \left( \frac{c^{\frac{N+1}{N+2}}}{y^{\frac{N+1}{N+2}}}(a^{N+1} + \frac{1}{a^{N+1}}) + \frac{y^{\frac{1}{N+2}}}{c^{\frac{1}{N+2}}} {\bf F}_{SU(2N+2)} \right) \nonumber \\ & + & \frac{b^{\frac{(N+2)(2N+3)}{N+3}}}{c^{\frac{1}{N+3}} y^{N+1}}(y^{N+1} + \frac{1}{y^{N+1}}) .
\eea

From this we see that this mass deformation should again map to the $5d$ mass deformation breaking $SU(2N+6)\rightarrow U(1)_b \times SU(2) \times SU(2N+4)$. As such, it is expected to lead to a flow from the $5d$ $SU(2N+6)$ SCFT to the $5d$ $SU(2)\times SU(2N+4)$ SCFT, and so we expect the resulting $3d$ theory to be the result of the compactification of the $SU(2)\times SU(2N+4)$ SCFT, where now the flux is in the $SU(2N+4)$ factor.

We can similarly consider the $(2N+2,2)$ split theory shown in figure \ref{QuiverSUNX23d} (b). Again we can analyze the perturbative and non-perturbative contributions. For the perturbative part we get very similar contributions to the previous case. Specifically, we have the flip fields contributing $2 a^{2(N+1)}x^2$ to the index. The $SU(N+1)$ mesons contribute $ a^{N+1}x^2 {\bf 2}_{SU(2)} ( \frac{b}{c} {\bf F}_{SU(2N+2)} + \frac{c}{b} \overline{\bf F}_{SU(2N+2)} )$ to the index, and the baryons contribute $a^{N+1}x^{N+1}( b^{N+1} + \frac{1}{b^{N+1}}) {\bf \Lambda}^{N+1}_{SU(2N+2)}$. There are also some perturbative contributions that did not exist in the previous case. For instance, as we now have two antifundamentals, we can form a flavor $SU(2)$ invariant from them, contract them with $N-1$ $SU(N+1)\times SU(N+1)$ bifundamentals to make an invariant under one $SU(N+1)$ gauge symmetry, and finally contract with $N-1$ $SU(N+1)\times SU(2N+2)$ bifundamentals to make an invariant under the other $SU(N+1)$ gauge symmetry. This contributes $n a^{N+1}x^{N+1}( c^2 b^{N-1} {\bf \Lambda}^{N-1}_{SU(2N+2)} + \frac{1}{c^2 b^{N-1}} \overline{\bf \Lambda}^{N-1}_{SU(2N+2)})$, where $n$ is some number that enters as there are two $SU(N+1)\times SU(N+1)$ bifundamentals.

We can next consider the non-perturbative contributions. Again, we shall not perform an exhaustive search, but rather content ourselves with looking at specific low-lying monopoles, notably, the ones with minimal charge under one or both groups. First we consider the minimal monopole with charge under only one group. Unlike in the previous case, now it is gauge charged. However, we can make an invariant by dressing it with $N-1$ $SU(N+1)\times SU(2N+2)$ bifundamentals. The resulting gauge invariant operator contributes to the index the term: $a^{N+1}x^{N+1}( \frac{c^2}{b^{N+3}} {\bf \Lambda}^{N-1}_{SU(2N+2)} + \frac{b^{N+3}}{c^2} \overline{\bf \Lambda}^{N-1}_{SU(2N+2)})$.

Next we consider the monopole with minimal charge under both groups. As in the previous case, this operator carries gauge charges under the $SU(N+1)\times SU(N+1)$ gauge symmetry, so to form an invariant we need to dress it with $N-1$ copies of the $SU(N+1)\times SU(N+1)$ bifundamentals. This gives an operator with $5d$ R-charge $0$, $U(1)_a$ charge $-(2N+2)$ and no charges under the other symmetries. The product of this operator with the flip fields then gives the marginal operators that are flavor symmetry singlets.

The states found so far are mostly consistent with the theory being the compactification of the $5d$ $SU(2)\times SO(4N+8)$ SCFT with minimal flux in its $SO(4N+8)$ global symmetry preserving $U(1)\times SU(2)\times SO(4N+4)$. First we note that we can identify $\frac{b}{c} {\bf F}_{SU(2N+2)} + \frac{c}{b} \overline{\bf F}_{SU(2N+2)}$ as the character of the vector of $SO(4N+4)$. The contributions of the flip fields and mesons then match the contributions we expect from the broken $SO(4N+8)$ currents. The remaining states we found can be identified as part of the contribution of the Higgs branch chiral ring operator, although the structure here is more complicated. Specifically, we expect it to be in the doublet of the $SU(2)$ and the spinor of $SO(4N+8)$. When we break $SO(4N+8)\rightarrow SU(2)\times SU(2)\times SO(4N+4) \rightarrow U(1)\times SU(2)\times SO(4N+4)$, the latter decomposes to the two spinors of $SO(4N+4)$, with one charged under the first $SU(2)$, while the other is charged under the second one. Further decomposing $SO(4N+4)\rightarrow U(1)\times SU(2N+2)$, as expected from the decomposition of the vector, we have that 
\bea \label{spinordecomp}
& & {\bf S}_{SO(4N+4)} \rightarrow \sum^{N+1}_{i=0} (\frac{c}{b})^{N+1-2i} {\bf \Lambda^{2i}}_{SU(2N+2)} \\ \nonumber
& & {\bf C}_{SO(4N+4)} \rightarrow \sum^{N}_{i=0} (\frac{c}{b})^{N-2i} {\bf \Lambda^{2i+1}}_{SU(2N+2)} 
\eea     
Looking at the previous terms we found, we see that they indeed match as being part of the spinor Higgs branch chiral ring operator. Specifically, the baryons have $5d$ R-charge $N+1$ and $U(1)_a$ charge $N+1$, which is consistent with the contribution of the part of the Higgs branch chiral ring operator in the doublet of the $SU(2)$ whose Cartan is the $U(1)$ carrying the flux. We interpret the term $b^{N+1} + \frac{1}{b^{N+1}}$ as the character of the doublet of the $SU(2)$ factor of the $5d$ SCFT global symmetry. Finally, the representations under the remaining flavor symmetries matches with part of the representations we expect from the decomposition of the $SO(4N+4)$ spinor, \eqref{spinordecomp}.

We can further consider the contributions of the basic $SU(N+1)$ monopoles and the other perturbative invariants we found. Assuming $n=1$, we can sum them to get
\bea
 & & a^{N+1}x^{N+1}( \frac{c^2}{b^{N+3}} {\bf \Lambda}^{N-1}_{SU(2N+2)} + \frac{b^{N+3}}{c^2} \overline{\bf \Lambda}^{N-1}_{SU(2N+2)}) \nonumber \\ & + & a^{N+1}x^{N+1}( c^2 b^{N-1} {\bf \Lambda}^{N-1}_{SU(2N+2)} + \frac{1}{c^2 b^{N-1}} \overline{\bf \Lambda}^{N-1}_{SU(2N+2)}) \nonumber \\ & = & a^{N+1}x^{N+1} (b^{N+1} + \frac{1}{b^{N+1}}) (\frac{c^2}{b^{2}} {\bf \Lambda}^{N-1}_{SU(2N+2)} + \frac{b^{2}}{c^2} \overline{\bf \Lambda}^{N-1}_{SU(2N+2)}) .
\eea
This is again consistent with being part of the contribution expected from the spinor chiral ring operator. Nevertheless, as we are not performing an explicit index computations, this does not show that we get the correct contributions to fully form the character of the spinor of $SO(4N+4)$. First, we did not show that $n=1$, that is the complete contribution of all multiplets to this term leaves just one of them uncanceled. Furthermore, we are missing additional terms to fully build the spinor, which presumably come from higher monopole operators or more complicated invariants. As we previously noted, we shall not perform an exhaustive search here and so content ourselves with finding a $5d$ interpretation for most of the basic operators. We do note that in cases where we have computed the full index, like $N=2$, it does confirm to our expectations.  

We can next consider mass deformations. Again to analyze this we first consider the embedding of the $3d$ global symmetry in the $5d$ global symmetry expected from the matching of the operators. Here we find

\bea
& & {\bf 2}_{SU(2)_{5d}} \rightarrow b^{N+1} + \frac{1}{b^{N+1}} , \\ \nonumber
& & {\bf V}_{SO(4N+8)} \rightarrow (a^{N+1} + \frac{1}{a^{N+1}}) {\bf 2}_{SU(2)_{3d}} + \frac{b}{c} {\bf F}_{SU(2N+2)} + \frac{c}{b} \overline{\bf F}_{SU(2N+2)} .
\eea

Next, we consider the mass deformations associated with integrating out the upper or lower flavors. Let's begin with the one associated with integrating out the upper flavor. We can again study it by decomposing: ${\bf F}_{SU(2N+2)}\rightarrow \frac{z}{b}{\bf F}_{SU(2N+1)} + \frac{b^{2N+2}}{z^{2N+2}}$, so that it resides inside $U(1)_b$. We then have that 

\bea
& & {\bf V}_{SO(4N+8)} \rightarrow (a^{N+1} + \frac{1}{a^{N+1}}) {\bf 2}_{SU(2)_{3d}} + \frac{b}{c} {\bf F}_{SU(2N+2)} + \frac{c}{b} \overline{\bf F}_{SU(2N+2)} \nonumber \\ & & \rightarrow (a^{N+1} + \frac{1}{a^{N+1}}) {\bf 2}_{SU(2)_{3d}} + \frac{z}{c} {\bf F}_{SU(2N+1)} + \frac{c}{z} \overline{\bf F}_{SU(2N+1)} + \frac{b^{2N+2}}{c} + \frac{c}{b^{2N+2}}. \nonumber \\  & &
\eea

We see that a real mass in $U(1)_b$ maps to the $5d$ mass deformation breaking $SU(2)\rightarrow U(1)$ and $SO(4N+8)\rightarrow U(1)\times SO(4N+6)$. This $5d$ mass deformation is the one leading to a flow from the $SU(2)\times SO(4N+8)$ SCFT to the $U(1)\times SO(4N+6)$ one. We further see that the flux is expected to map to a flux inside $SO(4N+6)$ such that its commutant is $U(1)\times SU(2)\times SO(4N+2)$.

We next consider the $3d$ real mass deformation associated with integrating out one of the lower flavors. To analyze it we take: ${\bf 2}_{SU(2)_{3d}}\rightarrow \frac{y}{c} + \frac{c}{y}$, so that the deformation resides inside $U(1)_c$. We then have that

\bea
& & {\bf V}_{SO(4N+8)} \rightarrow (a^{N+1} + \frac{1}{a^{N+1}}) {\bf 2}_{SU(2)_{3d}} + \frac{b}{c} {\bf F}_{SU(2N+2)} + \frac{c}{b} \overline{\bf F}_{SU(2N+2)} \nonumber \\ & & \rightarrow \frac{b^{\frac{N+1}{N+2}}y^{\frac{1}{N+2}}}{c} \left( \frac{b^{\frac{1}{N+2}}}{y^{\frac{1}{N+2}}} {\bf F}_{SU(2N+2)} + \frac{y^{\frac{N+1}{N+2}}}{b^{\frac{N+1}{N+2}}} (a^{N+1} + \frac{1}{a^{N+1}}) \right) \nonumber \\ & + & \frac{c}{b^{\frac{N+1}{N+2}}y^{\frac{1}{N+2}}} \left( \frac{y^{\frac{1}{N+2}}}{b^{\frac{1}{N+2}}} \overline{\bf F}_{SU(2N+2)} + \frac{b^{\frac{N+1}{N+2}}}{y^{\frac{N+1}{N+2}}} (a^{N+1} + \frac{1}{a^{N+1}}) \right) .
\eea  

We see that a real mass in $U(1)_c$ maps to the $5d$ mass deformation breaking $SO(4N+8)\rightarrow U(1)\times SU(2N+4)$. This $5d$ mass deformation is the one leading to a flow from the $SU(2)\times SO(4N+8)$ SCFT to the $SU(2)\times SU(2N+4)$ one. We further see that the flux is expected to map to a flux inside $SU(2N+4)$ such that its commutant is $U(1)^2\times SU(2N+2)$.

Considering everything noted so far we can form a conjecture for the $5d$ model corresponding to each of the $3d$ models. First, we consider the case of the $(N_f-2,2)$ split for $2\leq N_f \leq 2N+4$. In this case we expect the $3d$ model to correspond to the torus compactification of the $5d$ SCFT UV completing the $5d$ gauge theory $SU(N+1)_{N+3-\frac{N_f}{2}}+N_fF$. The global symmetry of this $5d$ SCFT contains $SO(2N_f)$, and the flux is of unit value and sits in a $U(1)$ whose commutant is $U(1)\times SU(2)\times SO(2N_f-4)$. This also fits with the demand that $N_f\geq 2$.

If we instead consider the $(N_f-1,1)$ split for $1\leq N_f \leq 2N+3$, then the $3d$ model should correspond to the torus compactification of the $5d$ SCFT UV completing the $5d$ gauge theory $SU(N+1)_{N+2-\frac{N_f}{2}}+N_fF$. The global symmetry of this $5d$ SCFT contains $SU(N_f+1)$, and the flux is of unit value and sits in a $U(1)$ whose commutant is $U(1)^2\times SU(N_f-1)$. This also fits with the demand that $N_f\geq 1$. 

Finally, for the $(N_f,0)$ split with $0\leq N_f \leq 2N+2$, the $3d$ model should correspond to the torus compactification of the $5d$ SCFT UV completing the $5d$ gauge theory $SU(N+1)_{N+1-\frac{N_f}{2}}+N_fF$. The global symmetry of this $5d$ SCFT contains an $SU(2)$, and the flux is of unit value in it.

Additionally, we also have the case of the $(2N+3,1)$ split, which as we mentioned, corresponds to the torus compactification of the $5d$ SCFT UV completing the $5d$ gauge theory $SU(N+1)_{0}+(2N+4)F$, with unit flux in a $U(1)$ subgroup of its $SU(2N+6)$ global symmetry, such that its commutant is $U(1)^2\times SU(2N+4)$. We further have the case of the $(2N+3,0)$ split, which as we mentioned, corresponds to the torus compactification of the $5d$ SCFT UV completing the $5d$ gauge theory $SU(N+1)_{\frac{1}{2}}+(2N+3)F$, with unit flux in a $U(1)$ subgroup of its $SU(2)$ factor in its $SU(2)\times SU(2N+4)$ global symmetry.


\section{Gluing together tubes of different types}
\label{additional}

So far we have constructed $3d$ theories corresponding to the compactification of $5d$ SCFTs on tubes with different values of flux. In some cases we had more than one $3d$ theory corresponding to the compactification of the same $5d$ SCFT on a tube but with different values of flux. We should then be able to glue these tubes together to form more general tubes. Here we shall discuss and test this with some examples.

\subsection{The case of the $SU(2)\times SU(8)$ SCFT}  

One case where we found two tubes is the case of the $SU(2)\times SU(2N+4)$ $5d$ SCFT, which is the UV completion of the $SU(N+1)_{\frac{1}{2}}+(2N+3)F$ gauge theory. In this case the $3d$ $SU-SU$ tubes that we previously discussed for both the $(2N+3,0)$ and $(2N+2,1)$ splits are associated with this $5d$ SCFT, where for the $(2N+3,0)$ split the flux is embedded in the $SU(2)$ factor while for the $(2N+2,1)$ split the flux is embedded in the $SU(2N+4)$ factor. In both tubes the puncture global symmetry is $SU(N+1)$ so we expect to be able to glue them together by gauging a diagonal $SU(N+1)$ subgroup. Here we shall discuss how this is done. For simplicity, we shall consider the specific case of $N=2$, with the generalization to generic $N$ being straightforward.

\begin{figure}
\center
\includegraphics[width=0.9\textwidth]{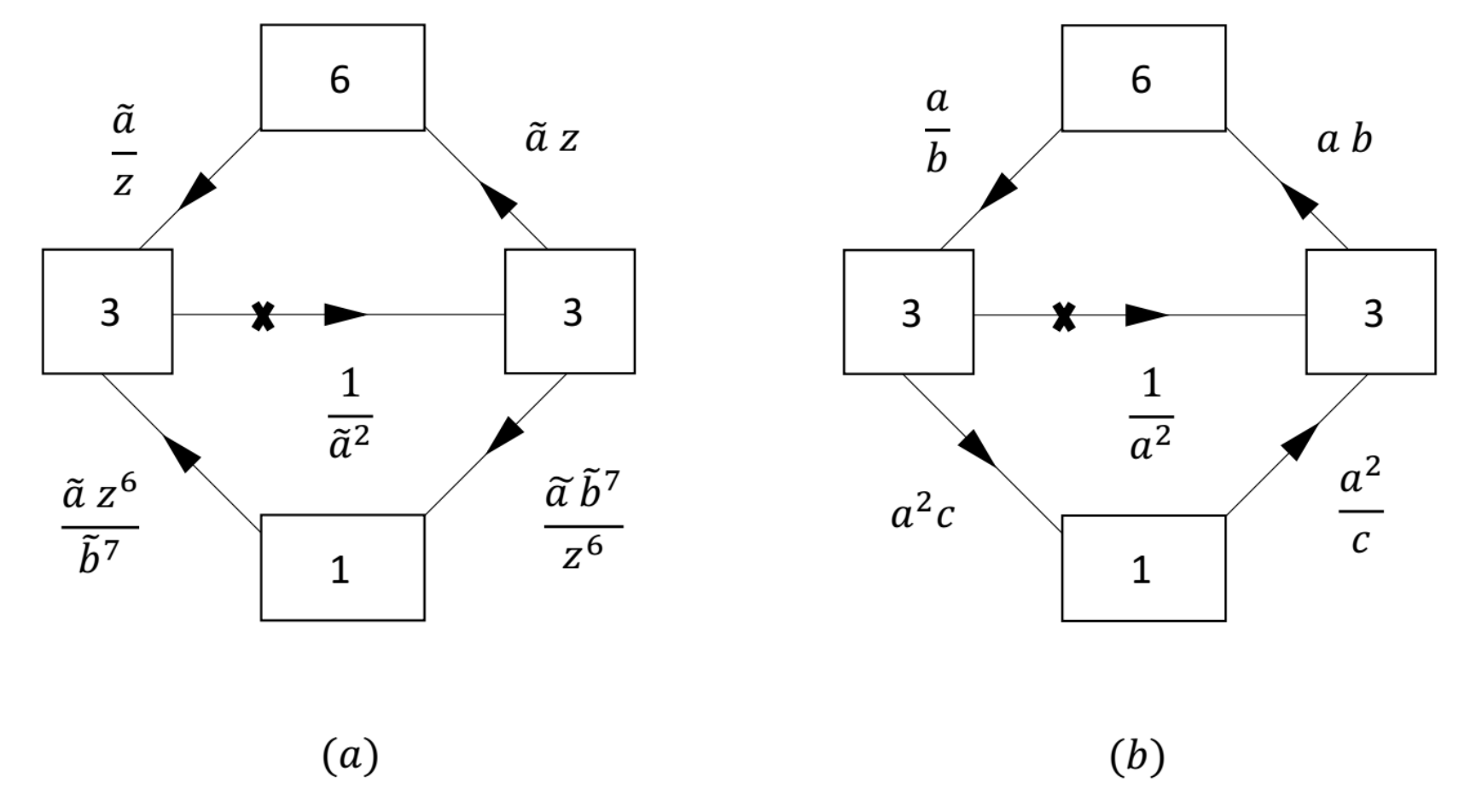} 
\caption{The $SU-SU$ tubes for the case of $N=2$ with (a) the (7,0) split and (b) the (6,1) split. Charges under the global symmetries are denoted using fugacities. In (a) we have also split the $7$ chirals on the two sides to $6+1$, and correspondingly only the $SU(6)\times U(1)_z$ subgroup of the $SU(7)$ global symmetry is manifest in the quiver.}
\label{SU2SU8Tubes}
\end{figure}

We first consider the two tubes, where for the $(7,0)$ split we further split the $7$ into $6+1$ to facilitate the gluing. The tubes are reproduced in figure \ref{SU2SU8Tubes}. Naively attempting to glue them we encounter the following problem. The representation of the $SU(3)$ flavors are different between the two sides making it not immediately obvious how these are to be glued. To get a better grip on this, we first consider how the symmetries of the two models are related. Specifically, the global symmetry of each of these tubes should be related to the $5d$ global symmetry, and the gluing should be done such that it is preserved. However, the $5d$ global symmetry is embedded differently in each of the two tubes, and in order to determine the correct gluing it is convenient to first determine the mapping.

Recall that we previously determined that for the $(7,0)$ split we should have that

\bea
& & {\bf 2}_{SU(2)_{5d}} \rightarrow \tilde{a}^{3} + \frac{1}{\tilde{a}^{3}} , \\ \nonumber
& & {\bf 8}_{SU(8)} \rightarrow \frac{z}{\tilde{b}^{\frac{7}{4}}} {\bf 6}_{SU(6)} + \frac{\tilde{b}^{\frac{21}{4}}}{z^3}(z^{3} + \frac{1}{z^{3}}) ,
\eea
while for the $(6,1)$ split we should have that
\bea
& & {\bf 2}_{SU(2)_{5d}} \rightarrow b^{3} + \frac{1}{b^{3}} , \\ \nonumber
& & {\bf 8}_{SU(8)} \rightarrow \frac{c^{\frac{3}{4}}}{b^{\frac{3}{4}}}(a^{3} + \frac{1}{a^{3}}) + \frac{b^{\frac{1}{4}}}{c^{\frac{1}{4}}} {\bf 6}_{SU(6)} .
\eea
From this we can find the mapping between the global symmetries of the two theories. There are several options and we shall take the mapping: $\tilde{a}\rightarrow b$, $z\rightarrow \frac{1}{a}$ and $\tilde{b}^7\rightarrow \frac{c}{a^4 b}$. Using this mapping, we see that the right $SU(3)\times SU(6)$ bifundamental in figure \ref{SU2SU8Tubes} (a) has exactly the opposite charges from that of the left $SU(3)\times SU(6)$ bifundamental in figure \ref{SU2SU8Tubes} (b). Similarly, the right $SU(3)$ fundamental in figure \ref{SU2SU8Tubes} (a) has exactly the same charges as that of the left $SU(3)$ fundamental in figure \ref{SU2SU8Tubes} (b). 

We can next consider gluing the two tubes. Up until now, all pairs of fields in the glued tubes had the same charges. This corresponds to the fact that the same component of the hypermultiplet receives Neumann boundary conditions on the two punctures. The gluing is then done by reintroducing the component receiving the Dirichlet boundary conditions and coupling it to the difference of the chirals on the two sides such so that the resulting F-term condition equates the two. However, here we also have a case where the two fields have opposite charges, implying that different components of the hypermultiplets receive the Neumann and Dirichlet boundary conditions on the two sides. In this case, when gluing we do not introduce any new field, but rather couple the fields at the two sides by a quadratic superpotential\footnote{If the field survives the reduction to $3d$ then it must receive Neumann boundary conditions at both the puncture and the domain wall. As such if we glue two punctures where different components receive Neumann boundary conditions then that difference should also translate to the boundary conditions at the domain wall. This implies that all component of the bulk hyper should receive a Dirichlet boundary conditions on at least one of the domain walls. As such the whole hyper should become massive in the $3d$ reduction, which is precisely achieved by the quadratic superpotential.}. The two types of gluings, with and without the added chirals, are refereed two as $\Phi$ and S gluing respectively in the context of compactifications of $6d$ SCFTs, see \cite{Gaiotto:2015usa,Razamat:2016dpl}. 

The difference in charges of the fields at the appropriate side of the tubes suggests that these tubes should be glued by a combination of S and $\Phi$ gluing. Specifically, we gauge the diagonal $SU(3)$ subgroup of the right $SU(3)$ group in figure \ref{SU2SU8Tubes} (a) and the left $SU(3)$ group in figure \ref{SU2SU8Tubes} (b), and to avoid gauge anomalies we also turn on a Chern--Simons term of level half. We also introduce the field $\Phi$, which is taken to be in the fundamental representation of the gauged $SU(3)$. We further introduce a superpotential coupling both the right $SU(3)$ fundamental in figure \ref{SU2SU8Tubes} (a) and the left $SU(3)$ fundamental in figure \ref{SU2SU8Tubes} (b) to the field $\Phi$, and another coupling of the right $SU(3)\times SU(6)$ bifundamental in figure \ref{SU2SU8Tubes} (a) and the left $SU(3)\times SU(6)$ bifundamental in figure \ref{SU2SU8Tubes} (b) together. Note that this is consistent with the charges of the fields.   

\begin{figure}
\center
\includegraphics[width=0.55\textwidth]{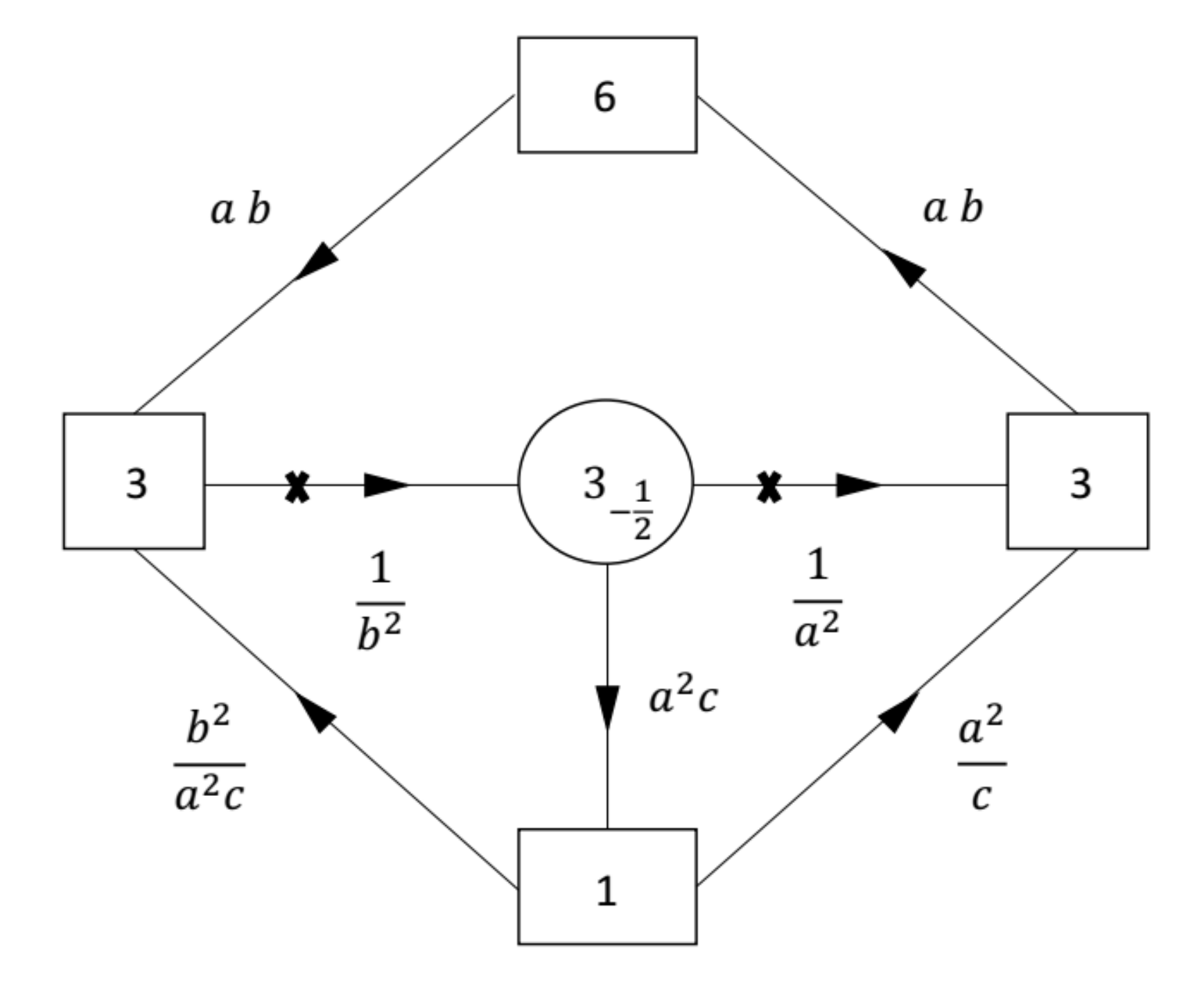} 
\caption{The tube we get by gluing the tube in figure \ref{SU2SU8Tubes} (a) with the one in figure \ref{SU2SU8Tubes} (b). As usual, we have denoted the charges under the non-R symmetries using fugacities.}
\label{SU2SU8GlTube}
\end{figure}

This gives the theory shown in figure \ref{SU2SU8GlTube}, which we then conjecture is the result of the compactification of the $SU(2)\times SU(8)$ $5d$ SCFT on a tube with flux $\frac{1}{2}(1,-1;1,-1,0,0,0,0,0,0)$.

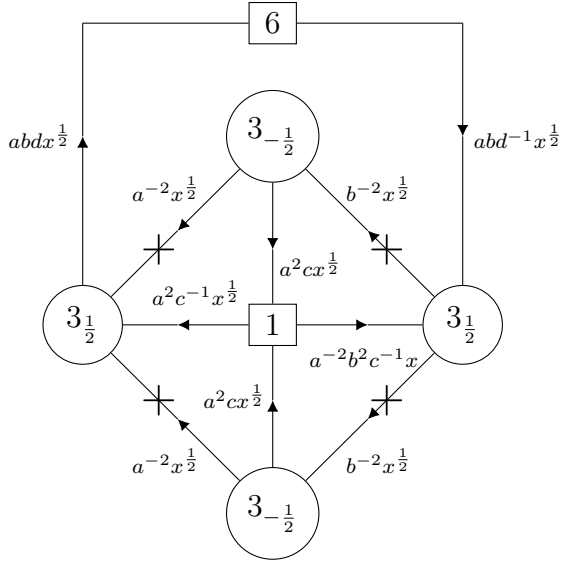
\begin{figure}
\center
\begin{tikzpicture}[baseline=0, font=\scriptsize]
\node[draw, circle] (a1) at (0,0) {\normalsize $ 3_{\frac{1}{2}}$};
\node[draw, circle] (a2) at (2.5,2.5) {\normalsize $ 3_{-\frac{1}{2}}$};
\node[draw, circle] (a3) at (5,0) {\normalsize $3_{\frac{1}{2}}$};
\node[draw, circle] (a4) at (2.5,-2.5) {\normalsize $3_{-\frac{1}{2}}$};
\node[draw, rectangle] (a5) at (2.5,0) {\normalsize $\, 1\,$};
\node[draw, rectangle] (a6) at (2.5,4) {\normalsize $\, 6\,$};
\draw[draw, solid] (a1)--(1.25,1.25);
\draw[draw, solid] (a2)--(3.75,1.25);
\draw[draw, solid,->] (a3)--(3.75,-1.25);
\draw[draw, solid,->] (a4)--(1.25,-1.25);
\draw[draw, solid,<-] (1.25,1.25)--(a2);
\draw[draw, solid,<-] (3.75,1.25)--(a3);
\draw[draw, solid] (3.75,-1.25)--(a4);
\draw[draw, solid] (1.25,-1.25)--(a1);
\draw[draw, solid] (a1)--(1.25,0);
\draw[draw, solid,<-] (1.25,0)--(a5);
\draw[draw, solid,->] (a2)--(2.5,1);
\draw[draw, solid] (2.5,1)--(a5);
\draw[draw, solid,->] (a5)--(3.75,0);
\draw[draw, solid] (3.75,0)--(a3);
\draw[draw, solid] (a5)--(2.5,-1);
\draw[draw, solid,<-] (2.5,-1)--(a4);

\draw[draw,solid,->] (a1) -- (0,2.5);
\draw[draw,solid] (0,2.5) -- (0,4);
\draw[draw,solid] (0,4) -- (a6);
\draw[draw,solid] (a6) -- (5,4);
\draw[draw,solid,->] (5,4) -- (5,2.5);
\draw[draw,solid] (5,2.5) -- (a3);

\node[thick,rotate=45] at (1,1) {\Large$\times$};
\node[thick,rotate=45] at (4,1) {\Large$\times$};
\node[thick,rotate=45] at (4,-1) {\Large$\times$};
\node[thick,rotate=45] at (1,-1) {\Large$\times$};

\node[above] at (1.5,0.125) {$a^2c^{-1}x^{\frac{1}{2}}$};
\node[above] at (3,0.5) {$a^2cx^{\frac{1}{2}}$};
\node[above] at (3.7,-0.7) {$a^{-2}b^2c^{-1}x$};
\node[above] at (2,-1.25) {$a^2cx^{\frac{1}{2}}$};
\node[above] at (1.1,1.5) {$a^{-2}x^{\frac{1}{2}}$};
\node[above] at (3.9,1.5) {$b^{-2}x^{\frac{1}{2}}$};
\node[below] at (1.1,-1.5) {$a^{-2}x^{\frac{1}{2}}$};
\node[below] at (3.9,-1.5) {$b^{-2}x^{\frac{1}{2}}$};
\node[left] at (0,2.5) {$abdx^{\frac{1}{2}}$};
\node[right] at (5,2.5) {$abd^{-1}x^{\frac{1}{2}}$};

\end{tikzpicture}
\caption{The quiver for the theory resulting from the compactification of the $5d$ $SU(2)\times SU(8)$ SCFT on a torus with flux $(1,-1;1,-1,0,0,0,0,0,0)$.}
\label{newSU2SU8torus}
\end{figure}

As a check of our construction, we can construct the torus model obtained by $\Phi$-gluing two copies of the tube in figure \ref{SU2SU8GlTube}. Again, to make sense of the theory we should turn on Chern--Simons interactions at the new $SU(3)$ gauge nodes that arise from the gluing, which we take to be at alternating levels $\pm\frac{1}{2}$. The resulting model is depicted in figure \ref{newSU2SU8torus}, where we also specify the CS levels of each gauge node, as well as a possible assignment of $U(1)$ charges and R-charges that is consistent with the superpotential, which for the moment we take to involve only the perturbative matter sector of the theory and we remain agnostic on possible monopole superpotentials. Notice that the R-symmetry that we are using is related to the $5d$ R-symmetry by the shifts $a\to a x^{\frac{1}{4}}$ and $b\to b x^{\frac{1}{4}}$.

The supersymmetric index of this theory reads
\bea
\mathcal{I}&=&1+2(a^6+b^6)x^{\frac{1}{2}}+\left(3(a^{12}+b^{12})+4a^6b^6+\frac{a^3bd}{c}{\bf 6}\right)x+\left(4(a^{18}+b^{18})+\right.\nn\\
&+&\left.6(a^{12}b^6+a^6b^{12})+a^3b^3\left(d^3+\frac{1}{d^3}\right){\bf 20}+2\frac{d}{c}(a^9b+a^3b^7){\bf 6}+\frac{a^3c}{bd}\overline{\bf 6}\right)x^{\frac{3}{2}}+\nn\\
&+&\left({\color{red}-4}{\color{blue}+2\left(d^6+\frac{1}{d^6}\right)}+5(a^{24}+b^{24})+8(a^{18}b^6+a^6b^{18})+9a^{12}b^{12}+\frac{b^4}{cd^2}\overline{\bf 15}+\right.\nn\\
&+&\left.2\left(d^3+\frac{1}{d^3}\right)(a^9b^3+a^3b^9){\bf 20}+3\frac{d}{c}(a^{15}b+a^3b^{13}){\bf 6}+2\frac{c}{d}\left(\frac{a^9}{b}+a^3b^5\right)\overline{\bf 6}+\right.\nn\\
&+&\left.\frac{a^6b^2d^2}{c^2}{\bf 21}+4\frac{a^9b^7d}{c}{\bf 6}\right)x^2+\cdots\,.
\eea

This theory is expected to be the result of the compactification of the $5d$ $SU(2)\times SU(8)$ SCFT on a torus with flux $(1,-1;1,-1,0,0,0,0,0,0)$, which preserves $U(1)^3\times SU(6)$. Instead, the global symmetry of our model is $U(1)^4\times SU(6)$, so we have one additional $U(1)$. Moreover, the $-4$ at order $x^2$ that we highlighted in red doesn't conform to our expectations from the $5d$ picture. In analogy with previous examples, we conjecture that this $U(1)$ is broken by a monopole superpotential consisting of all the $(1,1)$ monopoles, which should be properly dressed in order to be made gauge invariant. Specifically, the $(1,1)$ monopoles involving the left $SU(3)$ gauge node should be dressed with two copies of the associated bifundamental, while those involving the right $SU(3)$ should be dressed with three copies of the bifundamental. These give four gauge invariant monopole operators that have R-charge 2 under our R-symmetry, charge $\pm6$ under $U(1)_d$ and are uncharged under all the other abelian symmetries and the $SU(6)$ symmetry (their contribution in the index is highlighted in blue). We can turn them on leading to the breaking of $U(1)_d$. Accordingly setting $d=1$, the contribution of these monopoles cancels the $-4$, so that now our result for the index conforms to the $5d$ expectations. More precisely, the index of the model with the monopole superpotential is
\bea
\mathcal{I}&=&1+2(a^6+b^6)x^{\frac{1}{2}}+\left(3(a^{12}+b^{12})+4a^6b^6+\frac{a^3b}{c}{\bf 6}\right)x+\left(4(a^{18}+b^{18})+\right.\nn\\
&+&\left.6(a^{12}b^6+a^6b^{12})+2a^3b^3{\bf 20}+2\frac{1}{c}(a^9b+a^3b^7){\bf 6}+\frac{a^3c}{b}\overline{\bf 6}\right)x^{\frac{3}{2}}+\nn\\
&+&\left(5(a^{24}+b^{24})+8(a^{18}b^6+a^6b^{18})+9a^{12}b^{12}+\frac{b^4}{c}\overline{\bf 15}+\right.\nn\\
&+&\left.4(a^9b^3+a^3b^9){\bf 20}+3\frac{1}{c}(a^{15}b+a^3b^{13}){\bf 6}+2c\left(\frac{a^9}{b}+a^3b^5\right)\overline{\bf 6}+\right.\nn\\
&+&\left.\frac{a^6b^2}{c^2}{\bf 21}+4\frac{a^9b^7}{c}{\bf 6}\right)x^2+\cdots\,.
\label{indnewSU2SU8torus}
\eea

We can check that the spectrum of operators is consistent with the claim that this theory is the result of the compactification of the $5d$ $SU(2)\times SU(8)$ SCFT on a torus with a unit of flux for a $U(1)$ inside $SU(2)$ and for a $U(1)$ inside $SU(8)$ whose commutant is $U(1)^2\times SU(6)$. For example, the $SU(2)$ conserved current decomposes as
\be
{\bf 3}_{SU(2)}\to 1+b^6+\frac{1}{b^6}
\ee
and we can identify the second of these states in the index \eqref{indnewSU2SU8torus} as the contribution $2b^6x^{\frac{1}{2}}$. Instead, the $SU(8)$ conserved current decomposes as
\be
{\bf 63}_{SU(8)}\to \frac{b}{c}\left(a^3+\frac{1}{a^3}\right){\bf 6}_{SU(6)}+{\bf 35}_{SU(6)}+2+a^6+\frac{1}{a^6}+\frac{c}{b}\left(a^3+\frac{1}{a^3}\right)\overline{\bf 6}_{SU(6)}
\ee
and we can identify some of these states in the index \eqref{indnewSU2SU8torus} as the contributions $2a^6x^{\frac{1}{2}}$, $a^3bc^{-1}{\bf 6}x$ and $a^3b^{-1}c\overline{\bf 6}x^{\frac{3}{2}}$.

Remember that the set of the HB chiral ring generators of the $5d$ $SU(2)\times SU(8)$ SCFT, on top of the moment map for the global symmetry, contains also an operator in the $\bf 2$ of $SU(2)$, the $\bf 70$ of $SU(8)$ and the $\bf 4$ of $SU(2)_R$. We can also check the presence of some of the states coming from this operator according to the branching rule
\be
({\bf 2}_{SU(2)},{\bf 70}_{SU(8)})\to\left(b^3+\frac{1}{b^3}\right)\left[\left(a^3+\frac{1}{a^3}\right){\bf 20}_{SU(6)}+\frac{c}{b}{\bf 15}_{SU(6)}+\frac{b}{c}\overline{\bf 15}_{SU(6)}\right]\,.
\ee
In particular, we can see the states $a^3b^3{\bf 20}_{SU(6)}$ and $\frac{b^4}{c}\overline{\bf 15}_{SU(6)}$ which contribute with the terms $2a^3b^3{\bf 20}_{SU(6)}x^{\frac{3}{2}}$ and $\frac{b^4}{c}\overline{\bf 15}_{SU(6)} x^2$ respectively.

\subsection{The case of the $SU(2)\times SO(16)$ SCFT}

Next we consider the case of the $5d$ SCFTs UV completing the $5d$ gauge theories $SU(N+1)_{N+3-\frac{N_f}{2}}+N_f F$. Here we have presented two tubes, one involving two $SU$ type punctures and one involving an $SU$ and a $USp$ type punctures. We expect to be able to glue them together to get a new tube along the $SU$ type punctures of the two tubes. Here we shall analyze this case, where for simplicity we shall set $N=2$ and $N_f=8$. The generalization to generic cases should be straightforward.

\begin{figure}
\center
\includegraphics[width=0.85\textwidth]{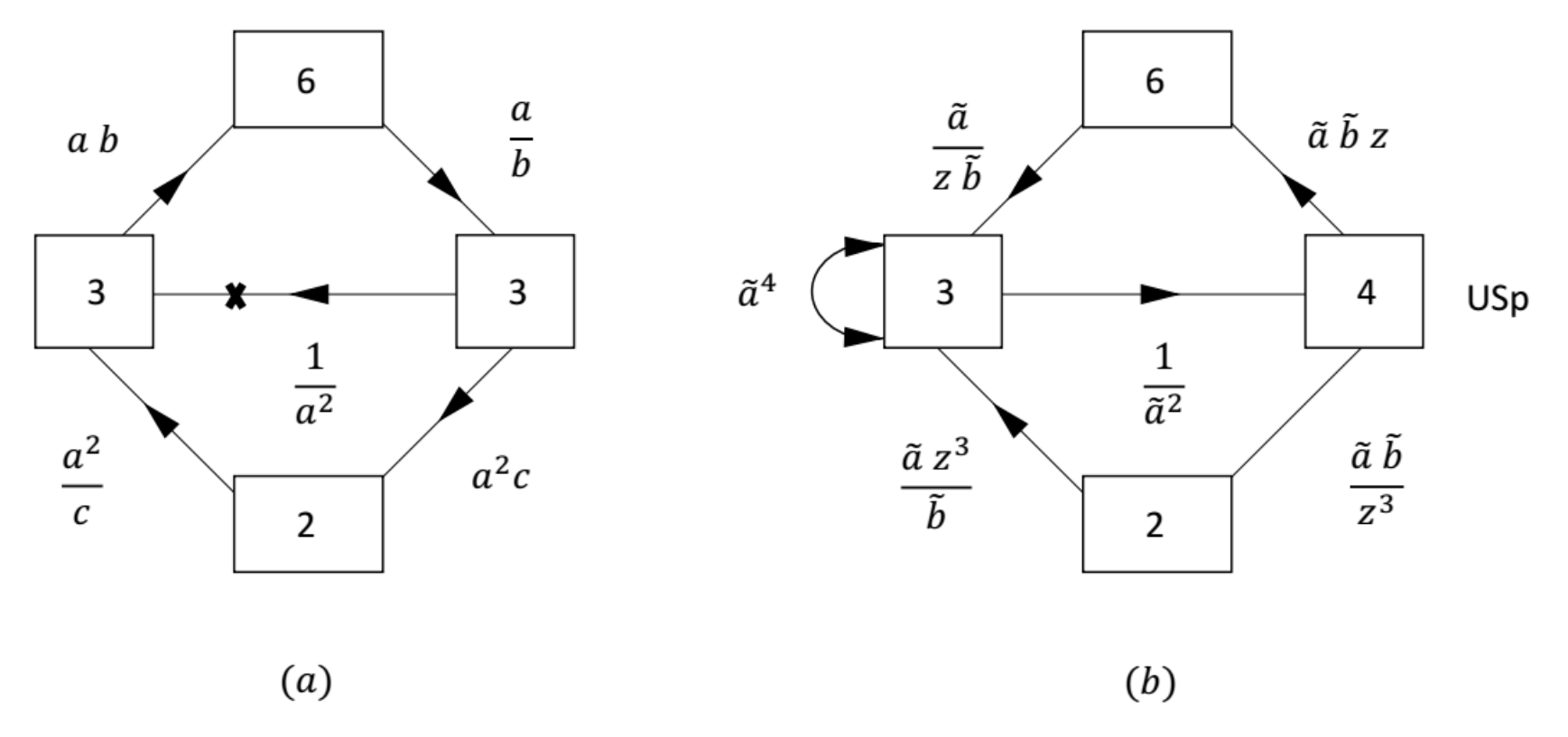} 
\caption{Two theories associated with the compactification of the $SU(2)\times SO(16)$ $5d$ SCFT on a tube with (a) two $SU(3)$ punctures and (b) an $SU(3)$ and a $USp(4)$ puncture. Charges under the global symmetries are denoted using fugacities. In (b) we have also split the $8$ chirals on the two sides to $6+2$, and correspondingly only the $SU(6)\times SU(2) \times U(1)_z$ subgroup of the $SU(8)$ global symmetry is manifest in the quiver.}
\label{SU2SO16Tubes}
\end{figure}

The analysis proceeds as in the previous case. We first write both tubes, where to facilitate the gluing we split the eight flavors of the $SU-USp$ tube as follows: ${\bf 8}_{SU(8)}\rightarrow z {\bf 6}_{SU(6)} + \frac{1}{z^3} {\bf 2}_{SU(2)}$. The resulting tubes are reproduced in figure \ref{SU2SO16Tubes}. We again encounter the issue that the $SU(3)$ representations of the flavors are different for the two tubes, and we expect that gluing of the two tubes should involve a combination of $\Phi$ and S gluing. To get a handle of it, it is convenient to first understand the symmetry mapping between the two theories.  

From the results of the previous sections, we see that the $3d$ global symmetries are related to the $5d$ ones by

\bea
& & {\bf 2}_{SU(2)_{5d}} \rightarrow \frac{\tilde{a}^{2}}{\tilde{b}^{4}} + \frac{\tilde{b}^{4}}{\tilde{a}^{2}} , \\ \nonumber
& & {\bf 16}_{SO(16)} \rightarrow \tilde{a} \tilde{b} z {\bf 6}_{SU(6)} + \frac{\tilde{a} \tilde{b}}{z^3}{\bf 2}_{SU(2)} + \frac{1}{\tilde{a} \tilde{b} z} \overline{\bf 6}_{SU(6)} + \frac{z^3}{\tilde{a} \tilde{b}}{\bf 2}_{SU(2)} ,
\eea
for the $SU-USp$ model, and

\bea
& & {\bf 2}_{SU(2)_{5d}} \rightarrow b^{3} + \frac{1}{b^{3}} , \\ \nonumber
& & {\bf 16}_{SO(16)} \rightarrow (a^{3} + \frac{1}{a^{3}}){\bf 2}_{SU(2)} + \frac{b}{c} {\bf 6}_{SU(6)} + \frac{c}{b} \overline{\bf 6}_{SU(6)} ,
\eea
for the $SU-SU$ one. 

From these we can determine the mapping of the symmetries. We again have several options, and we shall take: $b^3\rightarrow \frac{\tilde{b}^{4}}{\tilde{a}^{2}}$, $\frac{b}{c}\rightarrow \tilde{a} \tilde{b} z$, $a^3 \rightarrow \frac{\tilde{a} \tilde{b}}{z^3}$. Using this we see that the right $SU(3)\times SU(6)$ bifundamental in figure \ref{SU2SO16Tubes} (a) has exactly the same charges as that of the left $SU(3)\times SU(6)$ bifundamental in figure \ref{SU2SO16Tubes} (b). Similarly, the right $SU(3)$ fundamental in figure \ref{SU2SO16Tubes} (a) has exactly the opposite charges from that of the left $SU(3)$ fundamental in figure \ref{SU2SO16Tubes} (b). 

\begin{figure}
\center
\includegraphics[width=0.55\textwidth]{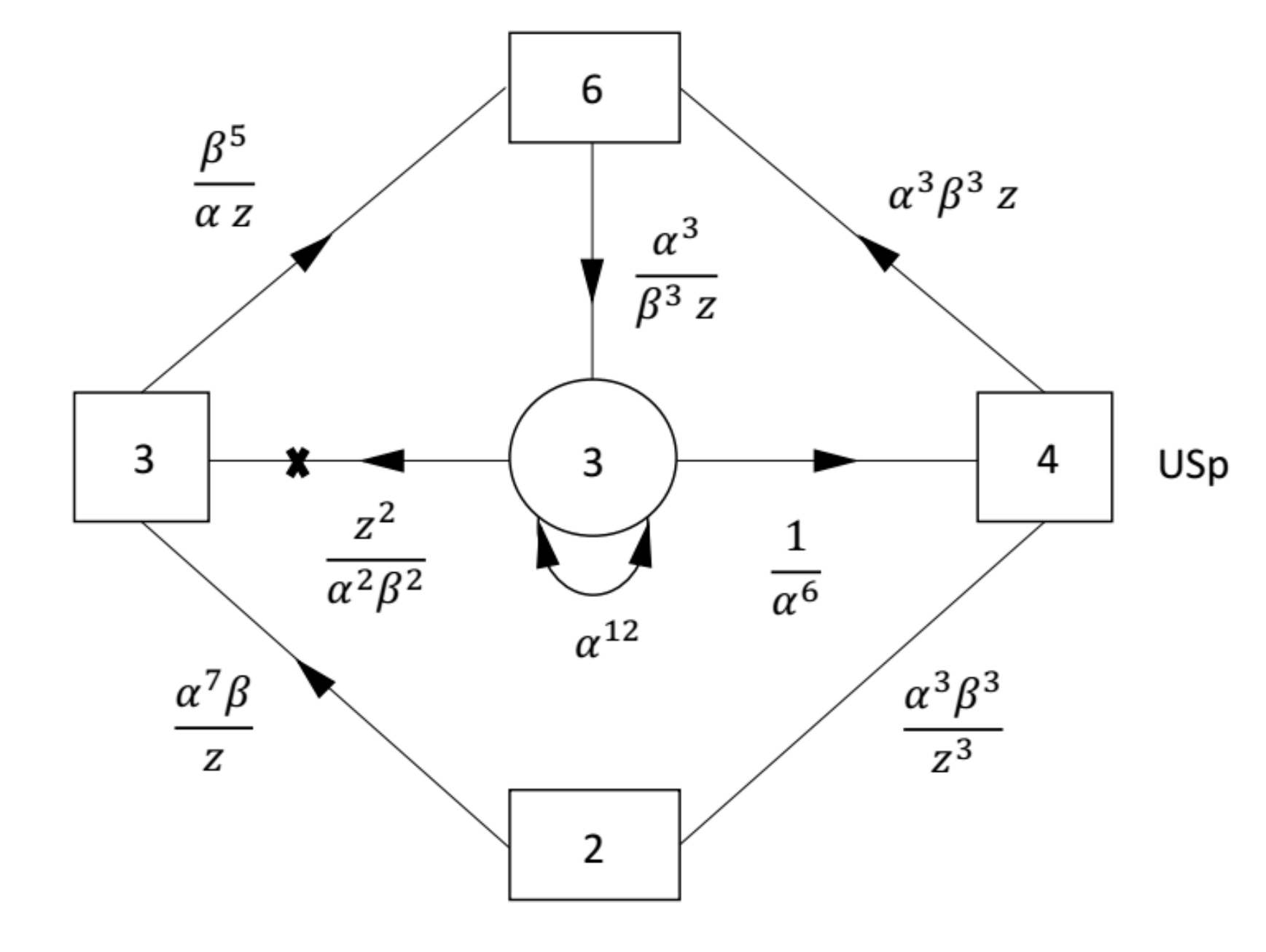} 
\caption{The tube we get by gluing the tube in figure \ref{SU2SO16Tubes} (a) with the one in figure \ref{SU2SO16Tubes} (b). As usual, we have denoted the charges under the non-R symmetries using fugacities.}
\label{SU2SO16GlTube}
\end{figure}

It is now straightforward to perform the gluing as in the previous subsection, using $\Phi$-gluing for fields with the same charges and S-gluing for fields with opposite charges. The resulting theory is shown in figure \ref{SU2SO16GlTube}, where we have redefined $\tilde{a}\rightarrow \alpha^3$ and $\tilde{b}\rightarrow \beta^3$ so as to avoid fractional powers. We conjecture that this theory is then associated with the compactification of the $SU(2)\times SO(16)$ $5d$ SCFT on a tube with flux $\frac{1}{4}(1,-1;3,3,1,1,1,1,1,1)$.

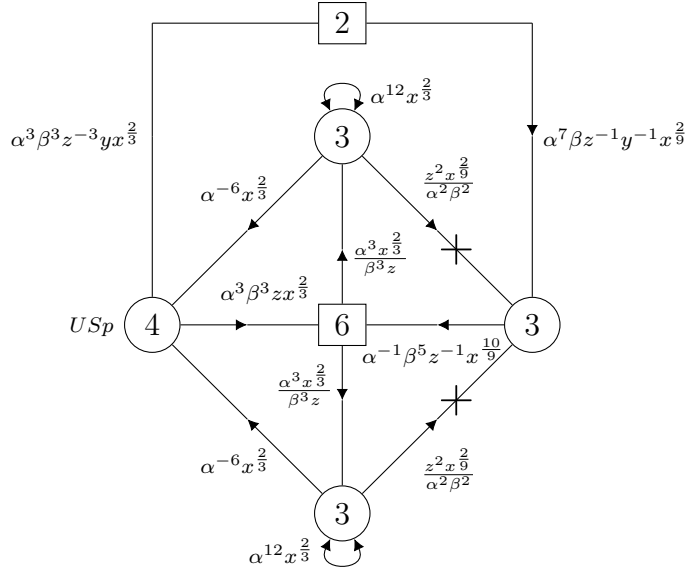
\begin{figure}
\center
\begin{tikzpicture}[baseline=0, font=\scriptsize]
\node[draw, circle] (a1) at (0,0) {\normalsize $4$};
\node[above] at (-0.8,-0.3) {$USp$};
\node[draw, circle] (a2) at (2.5,2.5) {\normalsize $ 3$};
\draw[black,solid,<->] (a2) edge [out=65,in=115,loop,looseness=4]  (a2);
\node[draw, circle] (a3) at (5,0) {\normalsize $3$};
\node[draw, circle] (a4) at (2.5,-2.5) {\normalsize $3$};
\draw[black,solid,<->] (a4) edge [out=65+180,in=115+180,loop,looseness=4]  (a4);
\node[draw, rectangle] (a5) at (2.5,0) {\normalsize $\, 6\,$};
ù\node[draw, rectangle] (a6) at (2.5,4) {\normalsize $\, 2\,$};
\draw[draw, solid] (a1)--(1.25,1.25);
\draw[draw, solid,->] (a2)--(3.75,1.25);
\draw[draw, solid] (a3)--(3.75,-1.25);
\draw[draw, solid,->] (a4)--(1.25,-1.25);
\draw[draw, solid,<-] (1.25,1.25)--(a2);
\draw[draw, solid] (3.75,1.25)--(a3);
\draw[draw, solid,<-] (3.75,-1.25)--(a4);
\draw[draw, solid] (1.25,-1.25)--(a1);
\draw[draw, solid,->] (a1)--(1.25,0);
\draw[draw, solid] (1.25,0)--(a5);
\draw[draw, solid] (a2)--(2.5,1);
\draw[draw, solid,<-] (2.5,1)--(a5);
\draw[draw, solid] (a5)--(3.75,0);
\draw[draw, solid,<-] (3.75,0)--(a3);
\draw[draw, solid,->] (a5)--(2.5,-1);
\draw[draw, solid] (2.5,-1)--(a4);

\draw[draw,solid] (a1) -- (0,2.5);
\draw[draw,solid] (0,2.5) -- (0,4);
\draw[draw,solid] (0,4) -- (a6);
\draw[draw,solid] (a6) -- (5,4);
\draw[draw,solid,->] (5,4) -- (5,2.5);
\draw[draw,solid] (5,2.5) -- (a3);

\node[thick,rotate=45] at (4,-1) {\Large$\times$};
\node[thick,rotate=45] at (4,1) {\Large$\times$};

\node[above] at (1.5,0.125) {$\ga^{3}\gb^3zx^{\frac{2}{3}}$};
\node[above] at (3,0.5) {$\frac{\ga^3x^{\frac{2}{3}}}{\gb^{3}z}$};
\node[above] at (3.7,-0.7) {$\ga^{-1}\gb^5z^{-1}x^{\frac{10}{9}}$};
\node[above] at (2,-1.25) {$\frac{\ga^3x^{\frac{2}{3}}}{\gb^{3}z}$};
\node[above] at (1.1,1.5) {$\ga^{-6}x^{\frac{2}{3}}$};
\node[above] at (3.9,1.5) {$\frac{z^2x^{\frac{2}{9}}}{\ga^{2}\gb^{2}}$};
\node[below] at (1.1,-1.5) {$\ga^{-6}x^{\frac{2}{3}}$};
\node[below] at (3.9,-1.5) {$\frac{z^2x^{\frac{2}{9}}}{\ga^{2}\gb^{2}}$};
\node[left] at (0,2.5) {$\ga^3\gb^3z^{-3}yx^{\frac{2}{3}}$};
\node[right] at (5,2.5) {$\ga^7\gb z^{-1}y^{-1}x^{\frac{2}{9}}$};
\node[above] at (2.5+0.8,2.5+0.3) {$\ga^{12}x^{\frac{2}{3}}$};
\node[above] at (2.5-0.8,-2.5-0.8) {$\ga^{12}x^{\frac{2}{3}}$};

\end{tikzpicture}
\caption{The quiver for the theory resulting from the compactification of the $5d$ $SU(2)\times SO(16)$ SCFT on a torus with flux $\frac{1}{2}(1,-1;3,3,1,1,1,1,1,1)$.}
\label{newSU2SO16torus}
\end{figure}

As a check of our construction, we can construct the torus model obtained by $\Phi$-gluing two copies of the tube in figure \ref{SU2SO16GlTube}. The resulting model is depicted in figure \ref{newSU2SO16torus}, where we also specify a possible assignment of $U(1)$ charges and R-charges that is consistent with the superpotential, which for the moment we take to involve only the perturbative matter sector of the theory and we remain agnostic on possible monopole superpotentials. Notice that the R-symmetry that we are using is related to the $5d$ R-symmetry by the shift $\ga\to \ga x^{\frac{1}{9}}$.

Let us now analyze the possible monopole superpotential. We expect that there should be one, since this should be the theory resulting from the compactification of the $5d$ $SU(2)\times SO(16)$ SCFT on a torus with flux $\frac{1}{2}(1,-1;3,3,1,1,1,1,1,1)$ which preserves a subgroup $U(1)^3\times SU(2)\times SU(6)$, while our model has an additional $U(1)$ symmetry. We observe that the $(1,1)$ monopole operators are gauge invariant, have R-charge 2 and have charge $\pm2$ under $U(1)_y$, while they are uncharged under all the other abelian symmetries. Hence, inserting them into the superpotential will break $U(1)_y$, which is precisely what we want.

Computing the index of the model with the monopole superpotential we find
\bea
\mathcal{I}&=&1+\frac{\ga^{12}}{\gb^{24}}x^{\frac{2}{3}}+\left(\frac{\ga^{24}}{\gb^{48}}+\ga^{6}\gb^6z^2{\bf 15}_{SU(6)}+2\frac{\ga^{15}z^3}{\gb^3}{\bf 2}_{SU(2)}+3\frac{\ga^6\gb^6}{z^6}+\right.\nn\\
&+&\left.2\frac{\ga^6\gb^6}{z^2}{\bf 2}_{SU(2)}{\bf 6}_{SU(6)}+3\frac{\ga^{15}}{\gb^3z}\overline{\bf 6}_{SU(6)}\right)x^{\frac{4}{3}}+\cdots\,.
\label{indnewSU2SO16torus}
\eea

We can check whether the spectrum of operators is consistent with the claim that this theory is the result of the compactification of the $5d$ $SU(2)\times SO(16)$ SCFT on a torus. Notice that the flux $\frac{1}{2}(1,-1;3,3,1,1,1,1,1,1)$ corresponds to flux $\frac{1}{2}$ for the $U(1)$ Cartan of $SU(2)$ and flux $\frac{1}{2}$ for a $U(1)$ whose commutant in $SO(16)$ is $U(1)\times SU(2)\times SU(6)$. If we look at the following decomposition of the $SU(2)$ conserved current with respect to its $U(1)$ subgroup:
\be
{\bf 3}_{SU(2)}\to 1+\frac{\ga^{12}}{\gb^{24}}+\frac{\gb^{24}}{\ga^{12}}\,,
\ee
we can immediately identify the second of these states in the index \eqref{indnewSU2SO16torus} as the contribution $\frac{\ga^{12}}{\gb^{24}}x^{\frac{1}{2}}$, which comes from the monopole with a unit of magnetic flux in the $USp(4)$ node. Observe that the multiplicity of such operator is one, which is consistent with the fact that the flux is half-integer.
Instead, the $SO(16)$ conserved current can be first decomposed under the $U(1)\times SU(8)$ subgroup and then under the $U(1)\times SU(2)\times SU(6)$ subgroup of the $SU(8)$ part
\bea
{\bf 120}_{SO(16)}&\to& q^2w^2{\bf 15}_{SU(6)}+\frac{q^2}{w^2}{\bf 2}_{SU(2)}{\bf 6}_{SU(6)}+\frac{q^2}{w^6}+w^4{\bf 2}_{SU(2)}{\bf 6}_{SU(6)}+{\bf 35}_{SU(6)}+2+\nn\\
&+&\frac{1}{w^4}{\bf 2}_{SU(2)}\overline{\bf 6}_{SU(6)}+\frac{1}{q^2w^2}\overline{\bf 15}_{SU(6)}+\frac{w^2}{q^2}{\bf 2}_{SU(2)}\overline{\bf 6}_{SU(6)}+\frac{w^6}{q^2}\,,
\eea
where the flux $\frac{1}{2}$ is inside $U(1)_w$. Again, we can immediately identify the states $w^6q^{-2}$, $q^2w^2{\bf 15}_{SU(6)}$ and $w^4{\bf 2}_{SU(2)}{\bf 6}_{SU(6)}$ with the contributions $3\ga^6\gb^6z^{-6}x^{\frac{4}{3}}$, $\ga^{6}\gb^6z^2{\bf 15}_{SU(6)}x^{\frac{4}{3}}$ and $2\ga^6\gb^6z^{-2}{\bf 2}_{SU(2)}{\bf 6}_{SU(6)}x^{\frac{4}{3}}$ in the index \eqref{indnewSU2SO16torus}, up to the map of the fugacities $q=\ga^{\frac{3}{2}}\gb^{\frac{3}{2}}z^{\frac{3}{2}}$ and $w=\ga^{\frac{3}{2}}\gb^{\frac{3}{2}}z^{-\frac{1}{2}}$. Observe that also in this case the multiplicities are consistent with the flux being $\frac{1}{2}$.

Remember that the set of the HB chiral ring generators of the $5d$ $SU(2)\times SO(16)$ SCFT, on top of the moment map for the global symmetry, contains also an operator in the $\bf 2$ of $SU(2)$, the $\bf 128$ of $SO(16)$ and the $\bf 4$ of $SU(2)_R$. We can also check the presence of some of the states coming from this operator according to the branching rule
\bea
({\bf 2}_{SU(2)},{\bf 128}_{SO(16)})&\to&\left(\frac{\ga^6}{\gb^{12}}+\frac{\gb^{12}}{\ga^6}\right)\left[\left(\frac{q^3}{w}+qw^5\right)\overline{\bf 6}_{SU(6)}+\left(q^3w^3+\frac{1}{q^3w^3}\right){\bf 2}_{SU(2)}+\right.\nn\\
&+&\left.\left(\frac{q}{w^3}+\frac{w^3}{q}\right){\bf 20}_{SU(6)}+qw{\bf 2}_{SU(2)}\overline{\bf 15}_{SU(6)}+\frac{1}{qw}{\bf 2}_{SU(2)}{\bf 15}_{SU(6)}+\right.\nn\\
&+&\left.\left(\frac{w}{q^3}+\frac{1}{qw^5}\right){\bf 6}_{SU(6)}\right]\,.\nn\\
\eea
In particular, we can see the states $\ga^6\gb^{-12}q^3w^3{\bf 2}_{SU(2)}$ and $\ga^6\gb^{-12}qw^5\overline{\bf 6}_{SU(6)}$ which contribute with the terms $2\ga^{15}\gb^{-3}z^3{\bf 2}_{SU(2)}x^{\frac{4}{3}}$ and $3\ga^{15}\gb^{-3}z^{-1}\overline{\bf 6}_{SU(6)}x^{\frac{4}{3}}$ to the index \eqref{indnewSU2SO16torus}.

\subsection{The case of the $SU(10)$ SCFT}

So far we have talked about gluing different tubes together. Here we want to consider gluing the same tube to itself but in such a way that we get a tube associated to a different flux. Specifically, we noted when discussing the $SU-SU$ tubes that the flux in these tubes is in an $SU(2)$ group that is usually embedded inside a larger group. As such, the Weyl symmetry of that larger group can be used to map that flux to an equivalent one that is embedded differently in the larger group. While the tube associated with that flux is equivalent to the original tube, gluing the two together should lead to a new tube preserving a different symmetry. This technique can then be used to generate many new tubes realizing more general fluxes. This method can be applied to most of the models we encountered here, but for simplicity we shall illustrate it for a specific example, that of the $SU(10)$ SCFT.

\begin{figure}
\center
\includegraphics[width=0.85\textwidth]{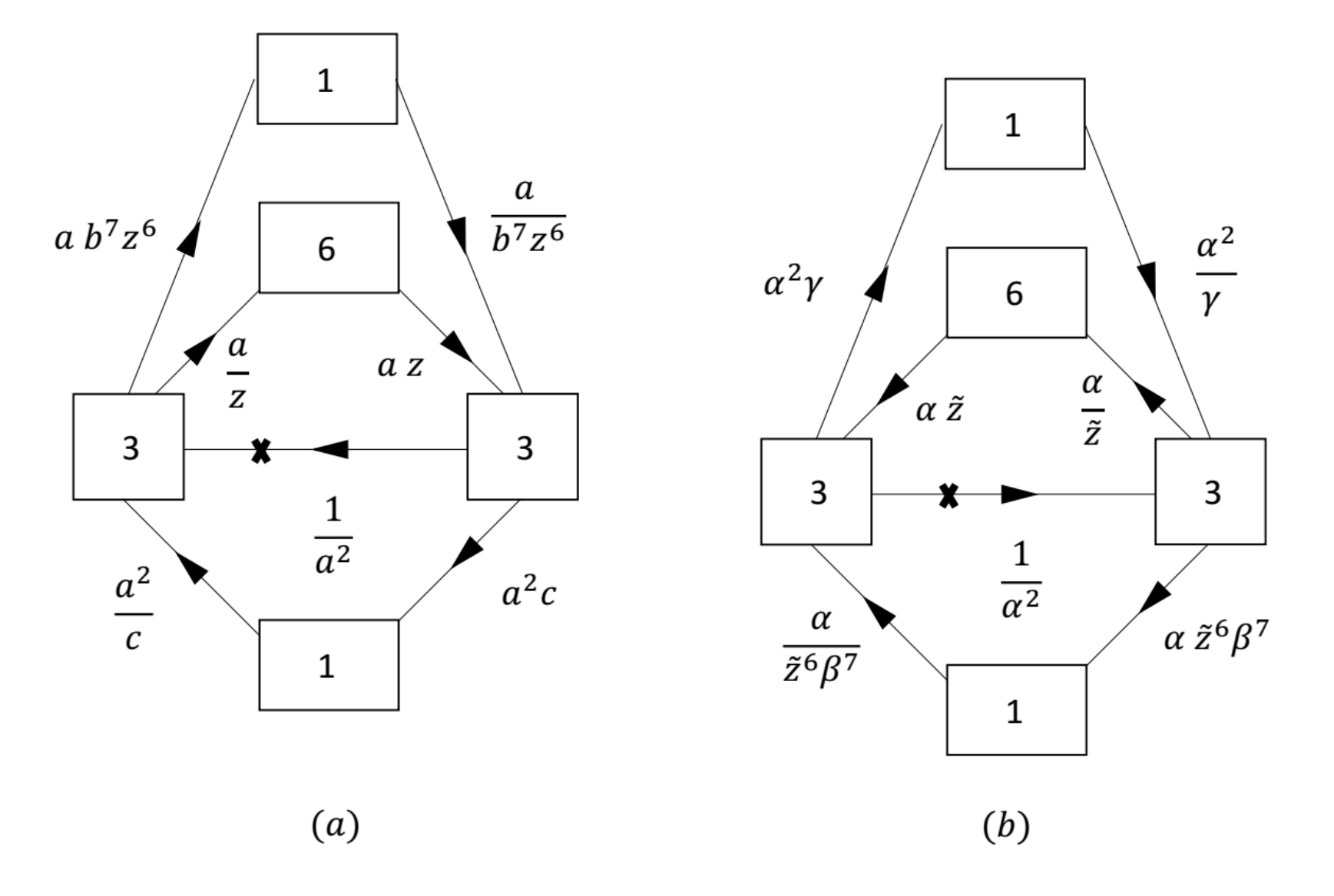} 
\caption{The theories associated with the compactification of the $SU(10)$ $5d$ SCFT on a tube. Here both (a) and (b) describe the same tube, differing only by a reflection. Charges under the global symmetries are denoted using fugacities. Here we have also split the $7$ chirals on the two sides to $6+1$, and correspondingly only the $SU(6)\times U(1)$ subgroup of the $SU(7)$ global symmetry is manifest in the quiver.}
\label{SU10Tubes}
\end{figure}

First consider the tube for the $SU(10)$ SCFT. For later convenience, we shall split the seven flavors of the tube as follows: ${\bf 7}_{SU(7)}\rightarrow \frac{1}{b z} {\bf 6}_{SU(6)} + b^6 z^6$. Furthermore, we shall take two identical tubes, shown in figure \ref{SU10Tubes}, where the tube in (b) is a mirror image of the one in (a). We have also used different fugacities for the two tubes to allow for the option of a different mapping of the symmetries between the two tubes. Recall that the $3d$ global symmetry is embedded in the $5d$ one as

\be
{\bf 10}_{SU(10)} \rightarrow \frac{1}{z c^{\frac{1}{5}} b^{\frac{7}{5}}} {\bf 6}_{SU(6)} + \frac{c^{\frac{4}{5}}}{b^{\frac{7}{5}}}( a^3 + \frac{1}{a^3}) + \frac{z^3 b^{\frac{28}{5}} }{c^{\frac{1}{5}}}( z^3 + \frac{1}{z^3}) ,
\ee
for the case in figure (a), and similarly

\be\label{SU10funddecomp}
{\bf 10}_{SU(10)} \rightarrow \frac{1}{\tilde{z} \gamma^{\frac{1}{5}} \beta^{\frac{7}{5}}} {\bf 6}_{SU(6)} + \frac{\gamma^{\frac{4}{5}}}{\beta^{\frac{7}{5}}}( \alpha^3 + \frac{1}{\alpha^3}) + \frac{\tilde{z}^3 \beta^{\frac{28}{5}} }{\gamma^{\frac{1}{5}}}( \tilde{z}^3 + \frac{1}{\tilde{z}^3}) ,
\ee
for the case in figure (b). 

From the symmetry structure we note that we can identify the two symmetries in various ways. The simplest way is to take $a\rightarrow \alpha$, $b\rightarrow \beta$, $c\rightarrow \gamma$ and $z\rightarrow \tilde{z}$. This just amounts to taking two identical copies of the tube, as can be seen by examining the charges of the fields. However, we can also map the symmetries using $a\rightarrow \tilde{z}$, $b^7\rightarrow \frac{\tilde{z} \gamma}{\alpha^4}$, $c\rightarrow \frac{\beta^7 \tilde{z}^4}{\alpha}$ and $z\rightarrow \alpha$. While this is an equivalent embedding, the resulting tubes differ by how the flux is embedded. Specifically, in the tube in figure \ref{SU10Tubes} (b) the flux is embedded in the $SU(2)$ spanned by $\alpha^3 + \frac{1}{\alpha^3}$, while for the one in figure \ref{SU10Tubes} (a) the flux is embedded in an $SU(2)$ that using that mapping is spanned by $\tilde{z}^3 + \frac{1}{\tilde{z}^3}$. As such the fluxes in the two tubes are equivalent, but embedded differently in the $SU(10)$ global symmetry. We previously noted that the flux associated with the tube in figure \ref{SU10Tubes} (a) should be $\frac{1}{2}(1,-1,0,0,0,0,0,0,0,0)$. As such the flux associated with the tube in figure \ref{SU10Tubes} (b) should be $\frac{1}{2}(0,0,1,-1,0,0,0,0,0,0)$.

\begin{figure}
\center
\includegraphics[width=0.55\textwidth]{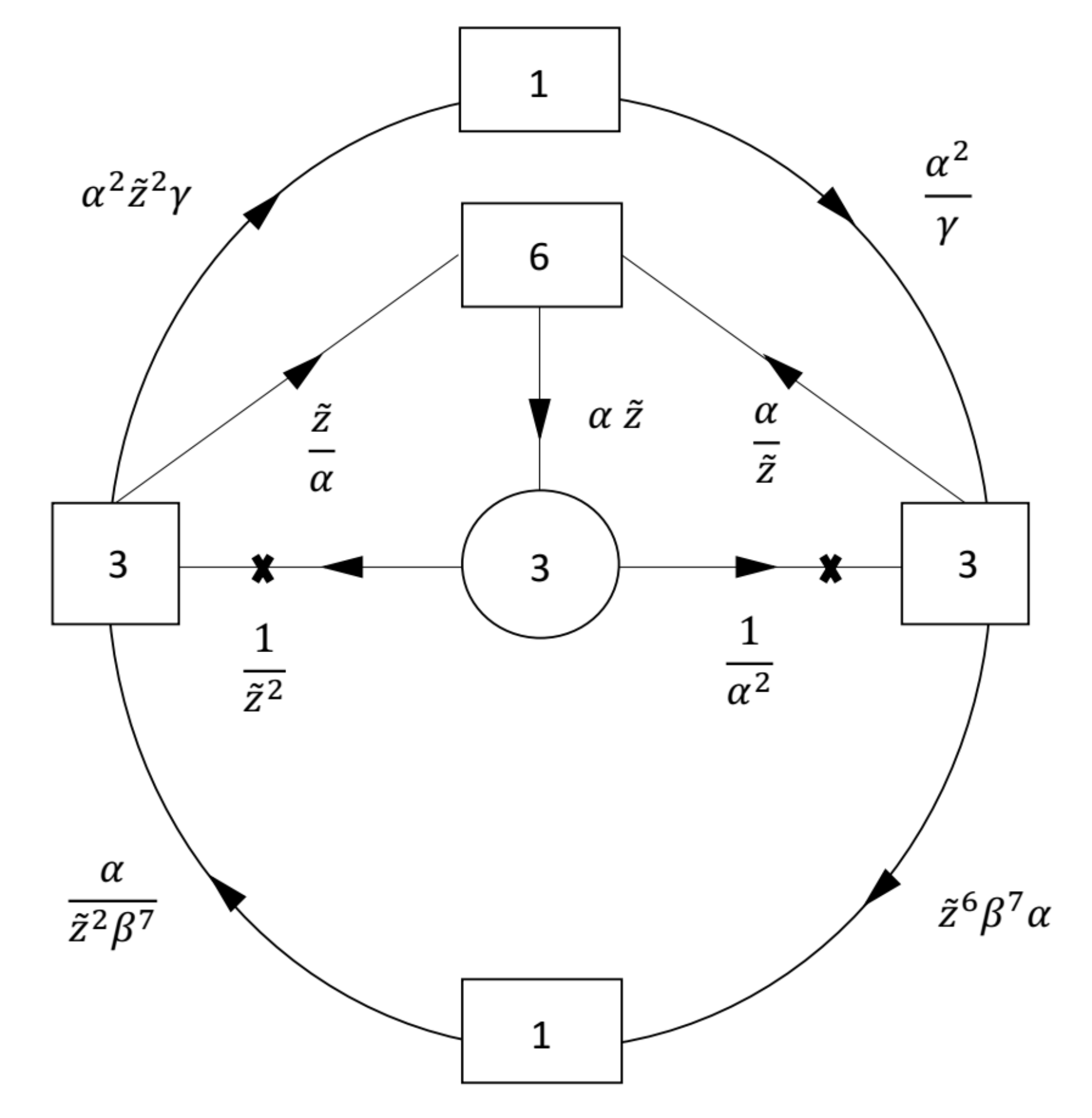} 
\caption{The tube we get by gluing the tube in figure \ref{SU10Tubes} (a) with the one in figure \ref{SU10Tubes} (b). As usual, we have denoted the charges under the non-R symmetries using fugacities.}
\label{SU10GenT}
\end{figure}

Next we can consider gluing the two tubes together. First we can note that as in the previous cases, some of the fields we need to glue carry the same charges, while others carry opposite charges. Specifically, the $SU(3)\times SU(6)$ bifundamental carry the same charges, while the two $SU(3)$ fundamentals carry opposite charges. As such the first should be glued by $\Phi$ gluing while the second should be glued using S gluing. Performing the gluing as we did previously, we get the tube in figure \ref{SU10GenT}. We conjecture that this theory is then associated with the compactification of the $SU(10)$ $5d$ SCFT on a tube with flux $\frac{1}{2}(1,-1,1,-1,0,0,0,0,0,0)$.

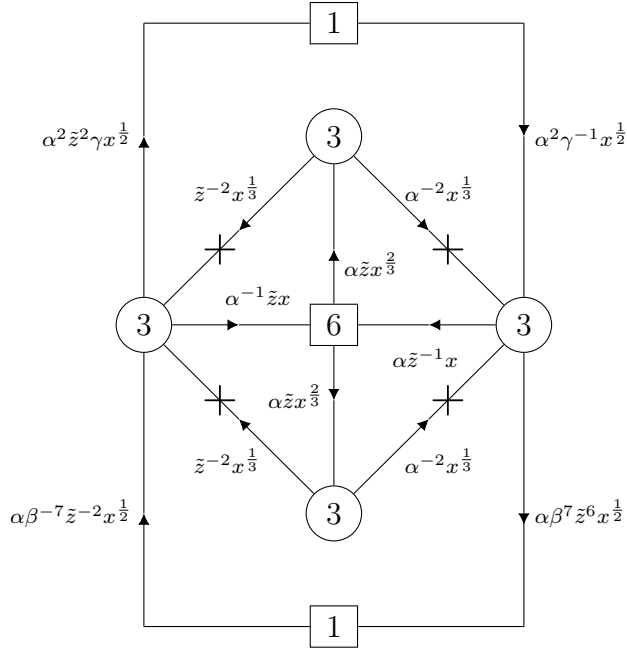
\begin{figure}
\center
\begin{tikzpicture}[baseline=0, font=\scriptsize]
\node[draw, circle] (a1) at (0,0) {\normalsize $ 3$};
\node[draw, circle] (a2) at (2.5,2.5) {\normalsize $ 3$};
\node[draw, circle] (a3) at (5,0) {\normalsize $3$};
\node[draw, circle] (a4) at (2.5,-2.5) {\normalsize $3$};
\node[draw, rectangle] (a5) at (2.5,0) {\normalsize $\, 6\,$};
\node[draw, rectangle] (a6) at (2.5,4) {\normalsize $\, 1\,$};
\node[draw, rectangle] (a7) at (2.5,-4) {\normalsize $\, 1\,$};
\draw[draw, solid] (a1)--(1.25,1.25);
\draw[draw, solid,->] (a2)--(3.75,1.25);
\draw[draw, solid] (a3)--(3.75,-1.25);
\draw[draw, solid,->] (a4)--(1.25,-1.25);
\draw[draw, solid,<-] (1.25,1.25)--(a2);
\draw[draw, solid] (3.75,1.25)--(a3);
\draw[draw, solid,<-] (3.75,-1.25)--(a4);
\draw[draw, solid] (1.25,-1.25)--(a1);
\draw[draw, solid,->] (a1)--(1.25,0);
\draw[draw, solid] (1.25,0)--(a5);
\draw[draw, solid] (a2)--(2.5,1);
\draw[draw, solid,<-] (2.5,1)--(a5);
\draw[draw, solid] (a5)--(3.75,0);
\draw[draw, solid,<-] (3.75,0)--(a3);
\draw[draw, solid,->] (a5)--(2.5,-1);
\draw[draw, solid] (2.5,-1)--(a4);

\draw[draw,solid,->] (a1) -- (0,2.5);
\draw[draw,solid] (0,2.5) -- (0,4);
\draw[draw,solid] (0,4) -- (a6);
\draw[draw,solid] (a6) -- (5,4);
\draw[draw,solid,->] (5,4) -- (5,2.5);
\draw[draw,solid] (5,2.5) -- (a3);

\draw[draw,solid] (a1) -- (0,-2.5);
\draw[draw,solid,<-] (0,-2.5) -- (0,-4);
\draw[draw,solid] (0,-4) -- (a7);
\draw[draw,solid] (a7) -- (5,-4);
\draw[draw,solid,-<] (5,-4) -- (5,-2.5);
\draw[draw,solid] (5,-2.5) -- (a3);

\node[thick,rotate=45] at (1,1) {\Large$\times$};
\node[thick,rotate=45] at (4,1) {\Large$\times$};
\node[thick,rotate=45] at (4,-1) {\Large$\times$};
\node[thick,rotate=45] at (1,-1) {\Large$\times$};

\node[above] at (1.5,0.125) {$\ga^{-1}\tilde{z}x$};
\node[above] at (3,0.5) {$\ga\tilde{z}x^{\frac{2}{3}}$};
\node[above] at (3.7,-0.7) {$\ga\tilde{z}^{-1}x$};
\node[above] at (2,-1.25) {$\ga\tilde{z}x^{\frac{2}{3}}$};
\node[above] at (1.1,1.5) {$\tilde{z}^{-2}x^{\frac{1}{3}}$};
\node[above] at (3.9,1.5) {$\ga^{-2}x^{\frac{1}{3}}$};
\node[below] at (1.1,-1.5) {$\tilde{z}^{-2}x^{\frac{1}{3}}$};
\node[below] at (3.9,-1.5) {$\ga^{-2}x^{\frac{1}{3}}$};
\node[left] at (0,2.5) {$\ga^2\tilde{z}^2\gc x^{\frac{1}{2}}$};
\node[right] at (5,2.5) {$\ga^2\gc^{-1}x^{\frac{1}{2}}$};
\node[left] at (0,-2.5) {$\ga\gb^{-7}\tilde{z}^{-2}x^{\frac{1}{2}}$};
\node[right] at (5,-2.5) {$\ga\gb^7\tilde{z}^6x^{\frac{1}{2}}$};
\end{tikzpicture}
\caption{The quiver for the theory resulting from the compactification of the $5d$ $SU(10)$ SCFT on a torus with flux $(1,-1,-1,-1,0,.\cdots,0)$.}
\label{newSU10torus}
\end{figure}

As a check of our construction, we can construct the torus model obtained by $\Phi$-gluing two copies of the tube in figure \ref{SU10GenT}. The resulting model is depicted in figure \ref{newSU10torus}, where we also specify a possible assignment of $U(1)$ charges and R-charges that is consistent with the superpotential. On top of the standard superpotential terms involving the matter fields we also have to consider a monopole superpotential, as standard in all of our models with more than 2 gauge nodes. Interestingly, in this case we find that the monopole superpotential we need to turn on is not the usual $(1,1)$ monopoles with minimal charges under adjacent groups, but rather consists of the basic monopoles of the upper and lower $SU(3)$ gauge nodes, which should be turned on linearly in the superpotential. This follows as one can check that using the charge assignments of figure \ref{newSU10torus} that we obtained from the gluing, such monopoles have R-charge 2 and are uncharged under all the global symmetries. The $(1,1)$ monopoles, properly dressed with one copy of the bifundamental to make them gauge invariant, are charged under a combination of the various $U(1)$ groups so introducing them would actually break some of the $5d$ symmetries. As such it seems that in this case the correct monopole superpotential involves the basic monopoles of the upper and lower $SU(3)$ gauge nodes, rather than the $(1,1)$ type monopoles. It would be interesting to find some general rule governing what type of monopoles is needed to be turned on, though we shall not pursue this here.

Notice also that the R-symmetry that we are using is related to the $5d$ R-symmetry by the shifts $\ga\to \ga x^{\frac{1}{6}}$, $\tilde{z}\to\tilde{z}x^{\frac{1}{6}}$, $\beta\to\beta x^{-\frac{2}{21}}$ and $\gamma\to\gamma x^{-\frac{1}{6}}$. The reason for this strange choice of mixing coefficients of the abelian symmetries with the R-symmetry is because in this way we have no mixing for the two combination of the $U(1)$'s that are expected to get enhanced to $SU(2)$, which is essential in order to correctly see $SU(2)$ characters in the index. Indeed, we can rewrite the decomposition \eqref{SU10funddecomp} as
\be\label{SU10funddecompnew}
{\bf 10}_{SU(10)} \rightarrow \frac{1}{\tilde{z} \gamma^{\frac{1}{5}} \beta^{\frac{7}{5}}} {\bf 6}_{SU(6)} + \alpha^{\frac{3}{2}}\tilde{z}^{3}\beta^{\frac{21}{10}}\gamma^{\frac{3}{10}}\left(\frac{\alpha^{\frac{3}{2}}\gamma^{\frac{1}{2}}}{\tilde{z}^3\beta^{\frac{7}{2}}}+\frac{\tilde{z}^3\beta^{\frac{7}{2}}}{\alpha^{\frac{3}{2}}\gamma^{\frac{1}{2}}}\right)+\frac{\beta^{\frac{21}{10}}\gamma^{\frac{3}{10}}}{\alpha^{\frac{3}{2}}}\left(\frac{\gamma^{\frac{1}{2}}}{\alpha^{\frac{3}{2}}\beta^{\frac{7}{2}}}+\frac{\alpha^{\frac{3}{2}}\beta^{\frac{7}{2}}}{\gamma^{\frac{1}{2}}}\right) \,,
\ee
from which we observe that we can reconstruct characters of $SU(2)_1\times SU(2)_2$ where the embedding is such that
\be\label{embeddingnewSU10}
{\bf 2}_{SU(2)_1}=\frac{\alpha^{\frac{3}{2}}\gamma^{\frac{1}{2}}}{\tilde{z}^3\beta^{\frac{7}{2}}}+\frac{\tilde{z}^3\beta^{\frac{7}{2}}}{\alpha^{\frac{3}{2}}\gamma^{\frac{1}{2}}},\qquad {\bf 2}_{SU(2)_2}=\frac{\gamma^{\frac{1}{2}}}{\alpha^{\frac{3}{2}}\beta^{\frac{7}{2}}}+\frac{\alpha^{\frac{3}{2}}\beta^{\frac{7}{2}}}{\gamma^{\frac{1}{2}}}\,.
\ee
Moreover, we can check that with our choice of the R-symmetry there is no mixing of it with the $U(1)^2$ Cartan of $SU(2)_1\times SU(2)_2$.

Computing the index of the model we find
\bea
\mathcal{I}&=&1+2\ga^3\tilde{z}^3\left(\frac{\alpha^{\frac{3}{2}}\gamma^{\frac{1}{2}}}{\tilde{z}^3\beta^{\frac{7}{2}}}+\frac{\tilde{z}^3\beta^{\frac{7}{2}}}{\alpha^{\frac{3}{2}}\gamma^{\frac{1}{2}}}\right)\left(\frac{\gamma^{\frac{1}{2}}}{\alpha^{\frac{3}{2}}\beta^{\frac{7}{2}}}+\frac{\alpha^{\frac{3}{2}}\beta^{\frac{7}{2}}}{\gamma^{\frac{1}{2}}}\right)x+\nn\\
&+&\left[\ga^{\frac{3}{2}}\tilde{z}^4\gb^{\frac{7}{2}}\gc^{\frac{1}{2}}\left(\frac{\alpha^{\frac{3}{2}}\gamma^{\frac{1}{2}}}{\tilde{z}^3\beta^{\frac{7}{2}}}+\frac{\tilde{z}^3\beta^{\frac{7}{2}}}{\alpha^{\frac{3}{2}}\gamma^{\frac{1}{2}}}\right)\overline{\bf 6}+\frac{\ga^{\frac{3}{2}}}{\tilde{z}\gb^{\frac{7}{2}}\gc^{\frac{1}{2}}}\left(\frac{\gamma^{\frac{1}{2}}}{\alpha^{\frac{3}{2}}\beta^{\frac{7}{2}}}+\frac{\alpha^{\frac{3}{2}}\beta^{\frac{7}{2}}}{\gamma^{\frac{1}{2}}}\right){\bf 6}\right]x^{\frac{3}{2}}+\nn\\
&+&\left[\frac{5}{\ga^6\tilde{z}^6}+\ga^6\tilde{z}^6\left(3\left(\left(\frac{\ga^3\gc}{\tilde{z}^6\gb^7}+\frac{\tilde{z}^6\gb^7}{\ga^3\gc}\right)\left(\frac{\ga^3\gb^7}{\gc}+\frac{\gc}{\ga^3\gb^7}\right)+1\right)+\right.\right.\nn\\
&+&\left.\left.2+\frac{\ga^3\gc}{\tilde{z}^6\gb^7}+\frac{\tilde{z}^6\gb^7}{\ga^3\gc}+\frac{\ga^3\gb^7}{\gc}+\frac{\gc}{\ga^3\gb^7}\right)+2\ga^3\tilde{z}^3{\bf 20}\right]x^2+\cdots\,.
\eea
We can notice that the index forms characters of $SU(2)_1\times SU(2)_2$ according to the embedding \eqref{embeddingnewSU10}, as expected
\bea
\mathcal{I}&=&1+2\ga^3\tilde{z}^3{\bf 2}_{SU(2)_1}{\bf 2}_{SU(2)_2}x+\left(\ga^{\frac{3}{2}}\tilde{z}^4\gb^{\frac{7}{2}}\gc^{\frac{1}{2}}{\bf 2}_{SU(2)_1}\overline{\bf 6}_{SU(6)}+\frac{\ga^{\frac{3}{2}}}{\tilde{z}\gb^{\frac{7}{2}}\gc^{\frac{1}{2}}}{\bf 2}_{SU(2)_2}{\bf 6}_{SU(6)}\right)x^{\frac{3}{2}}+\nn\\
&+&\left(\frac{5}{\ga^6\tilde{z}^6}+\ga^6\tilde{z}^6\left(3\left({\bf 3}_{SU(2)_1}{\bf 3}_{SU(2)_2}+1\right)+{\bf 3}_{SU(2)_1}+{\bf 3}_{SU(2)_2}\right)+2\ga^3\tilde{z}^3{\bf 20}_{SU(6)}\right)x^2+\cdots\,.\nn\\
\label{indnewSU10torus}
\eea
This is compatible with the claim that this is the theory associated to the compactification of the $5d$ $SU(10)$ SCFT on a torus of flux $(1,-,1-1,-1,0,.\cdots,0)$, which preserves the subgroup $U(1)^2\times SU(2)^2\times SU(6)$. Equivalently, this flux can be understood as a unit of flux for two $U(1)$'s inside $SU(10)$ which, in the parameterization of \eqref{SU10funddecomp}, correspond to $U(1)_\ga$ and $U(1)_{\tilde{z}}$.

We can check whether the spectrum of operators is also consistent with this claim. If we look at the following decomposition of the $SU(10)$ conserved current under the same subgroup under which we decomposed the $\bf 10$ in \eqref{SU10funddecomp}:
\bea
{\bf 99}_{SU(10)}&\to& {\bf 35}_{SU(6)}+2+{\bf 3}_{SU(2)_1}+{\bf 3}_{SU(2)_2}+\left(\ga^3\tilde{z}^3+\frac{1}{\ga^3\tilde{z}^3}\right){\bf 2}_{SU(2)_1}{\bf 2}_{SU(2)_2}+\nn\\
&+&\frac{1}{\tilde{z}^{\frac{5}{2}}\gc^{\frac{1}{2}}\gb^{\frac{7}{2}}}\left(\frac{1}{\ga^{\frac{3}{2}}\tilde{z}^{\frac{3}{2}}}{\bf 2}_{SU(2)_1}+\ga^{\frac{3}{2}}\tilde{z}^{\frac{3}{2}}{\bf 2}_{SU(2)_2}\right){\bf 6}_{SU(6)}+\nn\\
&+&\tilde{z}^{\frac{5}{2}}\gc^{\frac{1}{2}}\gb^{\frac{7}{2}}\left(\ga^{\frac{3}{2}}\tilde{z}^{\frac{3}{2}}{\bf 2}_{SU(2)_1}+\frac{1}{\ga^{\frac{3}{2}}\tilde{z}^{\frac{3}{2}}}{\bf 2}_{SU(2)_2}\right)\overline{\bf 6}_{SU(6)}\, .
\eea

We can immediately identify some of these states in the index \eqref{indnewSU10torus}. Specifically, the states $\ga^3\tilde{z}^3{\bf 2}_{SU(2)_1}{\bf 2}_{SU(2)_2}$, $\ga^{\frac{3}{2}}\tilde{z}^{4}\gc^{\frac{1}{2}}\gb^{\frac{7}{2}}{\bf 2}_{SU(2)_1}\overline{\bf 6}_{SU(6)}$ and $\frac{\ga^{\frac{3}{2}}}{\tilde{z}\gc^{\frac{1}{2}}\gb^{\frac{7}{2}}}{\bf 2}_{SU(2)_2}{\bf 6}_{SU(6)}$ give the contributions $2\ga^3\tilde{z}^3{\bf 2}_{SU(2)_1}{\bf 2}_{SU(2)_2}x$, $\ga^{\frac{3}{2}}\tilde{z}^4\gb^{\frac{7}{2}}\gc^{\frac{1}{2}}{\bf 2}_{SU(2)_1}\overline{\bf 6}_{SU(6)}x^{\frac{3}{2}}$ and $\frac{\ga^{\frac{3}{2}}}{\tilde{z}\gb^{\frac{7}{2}}\gc^{\frac{1}{2}}}{\bf 2}_{SU(2)_2}{\bf 6}_{SU(6)}x^{\frac{3}{2}}$, respectively. Similarly, the contribution $2 \ga^3\tilde{z}^3 {\bf 20}_{SU(6)} x^2$ can be identified as coming from the Higgs branch chiral ring generator in the ${\bf 252}$ of $SU(10)$, specifically, the state $\ga^3\tilde{z}^3 {\bf 20}_{SU(6)}$ in the decomposition.

\subsection{$SU(2)\times SO(16)$ SCFT revisited}

We have previously discussed how the $SU-USp$ and $SU-SU$ tubes can be glued together to form new tubes. Here we will be again interested in these tubes, but for a different purpose. Specifically, we would like here to glue $SU-USp$ tubes such that the resulting tube would be an $SU-SU$ tube with the same flux as the basic $SU-SU$ tube of the same theory discussed in subsection \ref{susu62}.

\begin{figure}
\center
\includegraphics[width=1.0\textwidth]{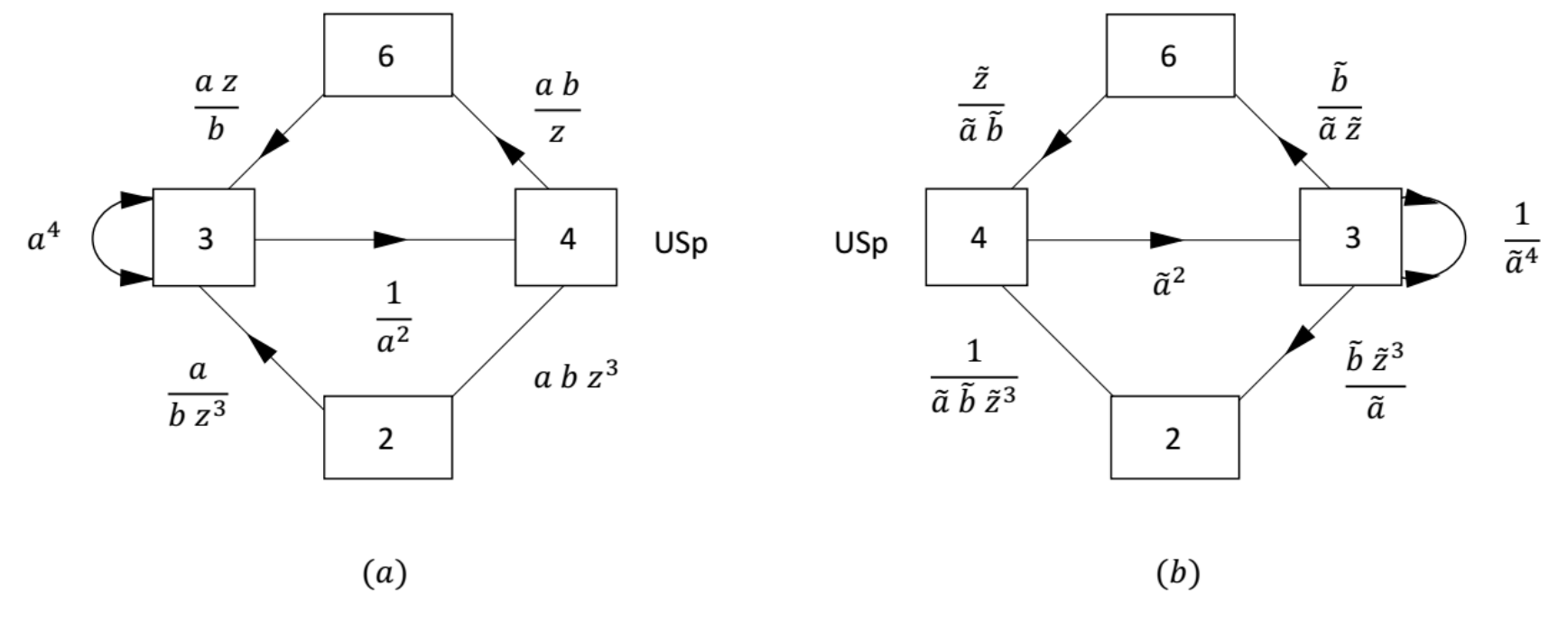} 
\caption{The theories associated with the compactification of the $SU(2)\times SO(16)$ $5d$ SCFT on a tube. Here both (a) and (b) describe the same tube, differing by charge conjugation. Charges under the global symmetries are denoted using fugacities. Here we have also split the $8$ chirals on the two sides to $6+2$, and correspondingly only the $SU(6)\times SU(2) \times U(1)$ subgroup of the $SU(8)$ global symmetry is manifest in the quiver.}
\label{SU2SO16TT}
\end{figure}

The analysis proceeds in a similar fashion as the previous cases. We begin by presenting the two $SU-USp$ tubes that we intend to glue to one another, which are shown in figure \ref{SU2SO16TT}. Here both tubes are just the basic $SU-USp$ tube that we presented in section \ref{SUUSp}, though we have separated the eight flavors to a group of six and two according to the breaking ${\bf 8}\rightarrow \frac{1}{z}{\bf 6} + z^3 {\bf 2}$. Additionally, for the tube in (b) we have taken the charge conjugate of the basic tube. The charge conjugation inverts the flux and also changes the signs of the two punctures as it inverts the charges of the fields. As such, the tube in (a) has flux $\frac{1}{4}(1,-1;1,1,1,1,1,1,1,1)$, while that in (b) has flux $\frac{1}{4}(-1,1;-1,-1,-1,-1,-1,-1,-1,-1)$.

We next want to act on the tube in (b) by the Weyl action acting on the two bottom flavors. To understand its action, we first remind the reader of the relation between the symmetries of the $3d$ tubes and those of the $5d$ SCFT:

\bea
& & {\bf 2}_{SU(2)_{5d}} \rightarrow \frac{a^2}{b^4} + \frac{b^4}{a^2} , \\ \nonumber
& & {\bf 16}_{SO(16)} \rightarrow\frac{a b}{z} {\bf 6}_{SU(6)} + a b z^3 {\bf 2}_{SU(2)_{3d}} + \frac{z}{a b} \overline{\bf 6}_{SU(6)} + \frac{1}{a b z^3} {\bf 2}_{SU(2)_{3d}} .
\eea

The operation that inverts the symmetries of the two flavors acted upon by $SU(2)_{3d}$ then acts as $\frac{a^2}{b^4}\rightarrow \frac{a^2}{b^4}$, $\frac{a b}{z}\rightarrow \frac{a b}{z}$, $a b z^3 \rightarrow \frac{1}{a b z^3}$. It is then straightforward to see that if we want to relate the two tubes by the action of this Weyl element then we need to take $\tilde{a}\rightarrow \frac{a^{\frac{2}{3}}}{z b^{\frac{1}{3}}}$, $\tilde{b}\rightarrow \frac{b^{\frac{5}{6}}}{z^{\frac{1}{2}} a^{\frac{1}{6}}}$ and $\tilde{z}\rightarrow \frac{a^{\frac{2}{3}}}{z b^{\frac{1}{3}}}$.

\begin{figure}
\center
\includegraphics[width=0.8\textwidth]{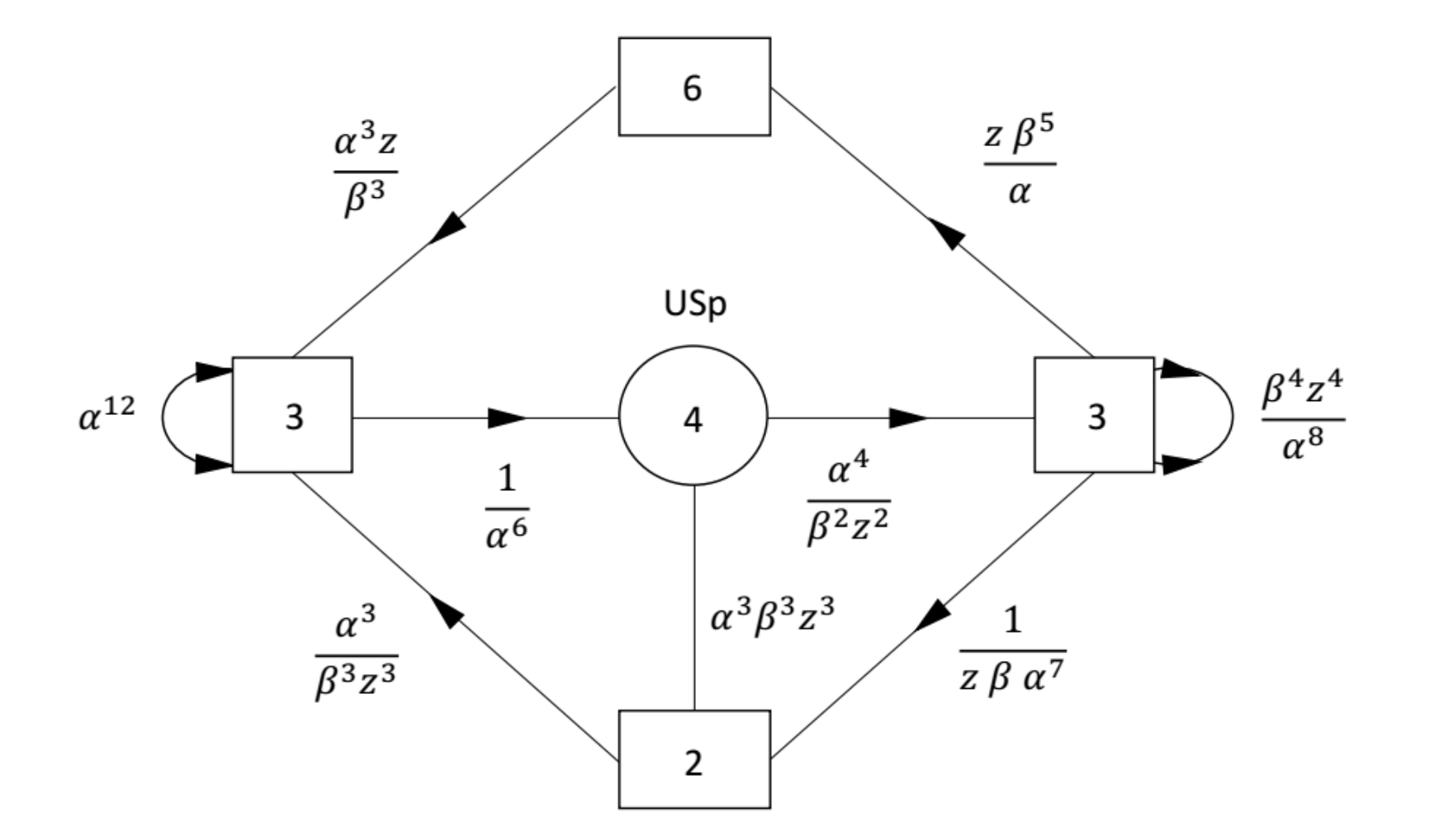} 
\caption{The tube we get by gluing the tube in figure \ref{SU2SO16TT} (a) with the one in figure \ref{SU2SO16TT} (b). As usual, we have denoted the charges under the non-R symmetries using fugacities. The flux of this tube is equivalent to the one used to construct the $(6,2)$ split model of section \ref{susu62} and in fact the two tube models are actually IR duals.}
\label{newtube}
\end{figure}

We can now glue the two tubes together. Using the relations of the symmetries that we found, one can see that the $USp(4)\times SU(6)$ bifundamentals of the two tubes carry opposite charges, while the $USp(4)\times SU(2)$ bifundamentals carry the same charges. As such the latter are glued by $\Phi$ gluing while the former are glued by S gluing. We end up with the theory in figure \ref{newtube}, where to avoid the square roots, we have defined $\alpha=a^{\frac{1}{3}}$, $\beta=b^{\frac{1}{3}}$.

The superpotential of the resulting tube model contains, on top of the standard terms involving the matter fields, also some monopole terms. Specifically, one can check that the assignment of charges given in figure \ref{newtube} that we obtained from the gluing is compatible with a monopole superpotential consisting of the basic $USp(4)$ monopoles turned on linearly. We also stress that such monopole superpotential is responsible for breaking a $U(1)$ global symmetry and leaving us only with the three abelian symmetries whose charge assignments are given in figure \ref{newtube}, in terms of the fugacities $\ga$, $\gb$, $z$. This turns out to be crucial to match the $5d$ predictions.

As we mentioned the flux associated with the tube in figure \ref{SU2SO16TT} (a) is $\frac{1}{4}(1,-1;1,1,1,1,1,1,1,1)$, while for the tube in figure \ref{SU2SO16TT} (b) we associate the flux $\frac{1}{4}(-1,1;1,1,-1,-1,-1,-1,-1,-1)$. Here two of the fluxes in the $SO(16)$ part have been inverted due to the action of the Weyl group. Summing these up we get the flux $\frac{1}{2}(0,0;1,1,0,0,0,0,0,0)$, which is the one we associate with the tube in figure \ref{newtube}. We note that this flux is the same as that of the basic $SU-SU$ tube associated with the same $5d$ SCFT, which led us to the $(6,2)$ split model of subsection \ref{susu62}. We thus expect to obtain IR dual theories when we use these tubes to construct closed surfaces.

\begin{figure}
	\center
\begin{tikzpicture}[baseline=0, font=\scriptsize]
\node[draw, circle] (a1) at (0,0) {\normalsize $3$};
\draw[black,solid,<->] (a1) edge [out=100,in=150,loop,looseness=4]  (a1);
\draw[black,solid,<->] (a1) edge [out=360-150,in=360-100,loop,looseness=4]  (a1);
\node[draw, circle] (a2) at (2.5,2.5) {\normalsize $ 4$};
\node[above] at (2.5,2.8) {$USp$};
\node[draw, circle] (a3) at (5,0) {\normalsize $3$};
\path[black,solid,-](a3)  edge [out=180-150,in=180-145,loop,looseness=4] (5.35,0.2) ;
\path[black,solid,>-<] (5.35,0.2) edge [out=180-145,in=180-105,loop,looseness=3] (5.1,0.4) ;
\path[black,solid,-] (5.1,0.4) edge [out=180-105,in=180-100,loop,looseness=4] (a3) ;
\path[black,solid,-](a3)  edge [out=-180+150,in=-180+145,loop,looseness=4] (5.35,-0.2) ;
\path[black,solid,>-<] (5.35,-0.2) edge [out=-180+145,in=-180+105,loop,looseness=3] (5.1,-0.4) ;
\path[black,solid,-] (5.1,-0.4) edge [out=-180+105,in=-180+100,loop,looseness=4] (a3) ;
\node[draw, circle] (a4) at (2.5,-2.5) {\normalsize $4$};
\node[below] at (2.5,-2.8) {$USp$};
\node[draw, rectangle] (a5) at (2.5,0) {\normalsize $\, 2\,$};
\node[draw, rectangle] (a6) at (2.5,4) {\normalsize $\, 6\,$};
\draw[draw, solid,->] (a1)--(1.25,1.25);
\draw[draw, solid,->] (a2)--(3.75,1.25);
\draw[draw, solid] (a3)--(3.75,-1.25);
\draw[draw, solid] (a4)--(1.25,-1.25);
\draw[draw, solid] (1.25,1.25)--(a2);
\draw[draw, solid] (3.75,1.25)--(a3);
\draw[draw, solid,<-] (3.75,-1.25)--(a4);
\draw[draw, solid,<-] (1.25,-1.25)--(a1);
\draw[draw, solid] (a1)--(1.25,0);
\draw[draw, solid,<-] (1.25,0)--(a5);
\draw[draw, solid] (a2)--(2.5,1);
\draw[draw, solid] (2.5,1)--(a5);
\draw[draw, solid] (a5)--(3.75,0);
\draw[draw, solid,<-] (3.75,0)--(a3);
\draw[draw, solid] (a5)--(2.5,-1);
\draw[draw, solid] (2.5,-1)--(a4);

\draw[draw,solid] (a1) -- (0,2.5);
\draw[draw,solid,<-] (0,2.5) -- (0,4);
\draw[draw,solid] (0,4) -- (a6);
\draw[draw,solid] (a6) -- (5,4);
\draw[draw,solid] (5,4) -- (5,2.5);
\draw[draw,solid,<-] (5,2.5) -- (a3);


\node[above] at (1.5,0.05) {$\frac{\ga^{3}}{\gb^3z^{3}}x^{\frac{4}{3}}$};
\node[above] at (3.25,0.5) {$\ga^3\gb^3z^3x^{\frac{1}{2}}$};
\node[above] at (3.7,-0.8) {$\frac{x^{\frac{4}{3}}}{\ga^7\gb z}$};
\node[above] at (1.85,-1.25) {$\ga^3\gb^{3}z^3x^{\frac{1}{2}}$};
\node[above] at (1.1,1.5) {$\ga^{-6}x^{\frac{1}{6}}$};
\node[above] at (3.9,1.5) {$\frac{\ga^4 x^{\frac{1}{6}}}{\gb^2z^2}$};
\node[below] at (1.1,-1.5) {$\ga^{-6}x^{\frac{1}{6}}$};
\node[below] at (3.9,-1.5) {$\frac{\ga^4 x^{\frac{1}{6}}}{\gb^2z^2}$};
\node[left] at (0,2.5) {$\frac{\ga^3zx^{\frac{5}{6}}}{\gb^3}$};
\node[right] at (5,2.5) {$\frac{\gb^5zx^{\frac{5}{6}}}{\ga}$};
\node[above left] at (-0.4,0.2) {$\ga^{12}x^{\frac{5}{3}}$};
\node[below left] at (-0.4,-0.2) {$\ga^{12}x^{\frac{5}{3}}$};
\node[above right] at (5.4,0.2) {$\frac{\gb^4z^4}{\ga^8}x^{\frac{5}{3}}$};
\node[below right] at (5.4,-0.2) {$\frac{\gb^4z^4}{\ga^8}x^{\frac{5}{3}}$};

\end{tikzpicture}
	\caption{The theory obtained from the compactification of the $SU(2)\times SO(16)$ $5d$ SCFT on a torus with flux $(0,0;1,1,0,0,0,0,0,0)$ preserving a $U(1)\times SU(2)^2\times SO(12)$ symmetry. This theory is obtained by gluing two copies of the tube model in figure \ref{newtube} and is dual to the $(6,2)$ model of subsection \ref{susu62}.}
	\label{newtorus}
\end{figure}
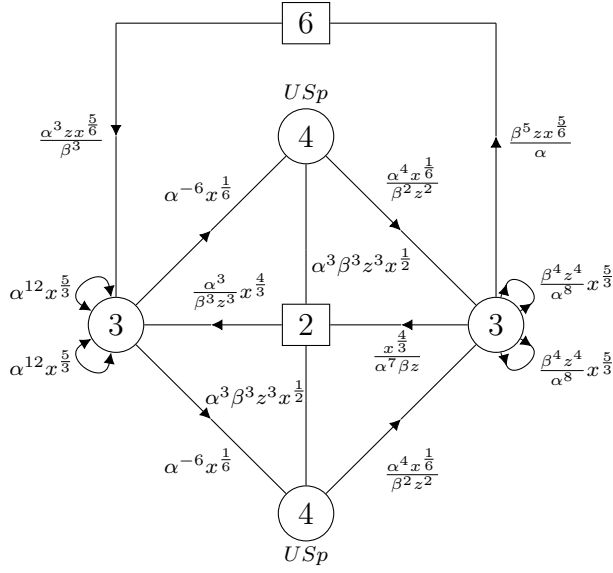

To check this construction, we can $\Phi$ glue two copies of the tube model in figure \ref{newtube} to form the theory corresponding to a torus, see figure \ref{newtorus}. We then expect this theory to be dual to the $(6,2)$ model of subsection \ref{susu62}, obtained by gluing two copies of the basic $SU-SU$ tube. Indeed, calculating the index of the torus theory of figure \ref{newtorus} we find the same expression as in Eq. \eqref{ind62}, under the identification $a=\alpha\beta z$, $b=\alpha^{-2}\beta^{4}$ and $c=\alpha^{-5}\beta z$ (where $a$, $b$ and $c$ here are those appearing in Eq. \eqref{ind62}). 

This result can be understood in terms of the more fundamental duality relating a $USp(4)$ gauge theory with 8 fundamental chirals and a linear monopole superpotential $\mathcal{W}=\mathfrak{M}$ to a Wess--Zumino (WZ) model with 28 chirals $M$ and superpotential $\hat{\mathcal{W}}=\mathrm{Pf}(M)$ \cite{Aharony:2013dha}. Recall indeed that in the model of figure \ref{newtorus} the basic monopoles of the two $USp(4)$ gauge groups are turned on in the superpotential, so we can apply the basic duality to both of them. When we do so, the dual fields become massive together with the two antisymmetrics at each $SU(3)$ gauge node, resulting in the $(6,2)$ model of figure \ref{QuiversSU3X23d} (b). Notice that the duality of \cite{Aharony:2013dha} holds also for generic rank, that is $USp(2N)$ with $2N+4$ fundamental chirals and $\mathcal{W}=\mathfrak{M}$ is dual to a WZ model of $(N+2)(2N+3)$ chirals $M$ with $\hat{\mathcal{W}}=\mathrm{Pf}(M)$, and this is compatible with our construction according to which the models of figures \ref{QuiversSU3X23d} (b) and \ref{newtorus} should be dual for any value of the ranks of the gauge groups $SU(N+1)$ and $USp(2N)$. This is another strong test of our proposal which is passed at any rank. From a different perspective, we can think of our construction as providing a geometric higher dimensional origin for the duality of \cite{Aharony:2013dha}.

\section{Conclusions}
\label{conc}

In this article we have studied the torus compactification of a family of $5d$ SCFTs with flux in their global symmetry leading to $3d$ $\mathcal{N}=2$ theories. The specific family of $5d$ SCFTs is the one UV completing the $5d$ gauge theories $SU(N+1)_k + N_f F$. This study relies on previous analysis of similar compactifications of $6d$ SCFTs to $4d$ and of the rank $1$ $5d$ $E_{N_f+1}$ SCFT, which is the $N=1$ case in the family, to $3d$. Specifically, we used various properties of the $5d$ SCFTs to relate the understanding of such compactifications to the properties of domain walls in $4d$. We then used the understanding of similar domain walls in $5d$, that appear in reductions of $6d$ SCFTs to $4d$, to conjecture the properties of these $4d$ domain walls, which were then used to formulate conjectures for the resulting $3d$ theories.

These conjectures were later tested in several ways, usually by the calculation of various partition functions of the $3d$ theory. Specifically, our conjectures can be used to generate many $3d$ theories that are proposed to be the reduction of the $5d$ SCFT on a torus with flux. As such these are expected to inherit the part of the global symmetry of the $5d$ SCFT not broken by the flux. However, in many cases this symmetry is not manifest in the $3d$ theory so the higher dimensional construction suggests that it is emergent in the IR. This can then be tested by the computation of the superconformal index and the central charges, which can be evaluated from the $S^3$ partition function. Additionally, the higher dimensional construction suggests the presence of operators that are expected to descend from the $5d$ energy-momentum and conserved current multiplets, as well as other Higgs branch chiral ring operators. Finding the presence of these operators then gives an additional consistency check.

Using these methods we were able to propose convincing models for the torus compactification of the classes of $5d$ SCFTs that we mentioned for $N_f+2|k|\geq 2N+2$ and for various values of flux. An interesting observation is that the different models are in many cases related to one another via real masses. Specifically, giving real masses to flavors in a given model leads in some cases to $3d$ models that are themselves compactifications of a $5d$ SCFT in this family but with smaller $N_f$. This is not surprising as the $5d$ SCFTs in the family are interrelated by real mass deformations in $5d$. As both the $5d$ and $3d$ mass deformations can be interpreted as vevs to background gauge fields, it is reasonable that these are related under dimensional reduction. We have also shown how this can be further motivated by understanding the precise relation between the $5d$ and $3d$ global symmetries. The consistency of the resulting flow pattern then provides further evidence for our proposal.

There are various interesting directions for further study. One is to try to exploit our results to also understand the compactification on more general surfaces, notably spheres with three punctures. The idea here is that the family of $5d$ SCFTs we discussed are related to one another by various flows. Some are the mass deformations that we mentioned previously, that preserve the rank of the $5d$ SCFT, but we also have Higgs branch flows that decrease $N$ and so the rank of the $5d$ SCFT. We have noted that $5d$ mass deformations appear to descend to $3d$ mass deformations, and it is interesting to also inquire as to the relation of Higgs branch flows across such dimensional reductions. This was studied in \cite{Razamat:2019mdt} in the context of the compactification of $6d$ SCFTs to $4d$, where it was found that such flows between the higher dimensional theories do reduce to flows between their compactification products up to certain subtleties that we shall next mention. Specifically, we can give a vev to a $4d$ operator which is a descendant of the $6d$ operator whose vev triggers the $6d$ flow, and that would lead to a new $4d$ theory, which is itself the result of the compactification of the $6d$ SCFT at the end of the $6d$ flow. The subtle issue here is that the compactification surface can have more punctures than the compactification surface appearing for the pair before the flow. It would be interesting to explore such relations also in the context of the compactification of $5d$ SCFTs. 

If these work out similarly to the $6d$ case, then they might also be exploited to understand compactifications on more general surfaces. Specifically, we pointed out that the $5d$ SCFTs in this family are interrelated by Higgs branch flows. Given that we now understand the reduction of such SCFTs on spheres with two punctures, we can then use this method to presumably generate $3d$ theories associated with the compactification on spheres with more than two punctures. This can be done by giving a vev to the operator descendant of the Higgs branch operator whose vev generates the $5d$ flow. This should then lead to a $3d$ model corresponding to the compactification of a $5d$ SCFT in this family, but with the rank decreased and on a Riemann surface with more than two punctures. Such a strategy was successfully employed to understand the compactification of various $6d$ SCFTs on spheres with three punctures in \cite{Razamat:2019ukg,Sabag:2020elc}, and it would be interesting to explore its application also to the case of compactifications of $5d$ SCFTs.       

Another interesting direction is instead to further explore the $3d$ consequences of our construction. Specifically, we expect our construction to lead to many cases of symmetry enhancement and duality, some of which have appeared in our discussion. It would be interesting to further explore this. For instance, we expect our construction to lead to many interesting dualities between different geometric constructions of the same surface. It would be interesting to systematically explore this, particularly, to see whether this leads to new $3d$ dualities.

Here we have explored the domain wall approach for the cases where the theories in the $4d$ bulk are either $SU(N+1)$ or $USp(2N)$ gauge groups, but have not looked at cases where the bulk gauge group is $SU(2)^N$. It would be interesting to also explore such cases. In the context of compactifications of $6d$ theories to $4d$, cases similar to this were studied in \cite{Kim:2018lfo}, and it led to interesting new $4d$ dualities. It would thus be very interesting to also explore this for the $3d$ case.

It would also be interesting to see if further checks of our proposal can be made. Some notable directions in this regard include the computation of partition functions of $5d$ SCFTs on a surface times a three manifold, which can in principle be evaluated using the techniques of \cite{Crichigno:2018adf,Hosseini:2018uzp,Hosseini:2021mnn}. These can then be compared against the partition functions of the $3d$ theories on the three manifold. Another interesting direction is to consider the discrete symmetries of these SCFTs, notably as these can have non-trivial 't Hooft anomalies. Various such aspects of $5d$ SCFTs were studied for instance in \cite{Morrison:2020ool,Albertini:2020mdx,Bhardwaj:2020phs,BenettiGenolini:2020doj,Apruzzi:2021vcu}. Here, our main interest is understanding how these behave under dimensional reductions. Specifically, it is known that anomalies in continuous symmetries are related between the parent and descendant theories in cases of compactifications, and it would be interesting to see if something similar also holds for the case of discrete symmetries. If true then this can be used to further test such proposals.

Finally, given that our results apply for the class of $5d$ SCFTs that UV complete the $SU(N+1)_k+N_fF$ theories of arbitrary rank, they could be used for the purposes of holography which holds at large $N$. It would indeed be interesting to understand the RG flow from five to three dimensions from a holographic perspective. Some works on the flow from AdS${}_6$ to AdS${}_4$ solutions have been done in \cite{Naka:2002jz,Bah:2018lyv,Hosseini:2018usu,Legramandi:2021aqv}. Our Lagrangian descriptions of the $3d$ theories obtained from the compactification of the $5d$ SCFTs could be useful for performing explicit computations of observables that can then be matched with the supergravity side.

\section*{Acknowledgments}
We would like to thank Yaron Oz and Shlomo Razamat for useful conversations. MS and GZ are supported in part by the ERC-STG grant 637844-HBQFTNCER and by the INFN. The research of MS is supported in part by the University of Milano-Bicocca grant 2016-ATESP0586, in part by the MIUR-PRIN contract 2017CC72MK003, in part by ERC Consolidator Grant \#864828 “Algebraic Foundations of Supersymmetric Quantum Field Theory (SCFTAlg)” and in part by the Simons Collaboration for the Nonperturbative Bootstrap under grant \#494786 from the Simons Foundation. GZ is also partially supported by the Simons Foundation grant 815892. OS is supported in part by Israel Science Foundation under grant no. 2289/18, by the I-CORE Program of the Planning and Budgeting Committee, by BSF grant no. 2018204, by grant No. I-1515-303./2019 from the GIF (the German-Israeli Foundation for Scientific Research and Development), by the Clore Scholars Programme and by the Mani L. Bhaumik Institute for Theoretical Physics at UCLA. 

\bibliographystyle{JHEP}
\bibliography{ref}

\end{document}